\documentclass[hyper,12pt,A4paper]{article}
\usepackage{jheppub}

\input xy
\xyoption{all}
\input xy
\xyoption{all}

\usepackage{graphicx}
\usepackage{latexsym,amsmath,amsfonts,amssymb}
\usepackage{mathrsfs}
\usepackage[makeroom]{cancel}
\usepackage{bbm}
\usepackage{bm}
\usepackage{subfigure}
\usepackage{paralist}
\usepackage{url}

\numberwithin{equation}{section}

\newcommand{\smat}[1]{{\small\begin{pmatrix} #1 \end{pmatrix}}}

\newcommand{\be}{\begin{equation}} 
\newcommand{\ee}{\end{equation}}
\newcommand{\bea}{\begin{equation} \begin{aligned}} \newcommand{\eea}{\end{aligned} \end{equation}}

\newcommand{\bit}{\begin{itemize}} 
\newcommand{\eit}{\end{itemize}}

\newcommand{\Z}{\mathbb{Z}}
\newcommand{\C}{\mathbb{C}}
\newcommand{\R}{\mathbb{R}}

\renewcommand{\t}{\widetilde }

\renewcommand{\d}{\delta }

\newcommand{\half}{{1\over 2}}

\newcommand{\CA}{\mathcal{A}}
\newcommand{\CB}{\mathcal{B}}
\newcommand{\CC}{\mathcal{C}}

\newcommand{\CE}{\mathcal{E}}
\newcommand{\CF}{\mathcal{F}}

\newcommand{\CH}{\mathcal{H}}
\newcommand{\CI}{\mathcal{I}}
\newcommand{\CJ}{\mathcal{J}}
\newcommand{\CK}{\mathcal{K}}

\newcommand{\CM}{\mathcal{M}}
\newcommand{\CN}{\mathcal{N}}
\newcommand{\CO}{\mathcal{O}}

\newcommand{\CQ}{\mathcal{Q}}

\newcommand{\CS}{\mathcal{S}}
\newcommand{\CT}{\mathcal{T}}

\newcommand{\CW}{\mathcal{W}}

\newcommand{\CZ}{\mathcal{Z}}

\newcommand{\FR}{\mathfrak{R}}

\newcommand{\GG}{\mathbf{G}}

\newcommand{\ov}{\over}

\newcommand{\Ext}{{\rm Ext}}

\newcommand{\bE}{{\bf E}}
\renewcommand{\P}{{\mathbb{P}}}

\newcommand{\h}{\widehat}

\newcommand{\MG}{{\mathbf X}} 
\newcommand{\FT}{{\mathcal{T}_\MG}} 
\newcommand{\CAT}{{\mathscr T}^{\rm BPS}_{\FT}}
\newcommand{\KK}{D_{S^1}}

\newcommand{\CUL}{\CM^C}

\newcommand{\TX}{{\Delta_{\MG}}}

\newcommand{\bd}{{\boldsymbol{\delta}}}

\title{On 5d SCFTs and their BPS quivers. \\ Part I: B-branes and brane tilings}

\author[\flat]{Cyril Closset,}
\author[\sharp]{Michele Del Zotto}

\affiliation[\flat]{ Mathematical Institute, University of Oxford\\ Woodstock Road, Oxford, OX2 6GG, United Kingdom}
\affiliation[\sharp]{Department of Mathematical Sciences and Centre for Particle Theory\\ Durham University, Durham DH1 3LE, UK}

\abstract{We study the spectrum of BPS particles on the Coulomb branch of  five-dimensional superconformal field theories (5d SCFTs)  compactified on a circle. By engineering these theories in M-theory on ${\mathbf X} \times S^1 $, for ${\mathbf X}$ an isolated Calabi-Yau threefold singularity, we naturally identify the BPS category of the 5d theory on a circle with the derived category of coherent sheaves on a resolution of ${\mathbf X}$. It follows that the BPS spectrum can be studied in terms of 5d BPS quivers, which are the  fractional-brane quivers for the singularity ${\mathbf X}$.  5d BPS quivers generalize the well-studied 4d BPS quivers for 4d $\mathcal{N}{=}2$ gauge theories that can be obtained from ${\mathbf X}$ in so-called geometric engineering limits. We study the interplay between 4d and 5d BPS quivers in detail. We particularly focus on examples when ${\mathbf X}$ is a toric singularity, in which case the 5d BPS quiver is given in terms of a brane tiling. For instance, the well-studied $Y^{p,q}$ brane tiling gives a 5d BPS quiver for the  $SU(p)_q$ 5d gauge theory. We present a conjecture about the structure of the BPS spectra of a wide class of models, which we test in the simple case of the 5d $SU(2)_0$ theory (more precisely, the $E_1$ SCFT). We also argue that 5d UV dualities can be realized in terms of mutation sequences on the BPS quivers, which are in turn interpreted as autoequivalences of the BPS category.}

\begin{document}
\maketitle

\section{Introduction}
The characterization of the Coulomb-branch BPS spectrum of 5d SCFTs is a challenging problem. Much of this challenge arises from the presence of BPS strings in five dimensions. In this work, we address a closely related (and, in a sense, more general) problem, by studying the BPS spectrum of 5d SCFTs compactified on a circle. In this context, the 5d BPS strings have an image in terms of ordinary 4d BPS particles in the 4d KK theory so obtained. The spectrum of BPS particles can be characterized in terms of the $\CN=4$ supersymmetric quantum mechanics (SQM) on their worldline. Often, such families of SQMs have a quiver description in terms of a so-called {\it 5d BPS quiver.} In this work, we exploit geometric engineering to derive such  quivers and to realize the BPS particles explicitly in terms of D-branes. We also use the quiver formalism to study the simplest examples of 5d RG flows, as well as 5d dualities, at the level of the BPS Hilbert space, and we present a conjecture about the structure of the full BPS spectrum for a large class of theories.

The superselection sector of the Hilbert space of the KK theory with zero string number proves to be a useful tool for understanding the dynamics of these systems, probing dualities, and unveiling the higher-dimensional origin of various phenomena proper to genuinely 4d theories.

\subsubsection*{5d SCFTs and M-theory}

Five-dimensional superconformal quantum field theories (5d SCFTs)  arise naturally within string theory. For instance, the 5d low-energy theory transverse to a canonical threefold singularity in M-theory (upon decoupling gravity) defines such a 5d fixed point \cite{Seiberg:1996bd, Morrison:1996xf, Douglas:1996xp, Intriligator:1997pq}. Conjecturally, every 5d SCFT, $\FT$, corresponds in this way to an isolated Calabi-Yau (CY) threefold singularity, $\MG$:
\be
\MG \qquad \longleftrightarrow\qquad  \FT~.
\ee
See {\it e.g.}~\cite{Cherkis:2014vfa, DelZotto:2017pti,Xie:2017pfl,Alexandrov:2017mgi,Jefferson:2017ahm,Jefferson:2018irk,Bhardwaj:2018yhy,Bhardwaj:2018vuu,Apruzzi:2018nre, Closset:2018bjz, Bhardwaj:2019jtr, Apruzzi:2019vpe, Apruzzi:2019opn, Apruzzi:2019enx, Bhardwaj:2019xeg, Saxena:2019wuy, Apruzzi:2019kgb}  for recent discussions. 
There is then an interesting correspondence between the geometry of $\MG$ and questions of 5d $\CN=1$ supersymmetric quantum field theory. 
In particular, the extended Coulomb branch (VEVs and real masses) of $\FT$ is identified with the extended K\"ahler cone of $\MG$.
A generic point on the extended Coulomb branch is then identified with a complete resolution of the singularity, $\pi : \h\MG \rightarrow  \MG$, with a particular choice of K\"ahler moduli. The low-energy physics at that point then captures the low-energy excitations of M-theory on the smooth local CY threefold~$\h\MG$. The low-energy BPS spectrum in 5d consists of both particles (wrapped M2-branes) and strings (wrapped M5-branes). 

Let us consider $\FT$ on a finite-size circle; this gives us a 4d $\CN=2$ Kaluza-Klein (KK) field theory, denoted by $\KK\FT$. By the IIA string theory/M-theory duality, this 4d $\CN=2$ KK theory is precisely engineered  by IIA string theory on $\MG$. In this paper, we are interested in the spectrum of half-BPS particles on the Coulomb branch of $\KK\FT$, taking full advantage of the IIA string-theory embedding.

\subsubsection*{5d BPS quivers and BPS category for $\KK\FT$}
We will identify a family of quivers, called {\it 5d BPS quivers} and denoted by $\CQ_\MG$, from which the full  BPS spectrum of $\KK\FT$ can be recovered, in principle, by standard representation-theoretic methods. The 5d BPS quiver is generally known as the {\it fractional-brane quiver} in the physics literature:
\be
\text{5d BPS quiver for} \; \KK\FT  \qquad \longleftrightarrow\qquad  \text{fractional-brane quiver for} \; \MG~.
\ee
  More generally, we identify the (triangulated) category of BPS states of $\KK\FT$, denoted by $\CAT$, with the derived category of coherent sheaves on $\h\MG$:
\be\label{CAT intro}
\CAT  \qquad \longleftrightarrow\qquad  D^b(\h\MG)~.
\ee
Indeed, the BPS particles correspond to D-branes wrapped over holomorphic cycles inside $\h\MG$, which are famously described as objects in the derived category of $\h\MG$. Importantly, the triangulated category $\CAT$ is independent of the Coulomb moduli and mass parameters; in particular, it could be computed at the superconformal point. Physically, the 5d BPS quiver is the abstract quiver corresponding to the low-energy $\CN=4$ supersymmetric quantum mechanics (SQM) on the worldline of D0-branes transverse to $\MG$. Mathematically, the BPS category \eqref{CAT intro} can be recovered as the derived category of quiver representations of $\CQ_\MG$. Of course, the interplay between quivers and BPS categories has a rich history, see {\it e.g.} \cite{Aspinwall:2009isa}.

A particularly simple class of examples, which we will study in detail, are the {\it toric} CY$_3$ singularities. In that case, 5d BPS quivers can be derived systematically from  the toric data of~$\MG$, using well-studied {\it brane tilings} techniques \cite{Hanany:2005ve, Franco:2005rj, Feng:2005gw}.

Whenever two 5d SCFTs can be related by an RG flow triggered by a mass deformation, corresponding to a partial resolution of the singularity, their 5d BPS quivers are related as well, by some 1d RG flow in the $\CN=4$ SQM triggered by taking the appropriate  Fayet-Iliopoulos (FI) parameters to be very large (and leading to a particular ``Higgsing'' of the 1d gauge group). For toric singularities, this RG flow can be most easily understood in terms of the brane tiling \cite{Franco:2005rj}.

In this paper, we take the first steps towards a systematic study of the BPS spectrum of $\KK\FT$ from their BPS quivers. In the simplest case of a free 5d hypermultiplet (corresponding to the conifold singularity), our results agree with recent results \cite{Banerjee:2019apt} that use distinct methods based on  spectral network~\cite{Gaiotto:2012rg, Eager:2016yxd, Banerjee:2018syt}. We also discuss the spectrum of the rank-one $E_1$ SCFT, and present a conjecture for the full spectrum in a so-called tame chamber, in a large class of examples.

\subsubsection*{Gauge-theory phases, 4d $\CN=2$ limits, subquivers and subcategories}
Oftentimes, the low-energy description on the extended Coulomb branch of $\CT_\MG$ can be written as the Coulomb-branch theory of a 5d $\CN=1$ non-abelian gauge theory, with the 5d inverse gauge couplings identified with massive-deformation parameters of the 5d SCFT---this we call a ``gauge-theory phase'' of $\FT$. Any 5d $\CN=1$ gauge theory consists of a vector multiplet in the adjoint representation of a gauge group $\GG$, and of hypermultiplets in representations $\FR$ of $\GG$, together with a choice of 5d Chern-Simons terms \cite{Seiberg:1996bd} and/or $\theta$ angles \cite{Douglas:1996xp, Bergman:2013ala}, depending on the choice of gauge group. Such a theory can be dimensionally reduced to a 4d $\CN=2$ gauge theory with the same gauge group and matter content.

Therefore, the 5d BPS quiver, describing the BPS states of the 5d theory on a finite-size circle in a gauge-theory phase, must be closely related to the 4d BPS quiver for the corresponding 4d $\CN=2$ gauge theory  \cite{Fiol:2000wx, Fiol:2000pd, Denef:2002ru, Cecotti:2010fi, Cecotti:2011rv,Alim:2011kw}. Indeed, historically speaking, some of the 4d BPS quivers studied {\it e.g.} in~\cite{Alim:2011kw} were first derived by ``decoupling some nodes'' from some fractional-brane quivers $\CQ_\MG$ \cite{Fiol:2000wx, Fiol:2000pd}, which is also how the BPS quiver for the 4d $\CN=2$ $T_N$ theory was derived in \cite{Alim:2011kw}. Building upon that idea, in this work,  we consider the full fractional-brane quiver $\CQ_\MG$ directly as a 5d BPS quiver.

We would like to insist on this elementary point: type IIA string theory on $\MG$ does not engineer a strictly 4d $\CN=2$ gauge theory, but instead it directly engineers the 5d SCFT on a circle, $\KK\FT$. This was already fully appreciated long ago \cite{Lawrence:1997jr} in 5d gauge-theory language. Incidentally, $\KK\FT$ does not always have a gauge-theory interpretation. For instance, the BPS spectrum of D-branes at the resolved $\C^3/\Z_3$ singularity \cite{Douglas:2000qw} corresponds to the BPS spectrum on the Coulomb branch of the rank-one 5d SCFT $E_0$~\cite{Morrison:1996xf} compactified on a circle, which does not have any gauge-theory phase.

By contrast to the 4d $\CN=2$ BPS quivers, 5d BPS quivers have additional structure corresponding to the richer charge lattice of the 5d theory. Consider a gauge-theory phase with gauge group $\GG= \prod_{k=1}^s G_k$ of total rank $r= \sum_k {\rm rank}(G_k)$, where each $G_k$ is a simple gauge group. The 5d SCFT has a flavor symmetry group of rank $f= f_{\rm 4d} + s$, where $f_{\rm 4d}$ would be the rank of the flavor group in 4d, and the additional $s$ factors in the maximal torus of the flavor symmetry group correspond to topological $U(1)$ symmetries in the gauge-theory phase, one for each factor $G_k$ (the symmetry at the fixed point could be further enhanced). While the charge lattice of the 4d $\CN=2$ gauge theory has dimension $D_{\rm 4d} = 2 r+ f_{\rm 4d}$, the charge  lattice of $\KK\FT$ has dimension $D= 2r +f +1$, corresponding to the electric and magnetic charges, the 5d flavor symmetry, and the KK momentum.

When a gauge-theory phase exists, the 4d $\CN=2$ gauge theory is obtained in a so-called {\it geometric engineering limit} \cite{Katz:1996fh}, which is a scaling limit in K\"ahler moduli space that decouples both the KK towers and the instanton particles \cite{Lawrence:1997jr, Chuang:2013wt}. At the level of BPS quivers, the 4d $\CN=2$ gauge-theory BPS quiver, with $D_{\rm 4d}$ nodes, is a subquiver of the 5d BPS quiver  $\CQ_\MG$ with $D$ nodes, which is simply obtained by removing the appropriate $s+1$ nodes that carry the KK and instanton charges.

At the level of the BPS categories, the strict 4d $\CN=2$ categories (see {\it e.g.} \cite{Cecotti:2012va, Chuang:2013wt}) are controlled subcategories \cite{LenzingICTP} of the 5d BPS category $\CAT$. These 4d BPS categories were studied from this same geometric point of view in \cite{Chuang:2013wt}.

More generally, 5d gauge-theory phases correspond to a choice of {\it heart} of the triangulated category $\CAT$; this correspond to a particular quiver $\CQ_\MG$, that contains the corresponding 4d $\CN=2$ quiver as a subquiver. There can be several distinct such choices, for different gauge-theory phases. In that language, the statement of ``UV duality'' between different 5d gauge theories \cite{Bergman:2013aca, Bergman:2014kza} is the statement that the two categories of quiver representations are derived equivalent. In the examples we studied, this is ensured by the fact that the two ``UV dual'' 5d BPS quivers are mutation equivalent.

Finally, we will also introduce the concept of an electric category for $\KK\FT$ when the latter possesses a gauge-theory phase, similarly to \cite{Cecotti:2012va}. The electric category is a subcategory of $\CAT$, which contains only mutually local objects and which essentially encodes the BPS states coming from BPS particles in 5d (as opposed to magnetic BPS strings). 

\subsubsection*{Outlook and summary}
This work initiates a systematic study of the BPS category $\CAT$ of $\KK\FT$, and of its associated 5d BPS quivers. There remains many interesting directions for further research. Here, let us just mention a few of them \cite{PROG}:
\bit
\item While this work is mostly focussed on the IIA engineering, much insight can be gained from considering the mirror CY geometry in IIB string theory, building upon \cite{Hori:2000ck}. For instance, this allows us to discuss more systematically the uplift of 4d $\CN=2$ theories to 5d SCFTs through their BPS quivers \cite{CDZtoappear}. 

\item A much more systematic understanding of the 5d BPS spectrum would be desirable, and the 5d BPS quiver framework is a promising avenue to achieve this goal. Relatedly, we can then proceed to study interesting invariants of the SCFTs from the knowledge of the BPS spectrum at a convenient point, such as, for instance, the $S^3 \times T^2$ partition function, along the lines of \cite{Iqbal:2012xm,Cordova:2015nma}. 

\item In this paper, we mostly discuss 5d field theories $\FT$ engineered by toric singularities, but the 5d BPS quiver perspective is completely general, and can shed much light onto the general classification of 5d SCFTs. Relatedly, it would be interesting to derive the 5d BPS quiver directly from a IIB $(p,q)$-web \cite{Aharony:1997ju, Aharony:1997bh} along the lines of \cite{Xie:2012dw,Xie:2012jd}, since many interesting 5d SCFTs can be more conveniently engineered in that way---see {\it e.g.}  \cite{DeWolfe:1999hj,Benini:2009gi,Bao:2013pwa, Zafrir:2014ywa,Hayashi:2014hfa, Zafrir:2015ftn,Hayashi:2016jak,Hayashi:2017btw,Hayashi:2018bkd, Hayashi:2018lyv,Hayashi:2019yxj} .

\item The M-theory geometric engineering picture is not {\it a priori} restricted to 5d SCFTs. One could similarly study interesting 3d $\CN=2$ field theories by considering M-theory at an isolated CY fourfold singularity. In that case, the BPS particles have an $\CN=2$ SQM on their worldline and the BPS quiver is a graded quiver with potential of the type studied in \cite{Franco:2015tna, Franco:2016nwv, Closset:2017yte, Franco:2017lpa, Eager:2018oww, Closset:2018axq}.
\eit

\medskip
\noindent
This paper is organized as follows. In section~\ref{sec:5dBPSbasics}, we discuss  in full generality the BPS category of the KK theory $\KK\FT$ and the associated 5d BPS quiver, and its relation to genuine 4d BPS quivers. In section~\ref{sec:branetilings}, we discuss 5d BPS quivers for 5d SCFTs engineered at toric singularities, which are encoded in brane tilings. In section~\ref{sec:rank1}, we illustrate our general methods with a study of the rank-one toric SCFTs and their 5d BPS quivers, the so-called toric del Pezzo quivers. In section~\ref{sec:rank1}, we study some rank-two examples and comment on the phenomenon of UV duality at the level of the BPS quivers. In section~\ref{sec:rankn}, we briefly discuss the BPS quivers associated to the 5d $SU(p)_q$ gauge theory. 
In section~\ref{sec:BPSspectrum}, we initiate the systematic study of the spectrum of $\KK\FT$ from the representation theory of the quiver. Some additional comments and computations are collected in Appendix.

\section{5d SCFTs on a circle, BPS categories and BPS quivers}\label{sec:5dBPSbasics}
Let us consider five-dimensional SCFTs corresponding to isolated CY$_3$ singularities, denoted by $\MG$, via geometric engineering in M-theory:
\be\label{eq:Mthyeng}
\text{M-theory on }\,  \mathbb R^{1,4} \times \MG\qquad  \longleftrightarrow\qquad \FT\, \in {\rm SCFT}_{5}\,.
\ee
To any complex codimension-three canonical Gorenstein threefold singularity $\MG$, one thus assigns a 5d SCFT $\FT$~\cite{Morrison:1996xf}. One can also consider any crepant (partial) resolution of the singularity:
\be
\pi\; : \h \MG \rightarrow \MG~,
\ee
which should be a deformation away from the SCFT.
The extended K\"ahler cone of the resolved singularity, $\h\CK(\MG)$, consisting of all possible resolutions indexed by their K\"ahler parameters \cite{Witten:1996qb}, is identified with the extended Coulomb branch of the SCFT $\CT_\MG$ (including both Coulomb-branch VEVs and mass parameters for flavor symmetries). 
Importantly, the theory $\FT$ is partly characterized by two non-negative integers $r$ and $f$, defined by:
\be\label{def r and f}
r+f= {\rm dim}\, H^{\rm cpt}_2(\h\MG, \Z)~, \qquad  r= {\rm dim}\, H^{\rm cpt}_4(\h\MG, \Z)~,
\ee
for any complete crepant resolution $\h\MG$. Here, $r$ is the number of exceptional four-cycles, and $r+f$ is the number of exceptional two-cycles. In terms of the SCFT $\FT$:
\begin{itemize}
\item $r$ is the rank of the SCFT $\FT$---{\it i.e.} the dimension of its Coulomb branch. 
\item  $f$ is the rank of the flavor symmetry $G_F$ of $\FT$.
\end{itemize}
Note that the 5d Coulomb branch is a real manifold, of real dimension $r$. The extended Coulomb branch has dimension $r+f$.
Note also that the full global symmetry of $\FT$ is $G_F \times SU(2)_R$, where the second factor is the $R$-symmetry. 
We refer to~\cite{Closset:2018bjz} for a recent review of the M-theory construction of $\FT$ and of its 5d gauge-theory phases. 

\subsection{Four-dimensional KK theory and IIA string-theory embedding}
In this work, we are interested in compactifying $\FT$ on a finite-size circle. The resulting theory is a four-dimensional Kaluza-Klein (KK) theory, with 4d $\CN=2$ supersymmetry, which we denote by $\KK\FT$.  We then have a direct correspondence between the 4d KK-theory and type-IIA string theory on~$\MG$ \cite{Lawrence:1997jr}:
\be
\begin{aligned}
\KK\FT\quad \,\,&\longleftrightarrow \,\, \quad\text{M-theory on }  S^1 \times \mathbb R^{1,3} \times \MG \\
& \longleftrightarrow\,\, \quad \text{IIA on } \mathbb R^{1,3} \times \MG
\end{aligned}
\ee
Therefore, on general grounds, the low-energy limit of type IIA on $\MG$ gives us a 4d $\CN=2$ supersymmetric KK field theory. We have the usual expressions for the IIA parameters:
\be\label{eq:IIAMthy}
R = \ell_s g_s~, \qquad \qquad  {\ell_p^3\ov R}= \ell_s^2 = \alpha'~,
\ee
with $R$ the M-theory circle radius and $\ell_p$ the 11d Planck length. 
To obtain genuine 4d $\CN=2$ theories and decouple the tower of KK modes (and, generally, instanton particles), one has to take an additional scaling limit in K\"ahler moduli space as we send $R \rightarrow 0$. For 4d $\CN=2$ gauge theories, this is generally called the geometric engineering limit \cite{Katz:1996fh}; see \cite{Lawrence:1997jr, Chuang:2013wt} for detailed accounts.

The KK theory $\KK \FT$ has a parameter space of complex dimension $r+f$, corresponding to a complexification of the extended K\"ahler cone $\h\CK(\MG)$. 
This parameter space is identified with the {\it extended Coulomb branch} of $\KK\FT$,  and denoted by $\CM^C$.  The proper Coulomb branch $\CB_r$ of the 4d KK theory, of complex dimension $r= {\rm rank}(\FT)$, is fibered over the mass-deformation parameters:
\be\label{ext CB MC of KKFT}
\CB_r \rightarrow \CM^C \rightarrow \{m \}~.
\ee
 The low-energy physics on $\h \MG$ in the large volume limit can be extracted from IIA supergravity, while worldsheet instantons contribute non-trivial corrections to the Coulomb-branch metric. The Seiberg-Witten (SW) geometry for the KK theory is captured by the versal deformations of the IIB mirror singularity $\MG^\vee$, where the holomorphic top-form plays the role of the SW differential. The mirror geometry is extremely useful in order to study the Coulomb-branch low-energy physics, as it does not receive quantum corrections.

\medskip

\noindent Explicit examples can be easily obtained in the case when $\MG$ (or $\h \MG$) is toric. Then, $\MG$ is defined in terms of a strictly convex (triangulated) toric diagram with $r$ internal points and $f+3$ external points. The Hori-Vafa (HV) mirror gives us an explicit description of the mirror geometry as a fibration of $\C^\ast \times \Sigma_r$ over a plane, with $\Sigma_r$ a Riemann surface of genus $r$~\cite{Hori:2000kt, Feng:2005gw}. We have the key identification:
\be
\text{Seiberg-Witten curve of}\;  \KK\FT  \qquad  \longleftrightarrow\qquad \text{curve} \; \Sigma_r \;  \text{in HV mirror} \;\MG^\vee~.
\ee
This type of KK theory SW curves are well-known: the first examples were obtained by Nekrasov, by uplifiting 4d $\CN=2$ gauge theory results to 5d using the relation between SW geometry and integrable systems~\cite{Nekrasov:1996cz}. Note that, in our discussion so far, we are not referring to any particular gauge theory description. Instead, we are considering the full extended Coulomb branch \eqref{ext CB MC of KKFT} of $\KK\FT$, not only the chambers corresponding to 5d gauge theory phases.

\subsection{BPS states in M-theory and type IIA}
The BPS states on the real Coulomb branch of the five-dimensional theory arise from wrapped M-branes inside $\h \MG$: M2-branes wrapped over a curve $\CC$ give rise to BPS particles, and M5-branes wrapped over a compact divisor $\bE$ (an exceptional four-cycle)  give rise to BPS strings, with mass and tension:
\be
M(\text{M2 on}\, \CC)= {{\rm vol}(\CC) \ov  {\ell_p}^3}~, \qquad T(\text{M5 on}\, \bE)= {{\rm vol}(\bE) \ov  {\ell_p}^6}~,
\ee
respectively. From \eqref{def r and f}, we see that there are $r+f$ distinct types of 5d BPS particles and $r$ distinct 5d BPS strings; the latter are magnetically charged under the $r$ 5d abelian dynamical gauge fields.  We stress here that the M2- and M5-branes have to wrap holomorphic cycles in the geometry, which are cycles calibrated by the K\"ahler form, in order to satisfy the half-BPS condition. Upon compactifications on a circle, they give rise to BPS particles from wrapped D2 and D4-branes, respectively. More precisely, the M2-brane can wrap the $S^1$, which corresponds to the four-dimensional instantons of $\KK\FT$, or it can remain transverse and give us the D2-brane state. Similarly, the M5-brane gives rise to both the D4-branes and to 4d BPS strings from wrapped NS5 branes. In addition, KK momentum along the circle gives rise to D0-branes. Semi-classically, therefore, we have the following scales:
\begin{itemize}\item BPS particle masses:
\be\label{eq:MparticlesIIA}
\begin{aligned}
&M(\text{D0 on pt})= {1\ov R}~,\\
&M(\text{D2 on } \CC)= {{\rm vol}(\CC) \ov  R \,\alpha^{\prime}}~,\\
&M(\text{D4 on } \bE)= {{\rm vol}(\bE) \ov  R\, {\alpha'}^2}~.
\end{aligned}
\ee
\item BPS string tensions:
\be\label{eq:MparticlesIIA2}
T(\text{NS5 on}\; \bE)= {{\rm vol}(\bE) \ov  (R  \, {\alpha'})^2}~.
\ee
\item Worldsheet instanton corrections:
\be\label{eq:MparticlesIIA3}\tau_0(\text{F1 on}\; \CC)={{\rm vol}(\CC) \ov  \alpha'}~.
\ee
\end{itemize}
Here, $\tau_0$ denotes the worldsheet instanton tension---in particular, it can be identified with a 4d inverse gauge coupling if $\CC$ corresponds to the base curve of a ruling of the exceptional locus of $\h \MG$. Note that, in equations \eqref{eq:MparticlesIIA}, \eqref{eq:MparticlesIIA2} and \eqref{eq:MparticlesIIA3}, the volumes ${\rm vol}(\CC)$ and ${\rm vol}(\bE)$ are understood to be of order $\alpha'$ and $(\alpha')^2$, respectively, but their precise form depends the extended Coulomb branch metric, which is quantum-corrected. For this reason, it is useful to turn to the mirror geometry, where the periods of the holomorphic top form compute the exact values of the D-brane central charges which control the masses of the corresponding BPS particles.

\subsection{The category of BPS states and 5d BPS quivers}
Having set the stage, we are now ready to discuss the BPS particles of the 4d $\CN=2$ KK theory $\KK\FT$. In the following, we introduce the notion of a category of BPS states for $\KK\FT$, denoted by $\CAT$. This is a very useful notion because it is purely algebraic. The actual BPS states are obtained by quantizing the moduli spaces of stable objects of  $\CAT$, for any given stability condition corresponding to a point on $\CM^C$. 

\subsubsection{5d BPS states and B-branes}
The BPS particles on the four-dimensional Coulomb branch of $\KK \CT_\MG$ are precisely the wrapped D0-, D2- and D4-brane states, which can form highly non-trivial bound states when the volumes ${\rm vol}(\CC)$,  ${\rm vol}(\bE)$ are small. 
In order to retain analytic control,  we consider the limit of vanshing string coupling, $g_s\rightarrow 0$.  This is the limit in which the D-branes wrapped over holomorphic cycles are accurately described as {\it B-branes}. We take this limit while keeping the radius $R$ finite. By \eqref{eq:IIAMthy}, this corresponds to taking $\alpha' \rightarrow \infty$. Fortunately, the B-branes are boundary conditions in the topological B-model, which does not depend on the  $\alpha'$ corrections \cite{Witten:1988xj}. At any given point on the extended Coulomb branch, $u \in \CM^C$, the B-branes $\CE$ are objects of the (bounded) derived category of (compactly supported) coherent sheaves on the corresponding crepant resolution, denoted by $\h\MG_u$ \cite{Kontsevich:1994dn, Sharpe:1999qz, Douglas:2000gi}:\footnote{We refer to \protect\cite{Sharpe:2003dr, Aspinwall:2004jr} for thorough reviews of B-branes.}
\be
\CE \in D^b(\h\MG_u)~.
\ee
This B-brane category gives us a standard notion of a {\it category of Coulomb-branch BPS states} of ${\rm D}_{S^1} \CT_\MG$. We write this as:
\be\label{key rel of cat}
\boxed{
\phantom{\Big|}\CAT  \quad  \cong \quad  D^b(\h\MG_u)~.}
\ee
In particular, this tells us that the category of BPS states of ${\rm D}_{S^1} \CT_\MG$ is a triangulated $3$-CY category \cite{Kontsevich:2006jb,Ginzburg:2006fu}.  Note that, while we indicated an explicit dependence on $u\in \CM^C$ in \eqref{key rel of cat}, the category $\CAT$ is independent of $u$, since $D^b(\h\MG)$ is itself independent of the K\"ahler parameters; it is also independent of the K\"ahler  chamber of $\h\MG$.\footnote{$\,$ In string theory, once we identify the B-brane category with the derived category, this is a consequence of the independence of the B-model on the K\"ahler moduli \protect\cite{Witten:1988xj}. 
That the derived category of $\h\MG$ stays the same upon flops was discussed in \protect\cite{1995alg.geom..6012B, bergh2002noncommutative}. }

 \medskip
 
 \noindent 
 Once we have identified an appropriate category, $\CAT$, for describing the BPS states of $\KK\FT$, the problem of determining the BPS spectrum of the theory, at any given point of its parameter space, can be translated into a problem of stability for the objects in the category \cite{Douglas:2000ah, Bridgeland:2007wo}. (This holds abstractly for any category of BPS states, independently of the particular geometric realization \eqref{key rel of cat} studied here.) Let us summarize the general procedure:

\paragraph{Charge lattice.}  First of all,  the Grothendieck group of the BPS category is identified with the charge lattice, $\Gamma$, of quantized conserved charges of the 4d $\CN=2$ KK theory:
\be
\boxed{\phantom{\Big|}
\Gamma \quad \cong \quad K_0\CAT \quad \cong \quad  \mathbb Z^D}~,
\ee
where:
\be\label{eq:quivernodes}
D = 2r + f +1~.
\ee
The $D$ generators consist of $r$ magnetic charges corresponding to the D4-brane charges, $r$ electric charges corresponding to the D2-brane charges associated to curves whose K\"ahler parameters give the Coulomb moduli of the 5d theory, $f$ flavor charges corresponding to the D2-brane charges associated to curves whose K\"ahler parameters give the deformations of the 5d theory, and the KK charge corresponding to the D0-brane charge.

\begin{figure}
\begin{center}
\includegraphics[scale=0.35]{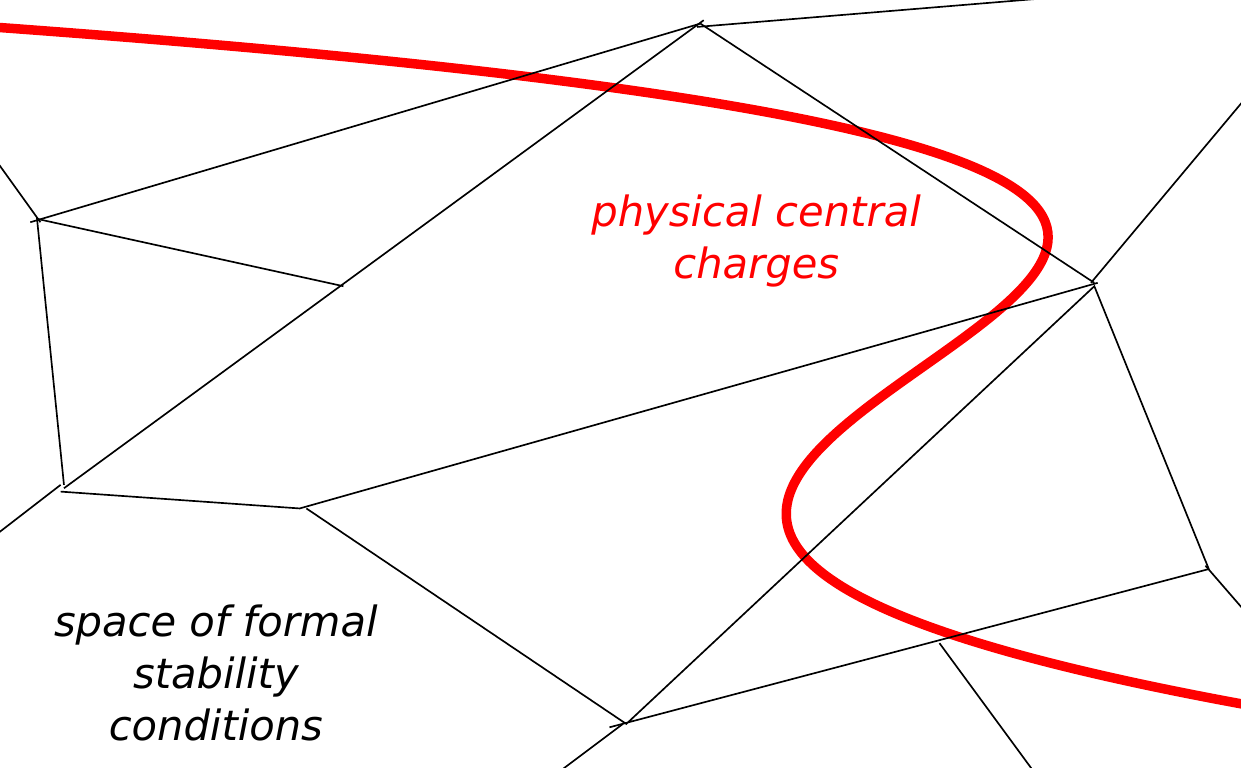}
\end{center}
\caption{Schematic representation of the quantum Schottcky problem: not all the possible assignments of formal stability conditions are physically realized. Since the space of stability conditions is divided into BPS chambers separated by walls of marginal stability, and since the physical central charges are on a locus of positive codimension, assigning an arbitrary central charge might lead to a formal BPS chamber which is not physically realized. Determining such a space from first principles is an interesting open problem.}\label{fig:QSP}
\end{figure}

\paragraph{Stability conditions and central charge.} The complex central charge $Z$ of the 4d $\CN = 2$ Poincar\'e superalgebra gives rise to a family of stability conditions (in the sense of \cite{Bridgeland:2007wo}), that are parametrized by points in $\CUL \times \mathbb R_+$, where $\R_+$ corresponds to the radius of the circle:
\be\label{Z 4d def}
Z: \CUL \times \mathbb R_+ \to \text{Hom}(\Gamma,\C) \; : \; (u,R) \mapsto Z(u,R\,; -)~.
\ee
For any given values of the parameters, $(u, R) \in \CUL \times \mathbb R_+$, the central charge assigns a complex number to every element of the charge lattice:
\be
Z(u,R\,; -) \; : \; \Gamma \rightarrow \C \; : \; \gamma \mapsto  Z(u,R\,; \gamma) \equiv Z_\gamma~,
\ee
where the dependence on the parameters is often kept implicit. Stability conditions are, in particular, \textit{linear maps}: $Z_{\gamma + \gamma'} = Z_\gamma + Z_{\gamma'}$. For this reason, a generic stability condition in $\text{Stab}(\CAT)$ is determined by $D$ complex numbers, corresponding to the values of the central charges for a basis of generators of the lattice $\Gamma$. Not all possible stability conditions correspond to physically realized central charges. The latter sit at a codimension $r$ locus in the space of stability conditions, as depicted in  Fig.~\ref{fig:QSP}; indeed, we have:
\be
D - 1 - \dim \CUL = r~.
\ee
The problem of discerning between physical and formal stability condition is the \textit{quantum Schottcky problem}.\footnote{For a description of this in terms of geometric engineering in IIB superstrings, see \protect\cite{Cecotti:2011gu,DelZotto:2011an}. The theories that are not affected by this problem are the so-called 4d $\CN=2$ complete SQFTs classified in~\protect\cite{Cecotti:2011rv}.} There is a deep analogy between this question and the Schottcky problem in algebraic geometry ({\it i.e.} the problem of discerning which Abelian varieties are Jacobian), rooted in special K\"ahler geometry---the 4d $\CN=2$ central charges are indeed periods of principally polarized Abelian varieties \cite{Cecotti:1990fq}.

\paragraph{BPS states and stable objects.} For every fixed charge $\gamma \in \Gamma$, the corresponding superselection sector of the Hilbert space of $\KK\FT$ is obtained by quantizing the moduli spaces $\CM(\gamma,Z)$ of the $Z$-stable objects $\CO \in \CAT$ such that $[\CO]=\gamma$, where $[\CO]$ is the class of $\CO$ in $K_0\CAT$. The moduli spaces $\CM(\gamma,Z)$ have a natural physical interpretation: these coincide with the Higgs branch moduli spaces for the supersymmetric quantum mechanics (SQM) with four supercharges (1d $\CN=4$) which govern the worldline of the corresponding BPS state. In particular, these moduli spaces are K\"ahler. The quantization procedure gives rise to a representation $\mathcal R_\gamma$ of $SU(2)_{\rm spin}\otimes SU(2)_R$ where $SU(2)_{\rm spin}$ is the little group of a particle in four-dimensions, and $SU(2)_R$ is the corresponding R-symmetry. These quantum numbers can be read off as the Lefschetz spin and the Hodge spin of the moduli spaces, respectively \cite{Cordova:2013bza,DelZotto:2014bga}:
\be
2 J_{\rm spin} = p + q - \dim \CM(\gamma,Z)~, \qquad   \qquad 2 J_R = p - q~,
\ee
where $(p,q)$ refer to the corresponding degree in an appropriate Dolbeault cohomology, $H^{p,q}(\CM(\gamma,Z))$, on the moduli space.\footnote{ The case in which $\CM(\gamma,Z)$ is a stack can be dealt with by replacing this cohomology theory with an appropriate Poincar\'e series (the so-called $\chi_y$ genus) that can be determined via localization \cite{Hori:2014tda,Cordova:2014oxa} or via the MPS Coulomb-branch formula \protect\cite{Manschot:2010qz,Manschot:2011xc,Sen:2011aa,Manschot:2012rx,Manschot:2013sya,Manschot:2013dua,Manschot:2014fua}.} Finally, the $SU(2)_\text{spin}\otimes SU(2)_R$ representation so obtained has to be tensored with a 4d $\CN=2$ half-hypermultiplet $(1/2,0)\oplus (0,1/2)$, which arises by quantizing the center-of-mass zero-modes of the brane stack \cite{Witten:1996qb}.

\medskip
\noindent As an example, if the moduli space $\CM(\gamma,Z_\gamma)$ is a point (i.e. the corresponding object is \textit{rigid}), $\mathcal R_\gamma = (0,0)$ and we obtain a half hypermultiplet. If $\CM(\gamma,Z_\gamma) \simeq \mathbb P^1$, then $\mathcal R_\gamma = (1/2,0)$ and:
\be
\big((1/2,0)\oplus (0,1/2)\big)\otimes (1/2,0) = (0,0) \oplus (1,0) \oplus (1/2,1/2)~,
\ee
which is a 4d $\CN=2$ vector multiplet.
\medskip

\noindent Let us now proceed by reviewing the interplay between fractional branes and quivers.

\subsubsection{5d BPS quivers as fractional brane quivers}
In all the examples considered in this paper, the supersymmetric quiver quantum mechanics governing the worldline of BPS particles are Lagrangian quiver gauge theories with gauge groups:
$$U(N_1) \times \cdots \times U(N_D)$$
The corresponding quiver diagram (with superpotential) is a BPS quiver for the theory $\KK\FT$, which we call a {\it 5d BPS quiver} to contrast it with the purely 4d BPS quivers studied {\it e.g.} in \cite{Alim:2011kw}.

The 5d BPS quiver for $\FT$, denoted by $\CQ_\MG$, can be immediately identified using the B-brane technology, as the quiver with superpotential that describes D-branes at the CY singularity $\MG$. Such quivers, generally known as {\it fractional-brane quivers,} have been studied extensively in the literature, starting with the study of D-branes at orbifolds \cite{Douglas:1996sw} --- see {\it e.g.} \cite{Douglas:2000qw,  Diaconescu:1997br, Kachru:1998ys, Klebanov:1998hh,Morrison:1998cs,  Denef:2002ru, Franco:2005rj} for a very partial list of reference.

\medskip
\noindent In type-II string theory, it is expected that any B-brane can be obtained as a bound state of a finite set of {\it simple objects,} $\CE_i \in D(\h\MG)$. Their defining property is that:
\be
{\rm Hom}(\CE_i, \CE_j) \cong \delta_{ij} \C~. 
\ee
These simple objects are the so-called {\it fractional branes} associated to the singularity. Simple objects are rigid by construction: the corresponding moduli space is a point. In addition, it is expected that the fractional branes are all approximately mutually BPS, meaning that there are quiver points $u^*\in \CUL$ such that the central charges $Z(u^*,R,[\CE_1]), \cdots, Z(u^*,R,[\CE_D])$ are all almost aligned. The fractional branes corresponding to a quiver point $u^*$ are also stable objects by construction. For the theories at hand, it is expected that the fractional branes become mutually BPS D-branes on the singular threefold, $\MG$---this corresponds to the ``fractionation'' of the D0-brane (skyscraper sheaf), when placed at the singularity, into a marginal bound state \cite{Douglas:1996sw}. Therefore, it is expected that 5d theories that can be constructed from $\MG$ always admit at least one quiver point. 
This is certainly the case for all the examples presented in this paper.\footnote{We would like to mention, however, that it is possible to generalize this construction by introducing fractional branes of different kinds, thus leading to more general quiver quantum mechanical models. The corresponding theory is under development; a proposal for the structure of the corresponding BPS categories has been discussed recently in the purely four-dimensional setting \protect\cite{Caorsi:2019vex}.}

\paragraph{Quiver from fractional branes: Dirac pairing and Euler form.} Given a collection of fractional branes, the quiver for the category is obtained as follows. The set of nodes of the quiver is in one-to-one correspondence with the simple fractional branes $\CE_1,\CE_2, \cdots,\CE_D$, and in between node $j$ and node $i$ we have the arrows determined by:
\be\label{eq:strings}
\# \text{ arrows } j \to i\;  \equiv \; \dim {\rm Hom}(\CE_i,\CE_j[1])~.
\ee
Since the category is 3-CY, we have the Serre isomorphism ${\rm Hom}(\CO,\widetilde{\CO}[3]) = D{\rm Hom}(\widetilde{\CO},\CO)$ for any pair of objects $\CO,\widetilde{\CO} \in\CAT$. Here $D = {\rm Hom(-,\C)}$ is the duality functor. For this reason, the Euler form of the BPS category, defined as:
\be\label{Euler pairing antisym}
\begin{aligned}
\langle [\CO] , [\widetilde{\CO}]\rangle_D \equiv \chi(\CO, \t \CO) &= \sum_{n=0}^3 (-1)^n \dim {\rm Hom}(\CO ,\widetilde{\CO}[n])\\
& =  \dim {\rm Hom}(\CO ,\widetilde{\CO}) -  \dim {\rm Hom}(\CO ,\widetilde{\CO}[1]) \\
&\qquad+  \dim {\rm Hom}(\widetilde{\CO},\CO [1]) -  \dim {\rm Hom}(\widetilde{\CO},\CO )~,
\end{aligned}
\ee
is antisymmetric and corresponds to the Dirac pairing on the charge lattice:
\be
\langle - , - \rangle_D \colon \Gamma \times \Gamma \to \mathbb Z~.
\ee
In particular, by definition, we have that:
\be\label{B from DX}
\langle \CE_i,\CE_j \rangle_D = B_{ij} = \# (\text{ arrows } i \to j )- \# (\text{ arrows } j \to i)~,
\ee
is the adjacency matrix of the underlying quiver. The quiver so obtained is denoted by $\CQ_\MG$. Importantly, we also have quiver relations generated by a superpotential $W$, which can be determined by studying the $A_\infty$ structure of the BPS category \cite{Herbst:2004jp, Aspinwall:2004bs}.\footnote{$\,$See \protect\cite{Closset:2017yte, Closset:2018axq} for a recent review with detailed computations,  in a slightly more general context.} In section \ref{sec:branetilings}, we will review some simpler technology to determine the quiver superpotential in the toric case \cite{Franco:2005sm}. 

For any such quiver with superpotential, $(\CQ_\MG, W)$, one defines its Jacobian algebra:
\be
\CJ \equiv { \C \CQ_\MG / (\partial W)}~,
\ee
consisting of the freely-generated path algebra of $\CQ_\MG$ divided by the ideal generated by the F-term relations.  The category of $\CJ$-modules, $\CJ\text{-mod}$,  is equivalent to the category of quiver representations.  One then obtains the relations:
\be\label{eq:quivercats}
\CAT\quad\cong\quad  D^b(\h\MG) \quad\cong\quad D^b(\CJ\text{-mod})~.
\ee
The second relation, between  $D^b(\h\MG)$ and the derived category of $\CJ$-modules, has been established in some generality; see {\it e.g.} \cite{Beilinson1978, Bondal1990}, \cite{Berenstein:2002fi}, as well as \cite{lam2014calabi} and references therein.~\footnote{A more precise version of this statement (and of the discussion below) would replace $\CJ$ with a dg-algebra associated with the quiver with superpotential \protect\cite{Ginzburg:2006fu, bergh2002noncommutative}. We wilfully skip over this subtlety, which does not really contain any new physics in the 3-CY case. (For a physics discussion in the n-CY case, see \protect\cite{Franco:2017lpa, Closset:2018axq}.)}

\medskip

\noindent  The fractional brane quivers, $(\CQ_\MG, W)$, were mostly studied by physicists in the context of D3-branes probing the CY$_3$ singularity $\MG$: performing three T-dualities along the worldvolume of the D3-branes, one obtains D0-branes in IIA. The underlying quiver theory is, of course, the same.
The Jacobian algebra $\mathcal J$ provides us with a non-commutative crepant resolution (NCCR) of the CY$_3$ singularity $\MG$, in the sense of Van den Bergh~\cite{bergh2002noncommutative}. In particular, the coordinate ring of $\MG$ appears as the center of $\CJ$:
\be
\MG \cong {\rm Spec} \, Z(\mathcal J)~.
\ee
In the literature on D3-branes at singularities, this construction is known as the ``mesonic moduli space'' of the 4d $\CN=1$ quiver gauge theory associated to $(\CQ_\MG,\CW)$; see {\it e.g.}~\cite{Forcella:2008bb}.

\paragraph{Hearts and BPS particles.} The stable objects of a triangulated category, $\mathscr T$, always belong to a {\it heart} for the corresponding t-structure \cite{Bridgeland:2005my, Bridgeland:2007wo}. One of the features of the hearts is that they are abelian categories, $\CA$, such that $\mathscr T \cong D^b\CA$. In particular, recall that an object $\CO \in \CA$ is stable iff for all its proper subobjects $0\neq \CS \subset \CO$ we have:
\be
0 < \arg Z([\CS]) < \arg Z([\CO]) \leq \pi~.
\ee

In the examples we consider in this paper, each quiver point determines, via the set of corresponding fractional branes, a canonical heart, namely the abelian category of quiver representations $\CJ\text{-mod} \equiv \text{Rep}(\CQ_\MG,W)$:
\be
\boxed{\phantom{\Big|}
\CA \quad\cong\quad \CJ\text{-mod}~.
}
\ee
Conversely, the full BPS category is recovered as the derived category of $\CJ$-modules, as in \eqref{eq:quivercats}. Whenever an object $\CO$ is a bound state of $N_1$ copies of $\CE_1$, $N_2$ copies of $\CE_2$, ..., and $N_D$ copies of $\CE_D$, we have
\be
[\CO] = N_1 [\CE_1] +  N_2 [\CE_2] + \cdots  N_D [\CE_D] 
\ee
and the resulting charge is 
\be
\gamma_\CO = [\CO] = (N_1,N_2,...,N_D) \in \Gamma
\ee 
which gets identified with the dimension vector of the corresponding quiver representation. All the $N_i$ above are positive numbers by construction (by the BPS condition, we have no  bound states of fractional branes and anti-branes). The choice of such abelian heart, $\CA \subset \CAT$, therefore corresponds to the choice of a splitting of the charge lattice in between particles (corresponding to objects in $\CA$) and antiparticles (corresponding to objects in $\CA[1]$). Indeed:
\be
[\CO[1]] = - [\CO] = - \gamma_\CO~,
\ee
is the charge of the antiparticle corresponding to the object $\CO$. By \textsc{CPT} invariance, it is enough to determine the stable objects in $\CA$ and the corresponding moduli spaces to completely determine the BPS spectrum. Each fractional brane, in particular, ends up contributing a full hypermultiplet to the BPS spectrum---we have a half-hyper with charge $\gamma_i \equiv [\CE_i]$ and a half-hyper with charge $-\gamma_i \equiv [\, \CE_i[1] \,]$.

\medskip
\paragraph{Relation to SQM.} Consider an object $\CO \in \CA$ with charge $\gamma_\CO = [\CO] = (N_1,...,N_D)$. The corresponding quiver quantum mechanics has a gauge group whose degrees of freedom correspond to strings stretched between the $N_i$ fractional branes $\CE_i$:
\be\label{UN quiver}
U(N_1) \times U(N_2) \times \cdots \times U(N_D)~,
\ee
and bifundamental matter corresponding to strings streched between the stacks of $N_i$ $\CE_i$ branes and $N_j$ $\CE_j$ branes, which are counted by \eqref{eq:strings}. Each $U(N_i)$ group has a corresponding FI term:
\be
\xi_i \equiv \text{Im} \left({Z(\CE_i) \over Z(\CO)}\right)~,
\ee
and, by King's theorem, the Higgs branch moduli space of this SQM is identified with the moduli space of semi-stable representations of $\CJ$ with dimension vector $\gamma_\CO$. Note the slight subtlety: the BPS particles correspond to \emph{stable} objects in the abelian heart $\CA$, and their quantum numbers are obtained by quantizing the corresponding \textit{semi-stable} moduli spaces.

\medskip
\paragraph{Mutations and autoequivalences of $\CAT$.} The description of the BPS category in terms of quiver SQM, and the associated representation theory of quivers with potentials, comes with an unnatural splitting of the BPS spectrum between particles and antiparticles \cite{Alim:2011ae}. There is a canonical choice dictated by the quiver point associated to the singularity, but a given model can admit several distinct quiver points.

 This corresponds to the fact that the BPS category $\CAT$ can have several different inequivalent hearts, related by autoequivalences of $\CAT$. Often, such autoequivalences factor via compositions of more elementary tilting functors, that correspond to elementary mutations of the underlying quivers.
 
Elementary mutations are interpreted at the level of the SQM as 1d Seiberg dualities, which in particular leave the underlying moduli spaces invariant. For this reason, even if the heart changes, the full BPS spectrum obtained by quantizing stable objects corresponding to a given point $u \in \CUL$ is the same.

Nevertheless, it can happen that a given quiver description is not enough to capture the structure of the spectrum over the full $\CUL$: to every path $p\colon [0,1] \to \CUL$ correspond an autoequivalence of $\CAT$ that often can be described in terms of a mutation sequence.  This remark is relevant in the context of 5d dualities, to which we now turn.

\subsubsection{Gauge-theory phases and 5d BPS quivers}\label{subsec: electric subcat}

\paragraph{Gauge subsectors and electric subcategories.} At the level of the corresponding triangulated BPS category $\CAT$, the fact that a given theory has a gauge subsector is understood in terms of a property of an Abelian heart---here we review some parts of \cite{Cecotti:2012va}. There is a collection of objects, $(W_\alpha)_{\alpha \in \Delta^+(G)} \in \CA$, labeled by $\Delta^+(G)$, the positive root lattice of the gauge group $G$, that are all simultaneously stable for some choice of central charges and satisfy the following properties:
\begin{itemize}
\item[G1.)] The dimension vectors $\boldsymbol{\delta_\alpha} \equiv \dim W_\alpha$ satisfy $\boldsymbol{\delta_{\alpha+\beta}} = \boldsymbol{\delta_{\alpha}}+\boldsymbol{\delta_{\beta}}$. In particular, we define  $\boldsymbol{\delta_i} = \boldsymbol{\delta_{\alpha_i}}$ where $\alpha_i$ are the simple roots in $\Delta^+(G)$;
\item[G2.)] All $W_\alpha$ have a moduli space $\mathcal M (\boldsymbol{\delta}_\alpha) \simeq \mathbb P^1$, and hence correspond to $\CN=2$ vector multiplets;
\item[G3.)] $\langle \boldsymbol{\delta_\alpha}, \boldsymbol{\delta_\beta} \rangle_D = 0$ for all $\alpha, \beta \in \Delta^+(G)$, and moreover we have the following integrality requirement (Dirac quantization): let us introduce the elementary electric charges (the elementary weights for representations of $G$):
\be
\mathbf q_i = \left(C(G)^T\right)^{-1}_{ij} \boldsymbol{\delta}_j  \in \Gamma \otimes \mathbb Q \qquad (\text{elementary electric charge})
\ee
where $C(G)$ is the Cartan matrix of the Lie group $G$. Then, the corresponding magnetic charges have to be integer valued:
\be
\mathbf{m}_i(\CO) \equiv \langle  [\CO] , \mathbf{q}_i \rangle_D \in \mathbb Z \qquad\,\text{for all } \CO \in \CAT~.
\ee
\end{itemize} 
Often, these gauge-group subsectors are realized in terms of a full subquiver of the appropriate kind for a quiver associated to the heart of the BPS category. These are the quivers of the form $A(1,1)\boxtimes G$ for the pure supersymmetric Yang-Mills theories with simply-laced simple gauge groups.
For non-simply-laced groups, the relevant quivers have been obtained in \cite{Cecotti:2012gh}. Having identified an abelian heart compatible with G1, G2 and G3, one can define the associated {\it electric subcategory}\footnote{Also known as the ``light subcategory,'' as these are precisely the states that remain light in the corresponding weak coupling limit; for a definition, see \protect\cite{Cecotti:2012va}.} as the abelian subcategory of the heart $\CA$ controlled by the magnetic charges $\mathbf{m}_i(-)$. We refer to Appendix \ref{app:control} for a review of controlled subcategories.
\medskip

It is interesting to remark that the objects in the electric category are precisely the ones that remain dynamical upon a weak-coupling deformation \cite{Cecotti:2012va}:
\be\label{Z def weak}
Z(\cdot) \mapsto Z(\cdot) - {1 \over g_{\rm 4d}^2} \mathbf{m}(\cdot)~,  \qquad g_{\rm 4d} \to 0~,
\ee
of the central charges. Since magnetic charges in five dimensions are carried by BPS strings, we are lead to identify the purely magnetic states with BPS monopole strings wrapped on $S^1$. For this reason, the count of the degeneracies of the purely five-dimensional BPS particles is encoded in the electric subcategory of the 5d BPS quiver.  Note that this is compatible with the above deformation because:
\be
{1\over g_{\rm 4d}^2} = {R \over g_{\rm 5d}^2}~,
\ee
and therefore, keeping $g_{5d}$ fixed, the limit \eqref{Z def weak} coincides with the limit $R\to \infty$.

\paragraph{5d dualities and autoequivalences of $\CAT$.} Two distinct five-dimensional gauge theories are ``UV dual'' provided that they both are gauge theory phases for the same 5d SCFT \cite{Bergman:2013aca, Bergman:2014kza, Closset:2018bjz}. If that is the case, the theory has to admit inequivalent gauge subsectors of the type characterized above, as well as representations corresponding to the correct matter content. On the other hand, the electric subcategories, as obtained by the above procedure for either gauge group, are expected to be equivalent. This is true even if we start from two different hearts $\CA$ and $\CA'$,  essentially because there can be no non-trivial wall-crossing amongst purely electric states (nor mutations of the underlying electric quiver). 

In the context of four-dimensional $\CN=2$ gauge theories, a similar phenomenon occurs in cases when we have an S-duality, which is suggestive of a possible five-dimensional origin for it. In the four-dimensional case, it has been understood \cite{Cecotti:2015hca,Caorsi:2016ebt} (see also \cite{Gang:2017ojg}) that such dualities are mapped to functorial autoequivalences of the underlying triangulated category, which is a useful perspective to determine the corresponding group structure for the S-duality group.

Here, we claim that this is the case for the five-dimensional UV dualities as well: obviously, a necessary condition for two inequivalent five-dimensional gauge theories to be dual is that there is one such equivalence at the level of the corresponding BPS categories.\footnote{ Notice that we are not claiming that there is necessarily an explicit overlap at the level of the corresponding BPS spectra.} In the cases studied in this paper, these functorial autoequivalences are nothing but tiltings, generated by appropriate sequences of elementary mutations.

\subsubsection{Relations between 5d and 4d BPS quivers} 

Interestingly, the radical of the Dirac pairing is a mutation invariant, and as such it coincides with an intrinsic property of the SCFT. It is simply the lattice of flavor charges:
\be\label{Gamma F def}
\Gamma_F \equiv \text{rad } \langle - , - \rangle_D \equiv \text{Ker } B\,\subseteq \Gamma
\ee
The charges in $\Gamma_F$ corresponds to D2-branes wrapping curves that do not intersect any exceptional divisor. These brane states are sometimes known as ``non-anomalous fractional branes'' in the T-dual context of  D3-brane at singularities---see {\it e.g.} \cite{Klebanov:1999rd,Klebanov:2000hb, Franco:2003ja}---because $B v=0$, for a dimension vector $v=(N_1, \cdots, N_D)$, is the condition for the vanishing of the 4d gauge anomalies.

It is convenient to organize $\Gamma_F$ as follows:
\be
\Gamma_F \equiv \mathbb Z \gamma_{KK} \oplus \bigoplus_{j=1}^{f} \mathbb Z \gamma_{f_j}
\ee
where $\gamma_{KK}$ is the KK charge and $\gamma_{f_j}$ are the flavor charges for the 5d SCFT. 
When $\FT$ has a gauge theory interpretation with a gauge group $G= \prod_{k=1}^s G_k$ with $s$ simple factors, we have a (topological) global symmetry $\prod_{k=1}^s U(1)_{T_k}$, and among the flavor charges $\gamma_{f_j}$ we could distinguish the topological charges $\gamma_{T_k}$ explicitly.

Considering the extended Coulomb branch of the KK theory $\KK\FT$, we can often find several inequivalent ways of taking the limit $R \to 0$ which produce a consistent genuinely four-dimensional theory.\footnote{$\,$ This is analogous to what already observed in the context of six-dimensional theories on $T^2$: the four-dimensional limits are analogous to Argyres-Douglas points along $\CUL$ \protect\cite{DelZotto:2015rca}.} If the four-dimensional theories so-obtained have a BPS quiver, the latter should correspond to a full subquiver for some element in the mutation class of the 5d BPS quiver $\CQ_\MG$. More precisely, such full subquiver corresponds to the quiver for a controlled subcategory (as reviewed in Appendix \ref{app:control}).

Naively, the control function associated to that subcategory is given by a projection onto states of vanishing KK charge: \be\lambda_{KK} (\gamma) = P_{\gamma_{KK}} \gamma~.\ee The objects in the resulting subcategory must indeed have zero KK charge, but more interesting combinations can be considered. For instance, if the 5d theory has a gauge theory phase with a simple gauge group $G$, then we can obtain the  4d BPS quiver for the corresponding gauge theory by means of two control functions: the projector $\lambda_{KK}$ above, to decouple the KK tower, and:
\be
\lambda_T(\gamma) = P_{\gamma_T} \gamma~,
\ee
 where $P_{\gamma_T}$ is a projector on the one-dimensional sublattice of $\Gamma$ generated by the topological $U(1)_T$ flavor charge, which corresponds to the instanton number associated to the gauge group $G$. Similarly, exploiting flavor symmetry Wilson lines on $S^1$, we could access more interesting controls corresponding to more general flavor charges.

Conversely, it is interesting to ask whether it is always possible to uplift a 4d BPS quiver to a consistent 5d BPS quiver by enlarging the corresponding charge lattice. This question will be addressed elsewhere \cite{CDZtoappear}. 



\section{Brane tilings and 5d BPS quivers}\label{sec:branetilings}
In this section, we study 5d BPS quivers for $\KK\FT$ when $\MG$ is toric. Then, a so-called toric quiver with superpotential $(\CQ_\MG, W)$ can be equivalently written in terms of a periodic tiling on the torus, known as a brane tiling, which is closely related to the mirror geometry~\cite{Feng:2005gw}.  See \cite{Kennaway:2007tq, Yamazaki:2008bt} for comprehensive reviews.~\footnote{Many of the brane-tiling computations for this work were algorithmized on {\it Mathematica}, building on a code written for~\protect\cite{Closset:2012ep}, which itself built on a package written by Jurgis Pasukonis for~\protect\cite{Davey:2009bp}.}

 \begin{figure}[t]
\begin{center}

\subfigure[\small $r{=}0$, $f{=}1$.]{\qquad
\raisebox{0.05cm}{\includegraphics[height=1.4cm]{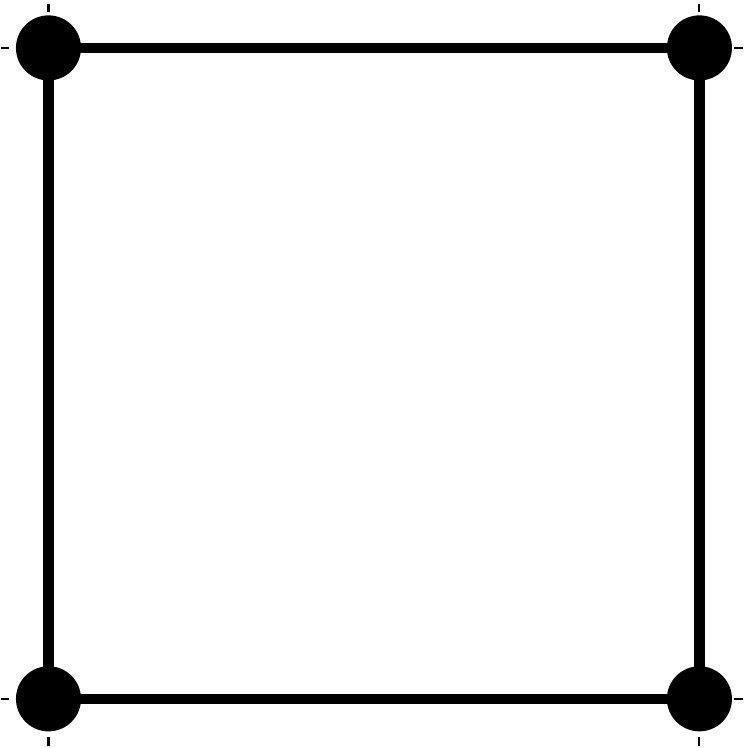}}\qquad\label{fig:C0 expl}}\;
\subfigure[\small $r{=}1$, $f{=}3$.]{
\includegraphics[height=2.6cm]{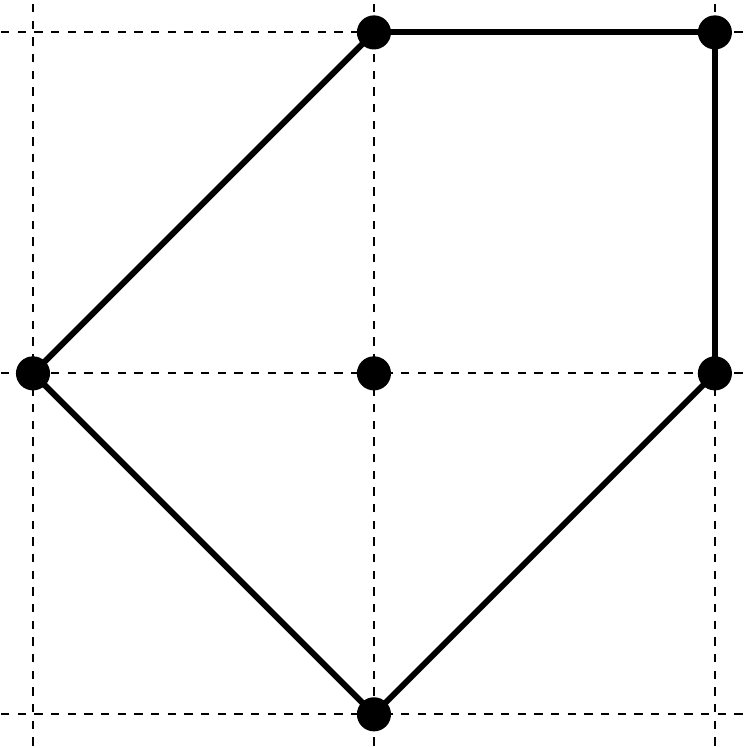}\label{fig:E3 sing 0}}\quad \;
\subfigure[\small  $r{=}2$, $f{=}3$.]{
\includegraphics[height=2.6cm]{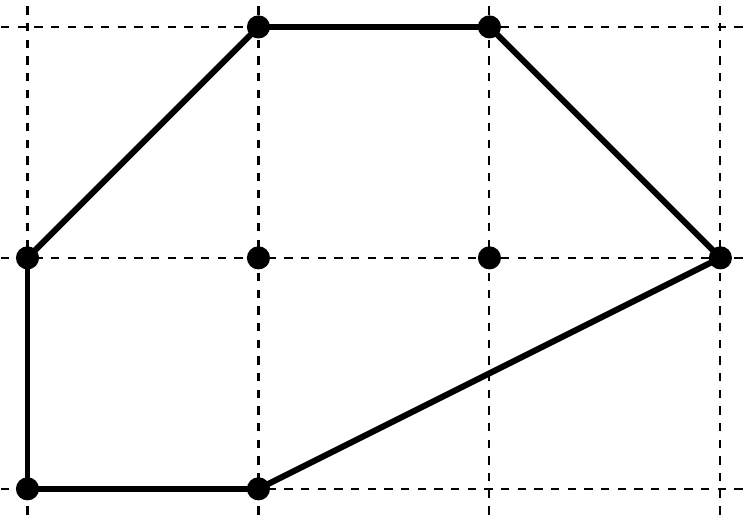}\label{fig:rk 2 exp}}\quad\,
\subfigure[\small  $r{=}3$, $f{=}1$.]{
\includegraphics[height=3.8cm]{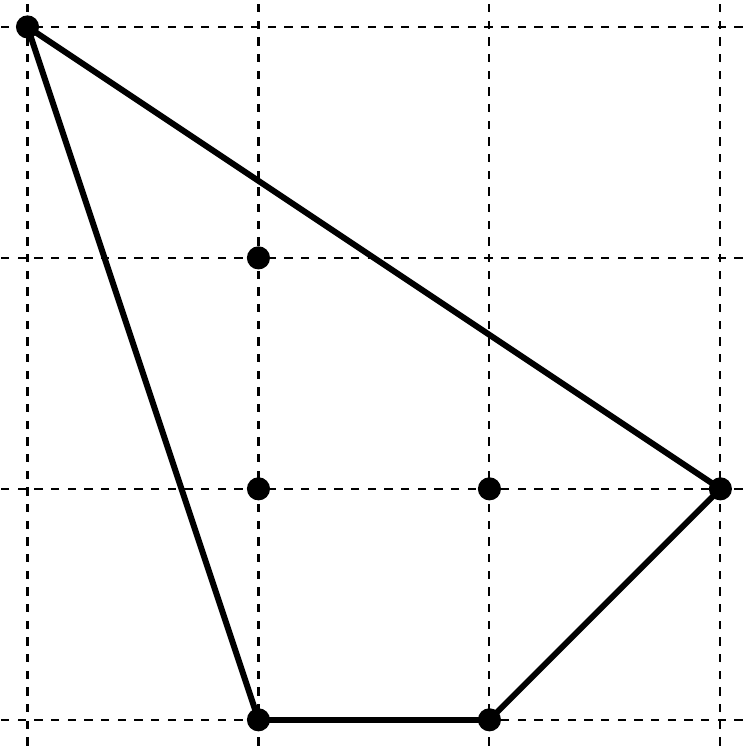}\label{fig:rk 3 exp}}
\caption{Some examples of toric diagrams. \label{fig:TDs expls}}
 \end{center}
 \end{figure}
\subsection{A lightning review of brane tilings}
Consider the toric diagram, $\TX \subset \Z^2$, for $\MG$ an isolated toric CY$_3$ singularity. It is a strictly convex polytope with $r$ internal points and $f+3$ external points, as shown in some examples in Fig.~\ref{fig:TDs expls}.
The corresponding quiver $\CQ_\MG$  has $D= 2 r+ f +1$ nodes.\footnote{We also have $2 {\rm Area}(\TX)=D$: there are $D$ simplices  in any complete triangulation of $\TX$.}

 A brane tiling (a.k.a. dimer model) is a bipartite graph on the torus, with an even number of nodes, $n_W$, half of them black and the other half whites. The nodes are connected by edges, which are the boundaries of faces. The brane tiling associated to a toric diagram $\TX$ has  $D$ faces. Since the tiling is on a torus, the number of tiling edges, $n_X$, is given by Euler's formula as $n_X= D+ n_W$. The brane tiling encodes a quiver $\CQ_\MG$ and its superpotential $W$. Each face corresponds to a quiver node, each edge corresponds to a quiver arrow, and each brane-tiling node corresponds to a superpotential term. The arrow directions are determined by the following traffic rule: circulation goes clockwise around white nodes, and counter-clockwise around black nodes. This assigns a direction to each brane tiling-edge, denoted by $\alpha$, and we will write the corresponding  quiver arrows as:
\be
X_\alpha = X_{ij}\; : (i) \longrightarrow (j)~.
\ee
Finally, we read off the superpotential terms from the brane-tiling nodes, as:
\be
W= \sum_{\text{white nodes}}  \prod_{\alpha }^{\boldsymbol{\circlearrowright}} X_\alpha\quad  -\sum_{\text{black nodes}}  \prod_{\alpha}^{\boldsymbol{\circlearrowleft}} X_\alpha~.
\ee
Here, the products are ordered products over the edges connected to the node, in the clockwise or anti-clockwise direction and with a $\pm$ sign for the white or black nodes, respectively.

\subsubsection{From $\CQ_\MG$ to  $\TX$: perfect matching variables and GLSM}
Given a brane tiling describing D0-branes at $\MG$, one can recover the toric diagram $\TX$ using the ``fast forward algorithm'' \cite{Franco:2005rj}. A {\it dimer} is a distinguished edge in a brane tiling, connecting a white and a black node. A {\it perfect matching} is a collection of dimers such that every node of the tiling is included only once. Then, a {\it dimer model} is just a brane tiling together with its perfect matchings, which are denoted by $\{p_k\}$. The perfect matching matrix, $P_{\alpha k}$, is defined as:
\be
P_{\alpha k} = \begin{cases} 1 & \text{if}\,\;  p_k\,\; \text{contains the edge} \,\; X_\alpha~,\\ 0 & \text{otherwise.}\end{cases}
\ee
The set of perfect matchings of given brane tiling can be derived using the Kasteleyn matrix; see \cite{Franco:2005rj} and references therein:
\be
K_{mn}(x,y) = \sum_{\beta \in \{m \rightarrow n\}} X_\beta \, x^{\langle X_\beta, \gamma_x\rangle} y^{\langle X_\beta, \gamma_x\rangle}~.
\ee
Here, the indices $m$ and $n$ run over the white and black nodes, respectively, $X_\beta$ are formal variables denoting the edges, and $x$ and $y$ are two more formal variables; $\gamma_x$ and $\gamma_y$ denote a choice of $(1,0)$ and $(0,1)$-cycles on the torus, and $\langle X_\beta, \gamma_x\rangle$ denotes the signed interesection number between the edge $X_\beta$ and the 1-cycle $\gamma$. Then, the perfect matchings are obtained by computing the permanent of the Kasteleyn matrix:
\be\label{perm K}
{\rm perm} \; K(x,y) = \sum_k p_k \; x^{\langle p_k, \gamma_x \rangle} y^{\langle p_k, \gamma_y \rangle}~,
\ee
where $p_k=\prod_\alpha X_\alpha^{P_{\alpha k}}$, formally. This procedure allows us to read off  the toric diagram, $\TX$, directly, since the expression \eqref{perm K} is the Newton polygon of $\TX$.~\footnote{That is, the  toric diagram consists of the points $w_k =(\langle p_k, \gamma_x \rangle, \langle p_k, \gamma_y \rangle) \in \Z^2$.}
This computation also naturally assigns perfect matchings to each point in the toric diagram. 
In a consistent brane tiling, each external point in $\TX$ corresponds to a single perfect matching, while the $r$ internal points can correspond to multiple perfect matchings.

Consider the $\CN=4$ SQM associated to $(\CQ_\MG, W)$ with gauge group $U(1)^D$, corresponding to the D0-brane representations of the quiver. This is a 1d gauged linear $\sigma$-model (GLSM). The isolated toric singularity, $\MG$, arises as the  supersymmetric moduli space:
\be\label{SQM mod space for X}
\MG \cong \Big\{X_\alpha \Big| F_\alpha= 0~,  \forall \alpha~, \;  \; D^a =0~, a=1, \cdots D\Big\}/U(1)^{D-1}~,
\ee
with the $F$-terms given by $F_\alpha= \partial_{X_\alpha} W$, and the standard $D$-term conditions for each $U(1)$ node in the gauge-theory quiver. (The diagonal $U(1)$ acts trivially.) 
The distinguishing feature of a toric quiver is that every quiver arrow $X_\alpha$ appears exactly twice in $W$, with opposite signs. This ensures that the $F$-term relations are of the form ``monomial$=$monomial.''  Given a dimer model, one assigns a complex variable, also denoted by $p_k$, to each perfect matching. Then, one can always solve the $F$-term relations in terms of these auxiliary variables, according to:
\be\label{X from p}
X_\alpha = \prod_k p_k^{P_{\alpha k}}~.
\ee
The moduli-space computation \eqref{SQM mod space for X} is then reduced to an ordinary K\"ahler quotient:
\be\label{KQ MG}
\MG \cong  \{p_k~, k=1, \cdots, c \}\, \slash\slash \, U(1)^{m+D-1}~,
\ee
with $m= c-D-2 \geq 0$ additional $U(1)$'s, corresponding to the redundancy in the solution \eqref{X from p}.

\subsubsection{From $\TX$ to $\CQ_\MG$: the fast inverse algorithm}
Given a toric singularity $\MG$ described by a strictly convex toric diagram $\TX$ with $r$ internal points, one can work out the corresponding brane tiling using the so-called ``fast inverse algorithm'' of Hanany and Vegh \cite{Hanany:2005ss}. The algorithm starts from a larger toric geometry for a non-isolated toric geometry, with a toric diagram $\TX_{n, m}$ which is a square of size $n\times m$ inside $\Z^2$, such that $\TX$ is contained within $\TX_{n,m}$.\footnote{The toric diagram $\TX_{n, m}$  corresponds to the underlying square grids shown in Fig.~\protect\ref{fig:TDs expls}, for instance.} The brane tiling for $\TX_{n,m}$ is well-known as an orbifold of the conifold. It consists of a square tiling similar to Fig.~\ref{fig:bt C0}, but with $D= 2nm$ distinct faces. The brane tiling for $\TX$ is the obtained by a systematic procedure that involves drawing the so-called zig-zag paths; let us simply refer to~\cite{Hanany:2005ss} for details and references. All the examples of quivers $\CQ_\MG$ discussed below can be obtained using that algorithm, combined with Higgsing/partial resolutions \cite{Franco:2005rj}.   See also~\cite{Davey:2009bp, Hanany:2012hi, Franco:2017jeo} for a systematic discussion of large families of toric geometries. 

It is important to note that the map from toric diagram to brane tiling is one-to-many; this corresponds to several ambiguities in the implementation of the fast inverse algorithm. However, any two distinct brane tilings obtained in this way, and the associated quivers with superpotential, are related by a series of quiver mutations. In the $\CN=4$ SQM language (after fixing a dimension vector for the quiver), a mutation is nothing but a Seiberg duality of the 1d $\CN=4$ supersymmetric gauge theory. Note also that, while there generally are an infinite number of  distinct quivers related to a given $(\CQ_\MG, W)$ by mutations, it appears that there only exists a finite number of toric quivers ({\it i.e.} arising from brane tilings) for a given $\MG$.

\subsection{Resolutions and Beilinson quivers}
Given a toric quiver $(\CQ_\MG, W)$, we consider the set of $\theta$-stability conditions, corresponding to the set of possible Fayet-Ilopolous (FI) parameters $\boldsymbol{\xi}= (\xi^a)$. Then, any crepant resolution can be realized as the K\"ahler quotient \eqref{KQ MG} with non-zero levels, with the D-term conditions  $D^a= \mu^a-\xi^a=0$.
This can also be formulated as a GIT quotient, corresponding to the moduli space of $\theta$-semistable representations with dimension vector ${\rm dim}(R)=(1, \cdots, 1) \equiv \alpha_{\rm D0}$. We then have:
\be\label{resolved X GLSM}
\h \MG\quad \cong\quad  \C^{m+D+2} \slash\slash_{\bf\xi} \, U(1)^{m+D-1}\quad \cong\quad  \CM_{\bf\theta}({\alpha_{\rm D0}})~.
\ee
From the point of view of the gauged $\CN=4$ SQM, a choice of $\theta$-stability is a choice of discretized FI parameters. We have $\boldsymbol{ \theta} = - \boldsymbol{ \xi} \in \Z^D$, with $\sum_a \xi_a=0$.\footnote{The sign is necessary to match between the most common physics and math notations.} Recall that a quiver representation $R$ is $\theta$-stable (resp., semi-stable) if, for any proper subrepresentation $0 \neq S \subset R$, we have $({\rm dim} S).\boldsymbol{\theta} >0$ (resp., $({\rm dim} S).\boldsymbol{\theta} \geq 0$). 
We refer to \cite{BenderMozgovoyII} for further details and references.

Given a perfect matching $p_k$, we define a quiver representation $R_{p_k}$ of dimension $\alpha_{\rm D0}$ such that $X_\alpha$ is given by $x_\alpha \in \C$ with:
\be
x_\alpha = \begin{cases}  0 & {\rm if} \; X_\alpha \in p_0~, \\ 1 & {\rm if} \; X_\alpha \notin p_0~.\end{cases}
\ee
For a given stability $\boldsymbol{\theta}$, the perfect matching $p_k$ is called $\theta$-stable if $R_{p_k} $ is $\theta$-stable \cite{BenderMozgovoyII}. 

To every perfect matching in the dimer model, we also associate a {\it Beilinson quiver},  denoted by $\CQ_\MG((p_k)$, which is the quiver obtained by removing the quiver arrows $X_\alpha$ that appear in $p_k$. Of particular interest to us will the Beilinson quivers associated to perfect matchings for internal points in $\TX$. They are quivers with relation (inherited from $W$) but without any closed loops \cite{Hanany:2006nm}. The $\theta$-stable representations, of dimension $\alpha_{\rm D0}$, of such ``internal'' Beilison quivers describe a D0-brane probing the exceptional locus of some (partial) resolution $\h \MG$ of the toric singularity.

\subsection{Brane charges and tilting bundles}\label{subsec:brane charge}
The fractional branes $\CE_i$ gives us a convenient basis of the charge lattice:
\be
 [\CE_i] \in \Gamma~.
\ee
In this basis, the charge of any BPS object $\CO$ is given by the dimension vector of the corresponding quiver representation---{\it i.e.} the ranks $N_i$ in the SQM description \eqref{UN quiver}, so that $[\CO]= \sum_i N_i [\CE_i]$. It is often useful, however, to also use the so-called {\it brane charge} basis, in terms of K-theory charges of the D-branes, $[\CO]= \sum_i q_i [\boldsymbol{C}_i]$, for some $q_i \in \Z$, in terms of the K-theory classes of D-branes wrapping compact cycles $\boldsymbol{C}_i$ on the resolved singularity.  For a given complete resolution $\h\MG$, we denote these classes by:
\be\label{K th charges gen}
\big([\boldsymbol{C}_i]\big) = \big([\bE_a]~, \,[\CC_k]~,\,  [{\rm pt}]\big)~, \qquad \quad  a=1, \cdots, r~, \qquad  k=1, \cdots, r+f~,
\ee
for the compact 4-cycles, 2-cycles and 0-cycle, respectively. Let $Q^\vee$ be the $SL(D, \Z)$ matrix that realizes this change of basis, namely: 
\be
[\CE_i]= \sum_{j=1}^D Q^\vee_{ij} [\boldsymbol{C}_j]~.
\ee
In the rest of this subsection, we explain how to determine $Q^\vee$ in the rank-one case \cite{Closset:2012ep}; the general case is similar but slightly more subtle, and a complete discussion is left for future work. Along the way, we introduce some auxiliary quantities implicitly related to a choice of resolution $\h\MG$, which we will use extensively in later sections. 

\subsubsection{Tilting collections of line bundles}
The fractional branes are often rather complicated objects in $D^b(\h\MG)$.
It is often fruitful to consider another set of objects, denoted by $L_i$, which are simpler coherent sheaves; see {\it e.g.}~\cite{Aspinwall:2008jk}.  They are dual to the fractional brane with respect to the Euler pairing:
\be
\langle L_i, \CE_j \rangle_D = \delta_{ij}~.
\ee
Note that this is not antisymmetric, in apparent contradiction with \eqref{Euler pairing antisym}. This is because the sheaves $L_i$ are not compactly supported; nonetheless, the Euler pairing can be extended to this case \cite{Chuang:2013wt}. For $\h\MG$ toric, the sheaves $L_i$ are line bundles.

In the toric case, the equivalence between the derived category of quiver representations, $D(\CJ\text{-mod})$, and the derived category of coherent sheaves on $\h\MG$  can be made completely explicit using the line bundles $L_i$. There exists a tilting equivalence~\cite{Aspinwall:2008jk, BenderMozgovoyII}:
\be\label{Psi map gen}
\Psi \; : \;  D(\CJ\text{-mod}) \rightarrow D^b({\rm coh} \, \h \MG)~,
\ee
with a tilting sheaf $\Psi(\CJ)$ given by a direct sum of line bundles over $\h\MG$ \cite{BenderMozgovoyII}:
\be
T=\Psi(\CJ)  \cong \oplus_{i=1}^D L_i~.
\ee
The line bundles $L_i$ form a tilting collection, which means that they satisfy:
\be
\Ext_{\h\MG}^n(L_i, L_j)=0 \quad  \text{for}  \quad  n>0~, \qquad \; \forall i, j~,
\ee
as coherent sheaves on $\h\MG$, and that they generate the full triangulated category $D^b(\h\MG)$.  Conversely, the quiver algebra $\CJ$ arises as the endomorphism algebra of the tilting sheaf.

\medskip

\noindent 
Given a dimer model of rank $r>0$, one can build a tilting collection of line bundles as follows \cite{Hanany:2006nm, BenderMozgovoyII}. Let us first choose an internal perfect matching, $p_0$, such that the nodes of the Beilinson quiver $\CQ_\MG(p_0)$ are partially ordered, with a single ``first node'' labelled by $i=1$, and a single ``last node'' labelled by $i=D$, and with all the nodes numbered such that any path in $\CQ_\MG$ goes from node $i$ to $j$ with $i<j$.\footnote{This first step is not necessary, but it gives us a very convenient setup for our purpose. See~\protect\cite{BenderMozgovoyII} for the more general algorithm. In later sections, we will always label the first and last nodes of $\CQ_\MG(p_0)$ as stated here, while we will sometimes pick some slightly different numbering of the other nodes. The rank-zero case, $r=0$, will be treated separately.}

The $\Psi$-map assigns to every quiver arrow $X_\alpha$ the formal sum of  all the perfect matchings in which $X_\alpha$ appears, $\Psi(X_\alpha)=\sum_k P_{\alpha k} p_k$. By linearity, it assigns a formal linear combination of perfect matchings to every path. When considering a specific stability condition, we also restrict the map to $\theta$-stable perfect matchings. 

Every perfect matching $p_k$ is associated to a toric divisor, corresponding to a node in the toric diagram and given by $D_k= \{ p_k=0 \}$ in the GLSM description \eqref{resolved X GLSM}.  Then, to every path:
\be
\boldsymbol{p} = X_{\alpha_1} X_{\alpha_2} \cdots X_{\alpha_n}~,   \qquad\qquad X_{\alpha_l} : i_l \rightarrow i_{l+1}~, 
\ee
the $\Psi$-map assigns a divisor over $\h\MG$ according to:
\be
\Psi_\theta(\boldsymbol{p}) = \sum_{l=1}^n \sum_{k | p_k \, \theta\text{-stab}}  P_{\alpha_l k} D_k~.
\ee
The tilting line bundles are then defined as follows. Given $\CQ_\MG(p_0)$ with the ordering of node as above, we choose a path $\boldsymbol{p}_i : 1 \rightarrow i$ for every node $i$. The line bundles are given by:
\be
L_i = \CO\left(\Psi_\theta(\boldsymbol{p}_i)\right)~, \qquad i=1, \cdots, D~.
\ee

\subsubsection{Exceptional locus and Beilinson quiver}
The choice of internal perfect matching $p_0$ gives implicitly a choice of partial resolution of $\MG$, corresponding to blowing up a single exceptional divisor associated to the internal point. 
In the special case $r=1$, the single exceptional divisor is a toric del Pezzo surfaces $\CS\cong dP_n$. More generally,  the exceptional locus $\CS$ of the partially-resolved singularity is a toric stack; see {\it e.g.} \cite{Herzog:2006bu}.

We expect that the Beilinson quiver algebra $\CQ_\MG(p_0)$ can be used to describe coherent sheaves on the exceptional locus $\CS$ \cite{Herzog:2006bu}, generalizing the well-understood del Pezzo case. 
Given the embedding $i: \CS \rightarrow \h\MG$, we then view the tilting line bundles $L_i$ as the pushforward of lines bundles ${\bf L}_i$ on $\CS$:
\be
L_i = i_\ast {\bf L}_i~.
\ee
The line bundles ${\bf L}_i$ should furnish an appropriate exceptional collections on $\CS$ \cite{Cachazo:2001sg, Herzog:2003zc, Herzog:2005sy, Herzog:2006bu}, which generate the derived category $D^b(\CS)$ of the stacky surface $\CS$.

\paragraph{The matrix $S^{-1}$.}  Let us define the pairing:
\be\label{Sinv def}
S^{-1}_{ij} = {\rm dim}\, {\rm Hom}({\bf L}_i, {\bf L}_j)= {\rm dim}\,  H^0(\CS, {\bf L}_i^\ast \otimes {\bf L}_j)~.
\ee
These dimensions can be computed in toric geometry, in principle, and they depend on the choice of partial resolution $\h \MG$ that contains $\CS$. They are most easily computed from the Beilinson quiver, as:
\be\label{Sinv from paths}
S_{ij}^{-1} = \# \; \text{of independent paths} \;\; \boldsymbol{p} : i \rightarrow j \; \; \text{in} \; \; \CQ_\MG(p_0)~.
\ee
Here, the relations amongst paths in the Beilinson quiver are inherited from the relations in the full quiver. 
Since $\CQ_\MG(p_0)$ is partially ordered, the matrix $S^{-1}$ is upper-triangular.

\paragraph{Fractional branes on $\CS$ and the matrix $S$.}
The fractional branes $\CE_i$ should also correspond to simple objects ${\bf E}_i$ in $D^b(\CS)$:
\be
\CE_i = i_\ast {\bf E}_i~.
\ee
The fractional branes ${\bf E}_i$ of course correspond to the quiver nodes of the Beilinson quiver $\CQ_\MG(p_0)$. 
Then, defining the Euler pairing on $\CS$ as:
\be
\chi_\CS({\bf E}, {\bf F})= {\rm dim}\,  {\rm Hom}({\bf E}, {\bf F}) - {\rm dim}\,  \Ext^1({\bf E}, {\bf F})+ {\rm dim}\,  \Ext^2({\bf E}, {\bf F})~,
\ee
for any pair ${\bf E}, {\bf F} \in D^b(\CS)$, we have:
\be
\chi_\CS({\bf L}_i, {\bf E}_j) = \delta_{ij}~,  \qquad \qquad S^{-1}_{ij} =\chi_\CS({\bf L}_i, {\bf L}_j)~.
\ee
The matrix $S$ (the inverse of $S^{-1}$) is defined as:
\be\label{mat S def}
S_{ij} = \chi_\CS({\bf E}_j, {\bf E}_i)~.
\ee
For compactly supported sheaves $\CE, \CF$ on $\h\MG$, with $\CE= i_\ast {\bf E}$ and  $\CF= i_\ast {\bf F}$, we have \cite{Aspinwall:2005ur}:
\be
\Ext^n_{\h \MG}(\CE, \CF) = \Ext^n_{\CS}({\bf E}, {\bf F}) \oplus  \Ext^{3-n}_{\CS}({\bf F}, {\bf E})~,
\ee
and therefore:
\be
\langle \CE, \CF\rangle_D= \chi_\CS(\CE, \CF)- \chi_\CS(\CF, \CE)~.
\ee
In particular, the signed adjacency matrix of the quiver $\CQ_\MG$, as defined in \eqref{B from DX}, can be recovered from the matrix $S$ in \eqref{mat S def}:
\be
B = S^T- S~.
\ee
 Note that $S$ depends on the choice of (partial) resolution implicit in the choice of internal perfect matching $p_0$, while $B$ can be computed directly in $D^b(\h \MG)$, and is thus independent of the choice of resolution.

\paragraph{The monodromy matrix $M$.}  Another interesting quantity is the monodromy matrix, defined as:
\be
M = (S^{-1})^T S~.
\ee
In the IIB mirror description, the fractional branes $\CE_i$ map to D3-branes wrapped over Lagrangian 3-cycles; in the Landau-Ginzburg (LG) description, they correspond to solitons interpolating between different LG vacua. The  matrix $M$ then encodes the monodromy amongst the vacua as we go once around the $W_{\rm LG}$-plane \cite{Cecotti:1992rm}. The eigenvectors of $M$ are the dimensions vectors of particularly interesting quiver representations. In particular, the eigenvectors $v_F$ with eigenvalue $1$ satisfy $B v_F= 0$. These are the D2-brane states that populate  the flavor charge lattice \eqref{Gamma F def}.

\subsubsection{Projective and simple objects, and their brane charges}\label{subsec:brane charge}
Given the Beilinson quiver $\CQ_\MG(p_0)$ as we just defined it, let $\CJ_{p_0} \subset \CJ$ be its Jacobian algebra (with the quiver relations inherited from $(\CQ_\MG, W)$). The fractional brane ${\bf E}_i$ correspond to simple $\CJ_{p_0}$-modules, consisting of a single node. The line bundles ${\bf L}_i$, on the other hand, correspond to the projective $\CJ_{p_0}$-modules, $P_i$, associated to a single node, generated by all the paths in $\CJ_{p_0}$ that ends at the node $i$ \cite{Aspinwall:2008jk}.

In the derived category $D(\CJ_{p_0}\text{-mod}) \subset D(\CJ\text{-mod})$, one can always find a projective resolution of the simple objects in terms of the $P_i$'s. At the level of the $K$-theory charges for the fractional branes, we then have:
\be
 [{\bf E}_i]= \sum_{j=1}^D S_{j i}\,  [{\bf L}_j]~.
 \qquad \Rightarrow \qquad [\CE_i]= \sum_{j=1}^D S_{j i}\,  [L_j]~.
\ee

Then, if we can assign a brane charge to each $L_i$, we directly find the brane charge for the fractional branes. Let us pick a K-theory basis $[\boldsymbol{C}_i]$ as in \eqref{K th charges gen}. Then, we denote by $Q$ and $Q^\vee$ the matrices of brane charges for the line bundles $\{L_i\}$ and for the fractional branes $\CE_i$, respectively, in that basis:
\be
[L_i]= \sum_{j=1}^D Q_{ij} [\boldsymbol{C}_j]~, \qquad\qquad
[\CE_i]= \sum_{j=1}^D Q^\vee_{ij} [\boldsymbol{C}_j]~.
\ee
We then have:
\be
Q^\vee= S^T Q~.
\ee
The non-trivial step, in this computation, is to write $[L_i]$ in a basis of compact cycles. This is easy to do for rank-one singularities, since $[L_i]= [{\bf L}_i]$ and the latter is readily understood as wrapping $\CS$ once. More generally, one has to translate between the stacky structure on $\CS$ (for a given $p_0$) and the completely resolved space with $r$ distinct four-cycles ${\bf E}_a$. For some of the higher-rank examples to be discussed below, we determined the brane charges by imposing various consistency conditions; we leave a more systematic discussion for future work.

\subsection{A first example: the free 5d $\CN=1$ hypermultiplet and  the conifold} 
 \begin{figure}
\begin{center}
\subfigure[\small Conifold toric diagram.]{
\raisebox{1cm}{\includegraphics[height=2.5cm]{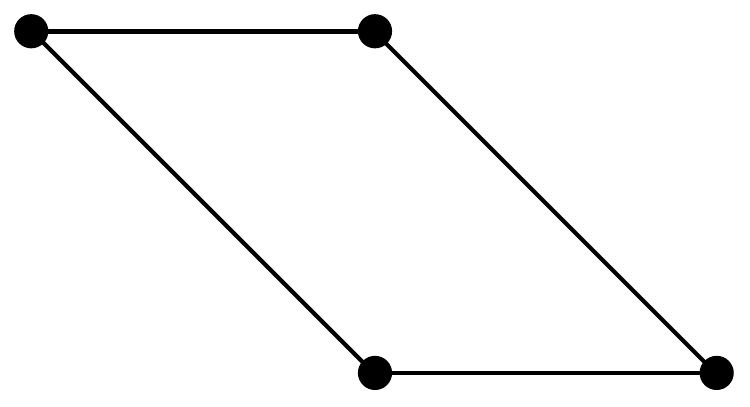}}\label{fig:TD C0}}\qquad
\subfigure[\small Conifold brane tiling.]{
\includegraphics[height=4.5cm]{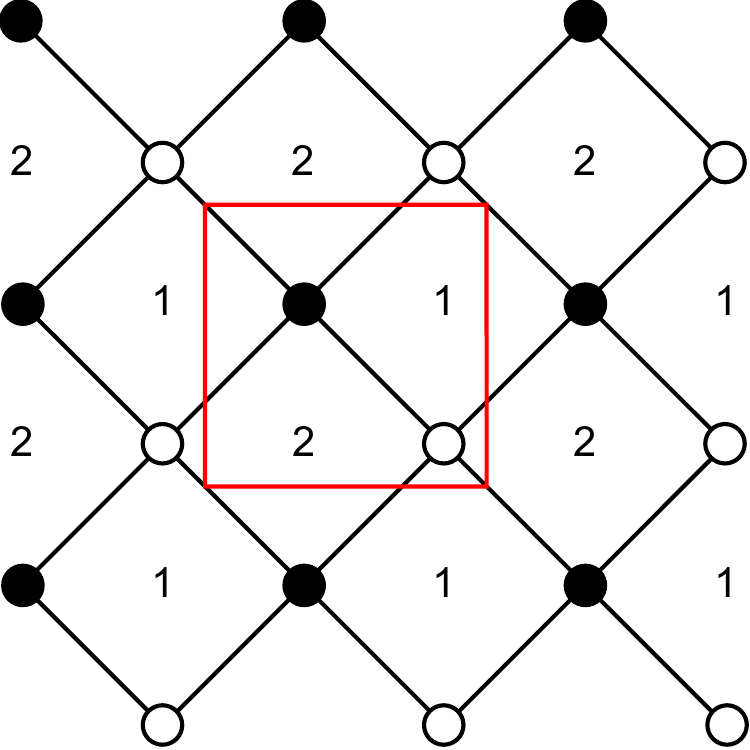}\label{fig:bt C0}}\qquad
\subfigure[\small $\CQ_\MG({\rm hyper})$]{\qquad\quad
\includegraphics[height=4.5cm]{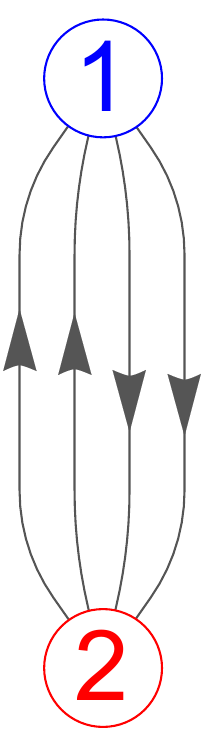}\label{fig:Q C0} \qquad}
\caption{Conifold singularity, associated brane tiling and quiver. The brane tiling is a double-periodic tiling on the plane or, equivalently, a tiling of the torus. The torus fundamental domain is shown is red.  \label{fig:conifold}}
 \end{center}
 \end{figure}
As a simple (and somewhat degenerate) example, consider the brane tiling shown in Fig.~\ref{fig:bt C0}. This corresponds to the  conifold singularity, $\CC_0$, whose toric diagram is shown in Fig.~\ref{fig:TD C0}. The corresponding 5d SCFT has rank $r=0$ and flavor rank $f=1$. In fact, this is simply the free hypermultiplet.\footnote{Formally, this can also be seen as the ``$SU(1)_0$'' gauge theory  \protect\cite{Closset:2018bjz} \label{foot:SU0}.} The 5d BPS quiver is the celebrated Klebanov-Witten quiver~\cite{Klebanov:1998hh} shown in Fig.~\ref{fig:Q C0}, with its superpotential:
\be
W= X_{12}^1 X_{21}^1 X_{12}^2 X_{21}^2-X_{12}^1 X_{21}^2 X_{12}^2 X_{21}^1~.
\ee
The two superpotential terms correspond to the two nodes, one white and one black, on the brane tiling of Fig.~\ref{fig:bt C0}.

The 5d BPS quiver of Fig.~\ref{fig:Q C0} is interpreted as the 5d uplift of the trivial single-node quiver for a free 4d $\CN=2$ hypermultiplet. The second node of the 5d BPS quiver carries the KK charge (indicated in red in Fig.~\ref{fig:Q C0}). While the free hypermultiplet is a rank-zero theory, we still have a one-dimensional extended Coulomb branch ($f=1$): the resolved conifold geometry corresponds to a massive hypermultiplet. It is well known, in that case, that the quiver representations of dimension vector  $\alpha_{\rm D0}= (1,1)$ correspond to the D0-brane probing the conifold. The massive hypermultiplet arises as a D2-brane wrapped over the exceptional curve $\CC$ inside the small resolution of the conifold.

In this degenerate ($r=0$) case, we need to choose {\it two} perfect matchings instead of one to obtain a Beilinson quiver, which is simply the Kronecker quiver (two arrows from node 1 to node 2 and no relations).~\footnote{This corresponds to having a exceptional locus of codimension 2 instead of codimension 1~\protect\cite{BenderMozgovoyII}.} Using the $\Psi$-map, one finds the tilting bundle~\cite{BenderMozgovoyII}:
\be
T= L_1 \oplus L_2~, \qquad L_1 \cong \CO~, \qquad L_2 \cong \CO(D_1)~.
\ee
 When restricted to the exceptional locus $\CC\cong \mathbb{P}^1$, this gives us the exceptional collection $\{{\bf L}_1, {\bf L}_1\}=\{\CO, \CO(1)\}$ of line bundles on $\mathbb{P}^1$. We can then directly compute:
\be
S^{-1} = \smat{1 & 2 \\ 0 & 1}~, \qquad S= \smat{1 & -2\\ 0 & 1}~, \qquad Q^\vee = \smat{1 & 0 \\ -1 & 1}~,
\ee
with the brane charge $Q^\vee$ written in the K-theory basis $([\CC], [{\rm pt}])$. We have  $[\rm pt]\cong [{\rm D0}] \cong [\CE_1] \oplus [\CE_2]$. The two fractional branes, corresponding to the two simple objects in the quiver algebra, carry the brane charges:
\be
[\CE_1] \cong [\CC]~, \qquad \qquad [\CE_2]\cong -[\CC]+ [\rm pt]~.
\ee 

In Appendix \ref{app:conifBPS}, we study in detail the structure of the BPS spectrum, via a direct analysis of the representation theory of the Jacobian algebra of the conifold. We obtain the following spectrum of stable particles for a stability condition $\arg Z_1 > \arg Z_2$:
\be\label{spectrum Coni}
\begin{aligned}
& \text{hypermultiplets: }\\
&\qquad [\CE_1] + n [{\rm D0}]~, &n \in \mathbb Z_{\geq 0}~,\\
&\qquad [\CE_2] + (m-1)  [{\rm D0}], \quad &m \in \mathbb Z_{\geq 1}~,\\
& \text{vector multiplets: }\\
&\qquad k  [{\rm D0}]~, &k \in \mathbb Z_{\geq 1}~.
\end{aligned}
\ee
Naively, one might have expected a different result for the D0-brane object, since the D0-brane moduli space is the whole resolved conifold, $\h \CC_0$; similarly, for $k$ D0-branes, we might have expected to find a moduli space ${\rm Sym}^k(\h \CC_0)$. Consider first the case $k=1$: we are quantizing a non-compact moduli space, and therefore we have to consider some appropriate cohomology theory, and as a first approximation we consider cohomology with compact support.~\footnote{$\,$ A finer treatment would likely require to use an appropriate $L^2$-type or intersection cohomology as suggested by the results of \protect\cite{MeinhardtReineke}.} This is the reason why, when quantizing the resolved conifold geometry, we only obtain the states corresponding to the exceptional $\mathbb P^1$ in $\h\CC_0$, namely a doublet of Lefshetz spin. Such a representation, tensored with the half-hypermultiplet, gives a short 4d $\CN=2$ vector multiplet. For $k>1$, we argue  in Appendix \ref{app:conifBPS}  that most of the configurations in ${\rm Sym}^k(\h \CC_0)$ are only marginally stable, and therefore cannot be counted when considering stable BPS states: the only stable configurations have a moduli space which equal just one copy of $\CC_0$ and therefore, once quantized, only contribute one vector multiplet each.

This BPS spectrum (and 5d BPS quiver) for the 5d free massive hypermultiplet on a circle agrees with the one determined by \cite{Banerjee:2019apt} using spectral network methods. In particular, the towers of  hypermultiplet states in \eqref{spectrum Coni} describe 4d hypermultiplets with $n$ units of momentum along the KK circle.\footnote{A possible explanation for the KK tower of vector multiplets we obtain may be in terms of the dual description of the conifold-- see footnote \protect\ref{foot:SU0}.}

\paragraph{Reduction to 4d.} In this case, the reduction to 4d is very simple. The 4d controlled subcategory is obtained by exploiting as a control function $\lambda(\CO) = P_{\gamma_{KK}} [\CO]$. This correspond to projecting out all the representations of the quiver in Figure \ref{fig:Q C0} that are supported on the node associated to $\CE_2$. The resulting quiver is the $A_1$ quiver, a quiver with a single node $\bullet$ and no arrows. This quiver in 4d has a unique stable representation, corresponding to the simple object $\CS
\cong\C$. The moduli space of said representation is a point, and therefore we obtain a half-hypermultiplet. Together with its CPT conjugate $\CS[1]$ we obtain, not surprisingly, the degrees of freedom of a full 4d $\CN=2$ hypermultiplet.

\section{Rank-one examples: del Pezzo singularities}\label{sec:rank1}
 \begin{figure}[t]
\begin{center}
\subfigure[\small $E_0$\, ($dP_0$)]{
\includegraphics[height=2.5cm]{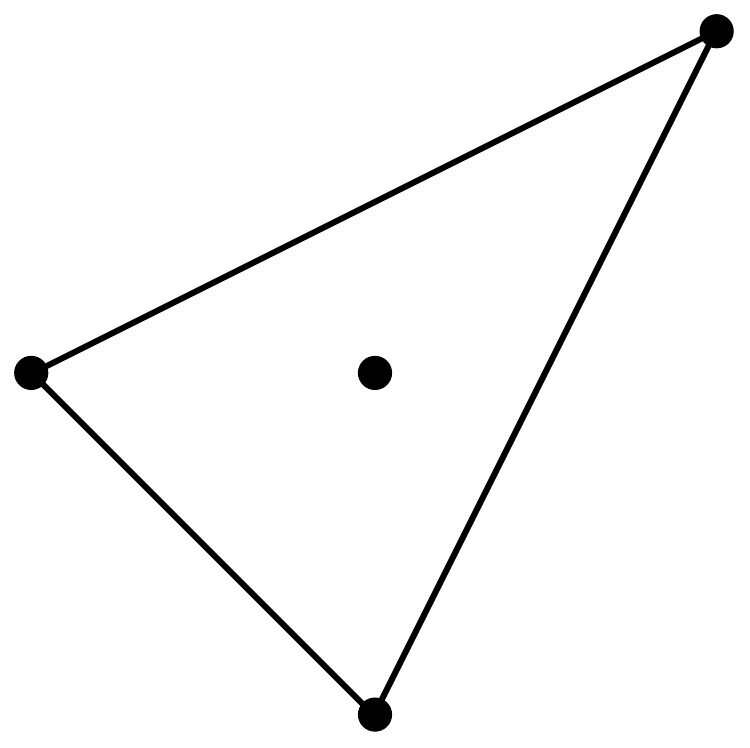}\label{fig:E0 sing}}\quad
\subfigure[\small $E_1$ \, ($\mathbb{F}_0$)]{
\includegraphics[height=2.5cm]{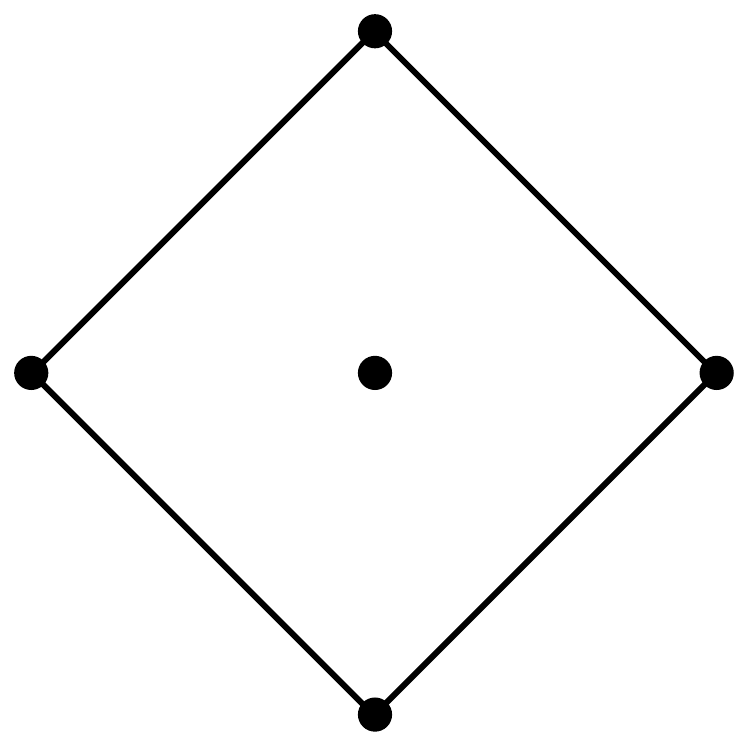}\label{fig:E1 sing}}\quad
\subfigure[\small $\widehat E_1$\, ($dP_1$)]{
\includegraphics[height=2.5cm]{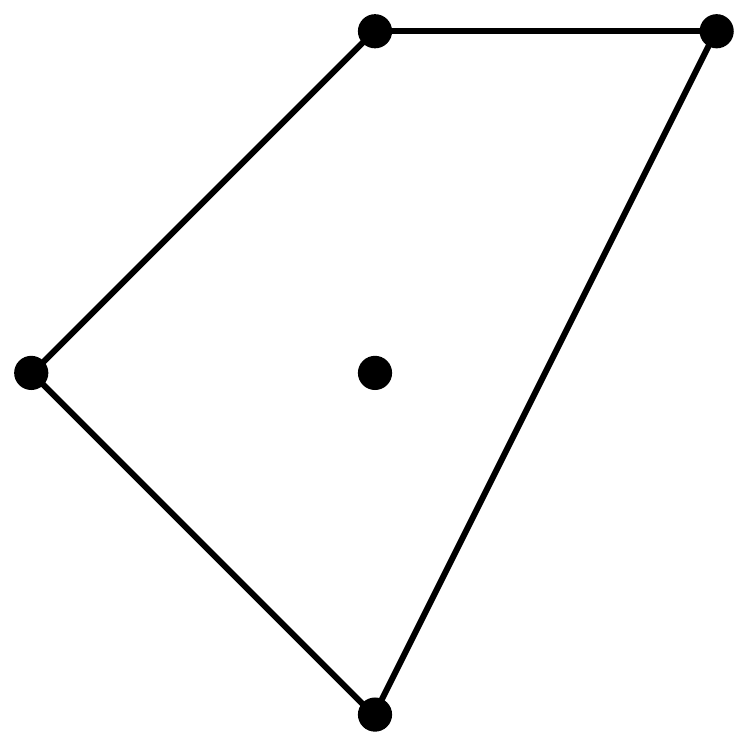}\label{fig:E1t sing}}\quad
\subfigure[\small $E_2$\, ($dP_2$)]{
\includegraphics[height=2.5cm]{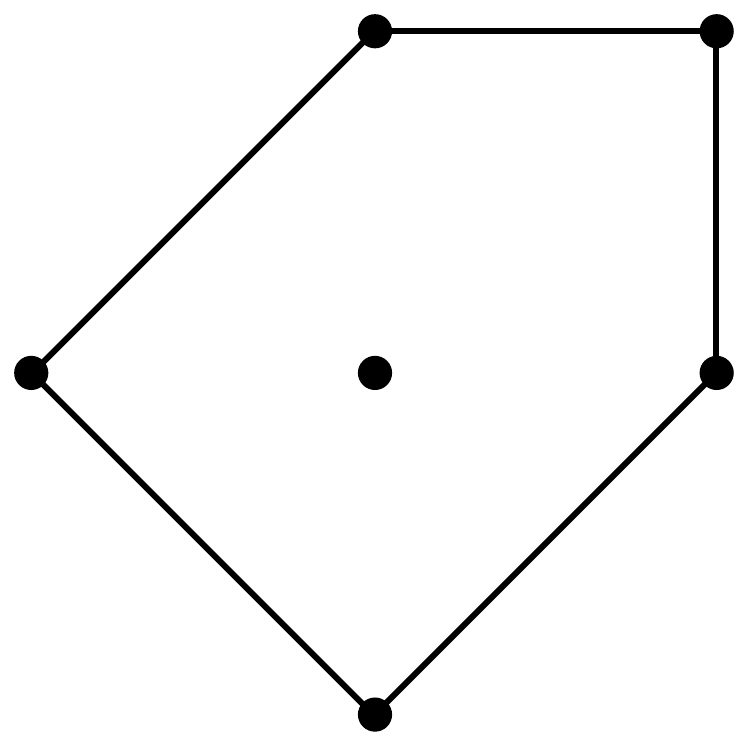}\label{fig:E2 sing}}\quad
\subfigure[\small $E_3$ \, ($dP_3$)]{
\includegraphics[height=2.5cm]{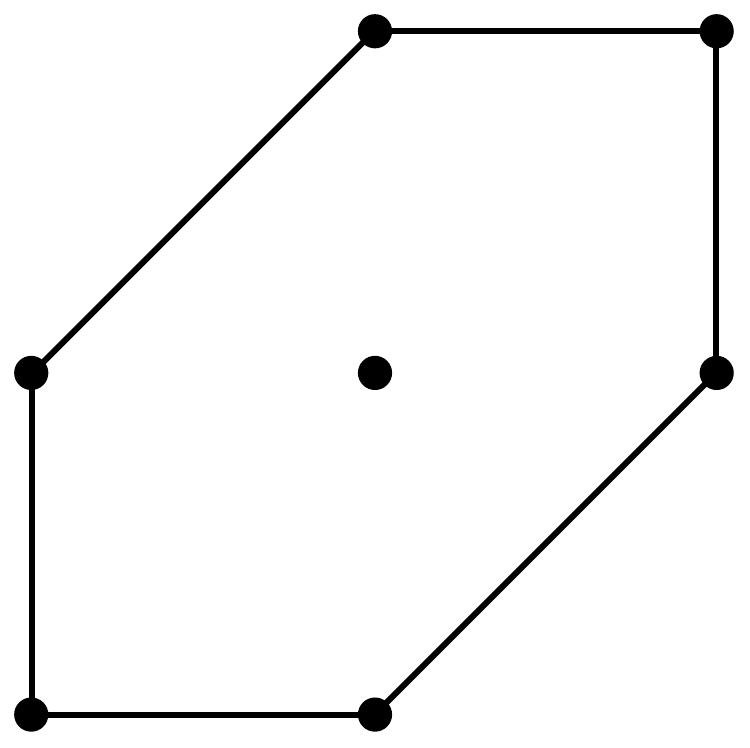}\label{fig:E3 sing}}\
\caption{Toric diagrams for the toric del Pezzo singularities. \label{fig:SU2 geoms}}
 \end{center}
 \end{figure}
The rank-one 5d SCFTs, denoted by $E_{N_f+1}$, correspond to the blow-down of a del Pezzo surface inside a Calabi-Yau manifold \cite{Morrison:1996xf, Douglas:1996xp}:
\be
\mathbb{P}^2\cong dP_0~, \qquad  \mathbb{F}_0 \cong \mathbb{P}^1 \times \mathbb{P}^1~, \qquad  dP_n~, \, n=1, \cdots, 8~,
\ee
where $dP_n$ denotes $\mathbb{P}^2$  blown-up at $n$ generic points. The resolved singularity:
\be
\h \MG ={ \rm Tot}(\CK \rightarrow dP_n)
\ee
admits a gauge theory interpretation as a Coulomb branch chamber of a 5d $SU(2)$ gauge theory with $N_f=n-1$  fundamental hypermultiplets. (For $N_f=0$, we have both $\mathbb{F}_0$ and $dP_1$, corresponding to $SU(2)_0$ and $SU(2)_\pi$, respectively.)
Amongst these, only the first five are toric---namely, $\mathbb{P}^2,  \mathbb{F}_0,  dP_1\cong \mathbb{F}_1, dP_2$ and $dP_3$. The toric diagrams of the corresponding CY$_3$ singularities $\MG$ are shown in Figure~\ref{fig:SU2 geoms}.
The quivers $\CQ_\MG$ are well-known---see {\it e.g.} \cite{Hanany:2001py, Cachazo:2001sg, Feng:2002fv, Wijnholt:2002qz, Franco:2002ae}. We discuss them in some detail in what follows, emphasizing the five-dimensional interpretation.

\subsection{The 5d $SU(2)$, $N_f=2$ gauge theory: local $dP_3$}
We start by looking at the ``largest'' of the toric singularities, the complex cone over $dP_3$, whose toric diagram is shown in Figure~\ref{fig:E3 sing}. 
  \begin{figure}[t]
\begin{center}
\subfigure[\small $dP_3$ toric diagram.]{
\includegraphics[height=5cm]{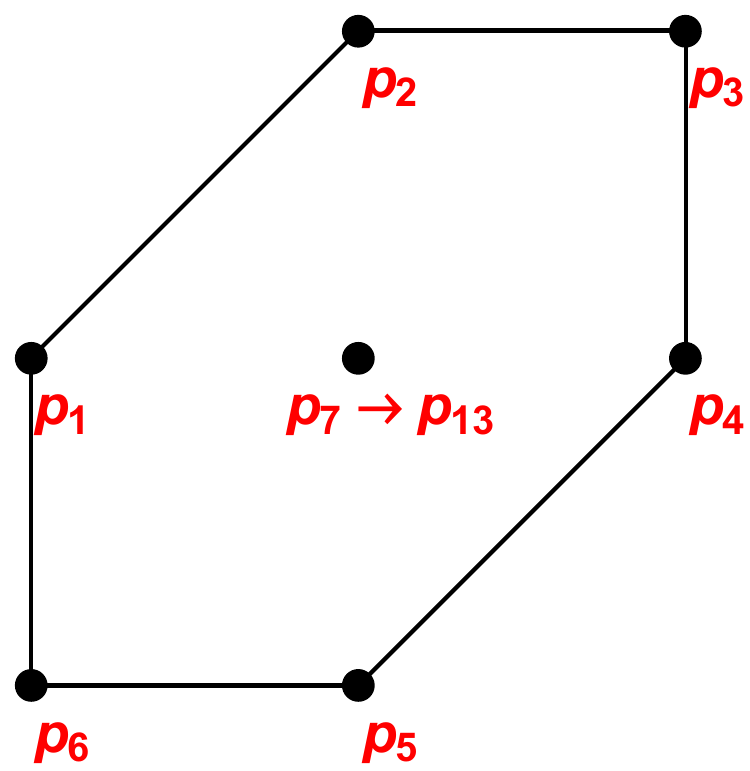}\label{fig:dP3 TD pms}}\qquad\qquad
\subfigure[\small  $dP_3$ brane tiling.]{
\includegraphics[height=5cm]{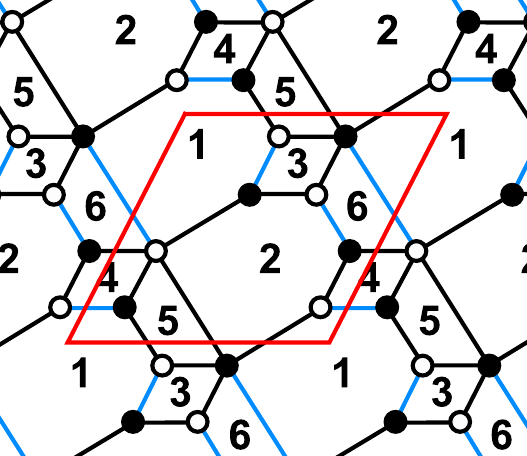}}\label{fig:dP3 bt}
\caption{Toric diagram and brane tiling for the $dP_3$ geometry.  The perfect matchings are associated to toric divisors as indicated on the toric diagram. The brane tiling consists of 6 faces, 14 edges and 8 vertices (4 white and 4 black); the torus fundamental domain is shown in red. The perfect matching $p_{13}$ is shown in light blue.  \label{fig:dP3 TD and bt}}
 \end{center}
 \end{figure} 
  \begin{figure}[t]
\begin{center}
\includegraphics[height=6cm]{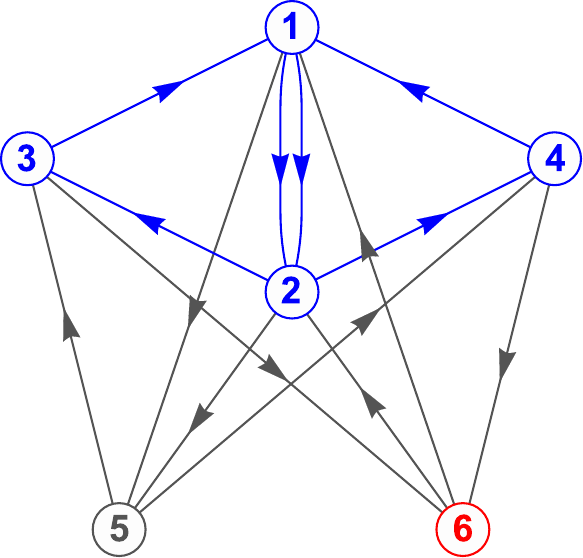}
\caption{ The $dP_3$ quiver,  $\CQ_{\MG(dP_3)}$, gives the 5d BPS quiver of the $SU(2)$ $N_f=2$ gauge theory. The 4d $\CN=2$ BPS subquiver is indicated in blue (nodes 1, 2, 3, 4). The red node (node 6) is the one that carries the KK charge. 
\label{fig:dP3 Q}}
 \end{center}
 \end{figure} 
  \begin{figure}[t]
\begin{center}
\subfigure[\small  Labelled toric diagram.]{
\raisebox{0.8cm}{\includegraphics[height=5.5cm]{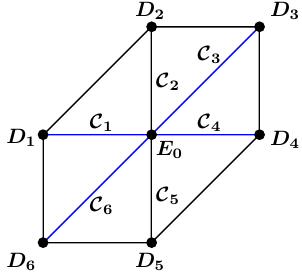}\label{fig:dP3 curves}}}\qquad\qquad
\subfigure[\small  $B\CQ_{\MG(dP_3)}(p_{13})$.]{
\includegraphics[width=4cm]{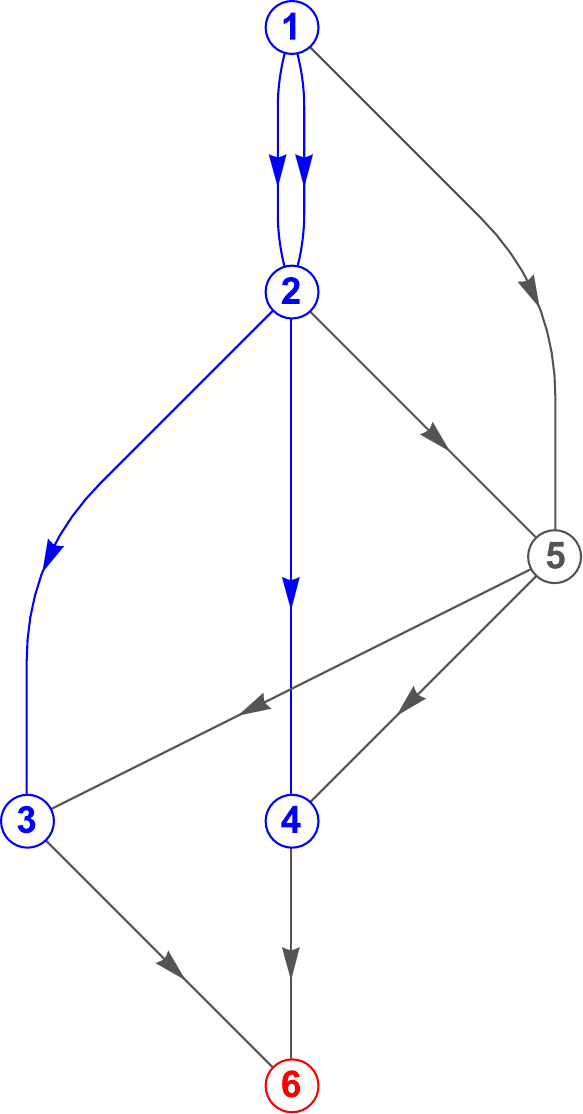}\label{fig:dP3 BQp13}}
\caption{\textsc{Left}: Toric diagram with the points and edges corresponding to divisors and curves, as indicated. The blue edges correspond to curves that give rise to perturbative states (hypermultiplets and W-bosons) in the 5d gauge theory, in the ``vertical'' S-duality frame. 
\textsc{Right}: The Beilinson quiver associated with the perfect matching $p_{13}$, with the partial order as displayed along the vertical axis. 
\label{fig:dP3 Q and BQ}}
 \end{center}
 \end{figure} 
The possible brane tilings for the $dP_3$ singularity are well-known. Let us consider the one shown in Figure~\ref{fig:dP3 bt}.~\footnote{That is sometimes called ``phase b'' of $dP_3$, or ``Model 10b'' in the notation of \protect\cite{Hanany:2012hi}.}  The corresponding quiver is shown in Figure~\ref{fig:dP3 Q}. The superpotential reads:
 \bea\label{W dP3}
&W_{dP_3}={\color{blue} X_{12}^1 X_{24} X_{41} -X_{12}^2 X_{23} X_{31}}  +X_{15} X_{53} X_{31} -X_{15} X_{54} X_{41}\cr
&\qquad\; +X_{62} X_{23} X_{36} - X_{62} X_{24} X_{46} + X_{25} X_{54} X_{46}  X_{61} X_{12}^2
  -X_{25} X_{53} X_{36}  X_{61} X_{12}^1~.
 \eea
The $dP_3$ brane tiling of Fig.~\ref{fig:dP3 bt} has 13 perfect matchings, corresponding to 6 external and 7 internal PMs, as indicated on the toric diagram in Fig.~\ref{fig:dP3 TD pms}. To an external PM $p_i$, we associate a non-compact toric divisor $D_i$. Note the linear equivalences:
\be
D_5\sim D_1+D_2-D_4~, \qquad D_6 \sim -D_1 +D_3 +D_4~.
\ee
We will take $\{D_1, D_2, D_3, D_4\}$ as a basis. We have the GLSM description:
\be
\begin{tabular}{l|ccccccc}
 & $p_1$ &$p_2$& $p_3$&$p_4$& $p_5$ & $p_6$& $p_0$\\
 \hline
$\CC_1$   & $-1$  & $1$  & $0$  & $0$  & $0$  &$1$ & $-1$ \\
$\CC_2$   & $1$  & $-1$  & $1$  & $0$  & $0$  &$0$ & $-1$ \\
$\CC_3$   & $0$  & $1$  & $-1$  & $1$  & $0$  &$0$ & $-1$\\
$\CC_4$   & $0$  & $0$  & $1$  & $-1$  & $1$  &$0$ & $-1$ \\
 \end{tabular}
\ee
The homogenous coordinates and the $U(1)$ actions are associated to divisors and curves, respectively, as indicated. We denote by $\bE_0$ the exceptional divisor of the resolved singularity, and by $\CC_i = D_i \cdot \bE_0$ the exceptional curves for a particular resolution, as indicated on Figure~\ref{fig:dP3 curves}. 

Let us now consider the Beilinson quiver associated to the internal perfect matching $p_0= p_{13}$, with:
\be\label{p13 dP3}
p_{13} = \{X_{31}, X_{41}, X_{61}, X_{62} \}~.
\ee
The partially ordered Beilinson quiver $B\CQ_\MG(p_{13})$ is displayed in Fig.~\ref{fig:dP3 BQp13}.
The perfect matching \eqref{p13 dP3} determines a tilting bundle $T= \oplus_{i=1}^6 L_i$ on $\h \MG$, with:
\bea
&L_1= \CO~, \qquad && L_4= \CO(D_1+D_2+D_3)~,\cr
&L_2=\CO(D_2 +2D_3+D_4)~,\qquad && L_5=\CO(D_3+D_4+D_5)~,\cr
& L_3=\CO(D_1+D_2+D_3+D_4)~, \qquad && L_6=\CO(D_1+D_2+2 D_3+D_4)~.
\eea
Note that, in the resolution shown in Fig.~\ref{fig:dP3 curves}, the toric divisors on $\h\MG$ reduce to toric divisors on the exceptional locus $\CS= dP_3$. One can then compute:
\be
(S^{-1}_{ij})={\rm Hom}({\bf L}_i,{\bf L}_j)=\smat{
 1 & 2 & 4 & 4 & 3 & 5 \\
 0 & 1 & 2 & 2 & 1 & 3 \\
 0 & 0 & 1 & 0 & 0 & 1 \\
 0 & 0 & 0 & 1 & 0 & 1 \\
 0 & 0 & 1 & 1 & 1 & 2 \\
 0 & 0 & 0 & 0 & 0 & 1 }~, \qquad
S= \smat{
 1 & -2 & 1 & 1 & -1 & 1 \\
 0 & 1 & -1 & -1 & -1 & 1 \\
 0 & 0 & 1 & 0 & 0 & -1 \\
 0 & 0 & 0 & 1 & 0 & -1 \\
 0 & 0 & -1 & -1 & 1 & 0 \\
 0 & 0 & 0 & 0 & 0 & 1 }~.
\ee
Note that $B=S^T- S$ correctly reproduces the adjacency matrix of the full quiver, $\CQ_{\MG(dP_3)}$.
Let us also write down the monodromy matrix:
\be
M = S^{-T} S=  \smat{1 & -2 & 1 & 1 & -1 & 1 \\
 2 & -3 & 1 & 1 & -3 & 3 \\
 4 & -6 & 2 & 1 & -5 & 5 \\
 4 & -6 & 1 & 2 & -5 & 5 \\
 3 & -5 & 1 & 1 & -3 & 4 \\
 5 & -7 & 1 & 1 & -6 & 7}~,
\ee
which has four Jordan blocks with eigenvalues $\{1,1,1,1\}$, and the four eigenvectors:
\bea\label{eigenvectors M dP3}
&v_{\rm KK} = (1,1,1,1,1,1)~, \qquad\qquad && v_{F_1}= (1,1,0,2,0,0)~, \cr
&v_{\CI} = (3,1,2,2,2,0)~, \quad &&v_{F_2}= (1,1,2,0,0,0)~.
\eea
The dimension vector $v_{\rm KK}$ corresponds to the D0-brane, which is always stable and can probe the full resolved singularity. The dimension vectors $v_\CI, v_{F_1}, v_{F_2}$, on the other hand, form a basis for the ``non-anomalous fractional branes'' as described after \eqref{Gamma F def}.  The quiver representations with these dimension vectors correspond to D-branes wrapping  exceptional curves dual to non-compact divisors.

We can compute the brane charges of all the fractional branes as explained in section~\ref{subsec:brane charge}.  
In the K-theory basis:
\be
\big([{\bf C}_i]\big)= \big([\bE_0]~, \; [\CC_1]~, \; [\CC_2]~, \; [\CC_3]~, \; [\CC_4]~, \; [{\rm pt}] \big)~,
\ee
we find:
\be
Q =\smat{  1 & 0 & 0 & 0 & 0 & 0 \\
 1 & 0 & 0 & 1 & 1 & 0 \\
 1 & 0 & 1 & 2 & 1 & 0 \\
 1 & 1 & 1 & 1 & 1 & 0 \\
 1 & 1 & 1 & 1 & 0 & 0 \\
 1 & 1 & 1 & 2 & 1 & 1 }~,\qquad\qquad
Q^\vee= \smat{ 1 & 0 & 0 & 0 & 0 & 0 \\
 -1 & 0 & 0 & 1 & 1 & 0 \\
 0 & -1 & 0 & 0 & 0 & 0 \\
 0 & 0 & 0 & -1 & 0 & 0 \\
 -1 & 1 & 1 & 0 & -1 & 0 \\
 1 & 0 & -1 & 0 & 0 & 1}~.
\ee
In other words, we have:
\bea\label{K of B dP3}
&K[\CE_1] \cong [\bE_0]~, \qquad && K[\CE_4] \cong -[\CC_3]~, \cr
&K[\CE_2] \cong -[\bE_0]+ [\CC_3]+ [\CC_4]~, \qquad && K[\CE_5] \cong -[\bE_0]+ [\CC_1]+ [\CC_2]- [\CC_4]~, \cr
&K[\CE_3] \cong -[\CC_1]~,  \qquad\quad && K[\CE_6] \cong [\bE_0]- [\CC_2]+[{\rm pt}]~.
\eea

\paragraph{Geometric-engineering consistency check.} Let us compare the brane charges \eqref{K of B dP3} to the Coulomb-branch gauge-theory description. For an $SU(2)$ gauge theory with $N_f$ flavors, we expect BPS states associated to the monopole and the dyon,  in addition to $N_f$ electrically-charged BPS particles which also carry flavor charges. In addition, the 5d theory on $\R^4 \times S^1$ contains BPS states charged under the KK symmetry, $U(1)_{\rm KK}$, and the instanton symmetry, $U(1)_{\CI}$. We know that the monopole correspond to a D4-brane wrapping the exceptional divisor, $\bE_0$; therefore we can identify the simple object $e_1$ (the first node in the quiver) with the monopole state. Moreover, in the S-duality frame associated to the vertical direction of the toric diagram, the M2-branes wrapping the curves give rise to the following five-dimensional particles \cite{Closset:2018bjz}:
\bea\label{eq:statesdp3}
&W \; &&: \; \; \CC_3+ \CC_4~, \qquad \quad && M(W) =  2 \varphi~,  \cr
& \CH_{1,1}\; &&: \;\; \CC_3~, && M(\CH_{1,1}) = \varphi-m_1~, \cr
& \CH_{2,1}\; &&: \;\; \CC_4~, && M(\CH_{2,1}) = \varphi+m_1~, \cr
& \CH_{1,2}\; &&: \;\; \CC_1~, && M(\CH_{2,1}) = \varphi-m_2~, \cr
& \CH_{2,2}\; &&: \;\; \CC_6~, && M(\CH_{2,2}) = \varphi+m_2~, \cr
& \CI_{1}\; &&: \;\; \CC_5~, && M(\CI_1) = h_0 + \varphi- m_1 -m_2~, \cr
& \CI_{2}\; &&: \;\; \CC_2~, && M(\CI_2) = h_0 + \varphi~. \cr
\eea
Here, $W$, $\CH$ and $\CI$ denote the W-boson, hypermultiplet and instanton particles, respectively, and the five-dimensional real masses $M$ are given in terms of the 5d Coulomb branch VEV $\varphi$, the hypermultiplet masses $m_1, m_2$ and the inverse gauge coupling $h_0$, as reviewed in~\cite{Closset:2018bjz}. These wrapped M2-branes become D2-branes wrapped on the same curves in type IIA on $\R^4 \times \h \MG$.
Therefore, we see from \eqref{K of B dP3} that the fractional branes $\CE_1$ and $\CE_2$ can be identified with the monopole and with the dyon, respectively, and that the fractional branes $\CE_3$ and $\CE_4$ can be identified with hypermultiplet particles. The four quiver nodes $e_1, e_2, e_3, e_4$ form a subquiver, shown in blue in Fig.~\ref{fig:dP3 Q}, which is precisely the BPS quiver of the 4d $\CN=2$ $SU(2)$ $N_f=2$ gauge theory \cite{Alim:2011kw}---this include the correct superpotential, corresponding to the first two terms in \eqref{W dP3}. In addition, we have the fractional branes $\CE_5$ and $\CE_6$ which carry KK and instanton charge, and which complete the 4d quiver into a consistent 5d BPS quiver with superpotential, as determined by the brane tiling. 

Note also that the eigenvectors \eqref{eigenvectors M dP3} of the monodromy matrix correspond to the following brane charges:
\bea
&[v_{\rm KK}]  \cong [{\rm pt}]~,   \cr
&[v_{\CI} ]\cong  2[\CC_2]-[\CC_3]- [\CC_4]~,   \;\; && (M= 2h_0)~,\cr
& [v_{F_1}] \cong  -[\CC_3]+ [\CC_4]~,\;\; \qquad &&(M=2 m_1)~, \cr
    &[v_{F_2}] \cong  -[\CC_1]+ [\CC_6]~, \; \; &&(M=2 m_2)~. 
\eea
Here, we indicated the curves wrapped by the ``non-anomalous fractional branes,'' together with the (formal) mass of the corresponding 5d particle, in the 5d gauge-theory notation. 
 
\paragraph{Electric subcategory.} For this $dP_3$ quiver, the representation with dimension vector:
\be
\boldsymbol{\delta} = [\CE_1] + [\CE_2] = (1,1,0,0,0,0)
\ee
has a moduli space $\mathcal M \simeq \P_1$, and it is stable whenever:
\be\arg Z(\CE_2) < \arg Z(\CE_1)\,.\ee
Therefore, in this case, we have a vector multiplet in the spectrum and the corresponding elementary electric charge is $\mathbf{q} = {1\over 2} \boldsymbol{\delta}$. This corresponds to the reduction of the 5d W-boson we identified above. For a generic representation $\CO$ with dimension vector $(N_1,N_2, ...,N_6)$ the magnetic charge is
\be
\mathbf{m}(\CO) = \langle [\CO],\mathbf{q} \rangle_D = N_1 - N_2 + N_6 - N_5~,
\ee
in perfect agreement with \eqref{K of B dP3}. The controlled subcategory with control function $\mathbf{m}$ contains the quiver representations such that:
\be
N_1 + N_6 = N_2 + N_5\,.
\ee 
This category has several components, characterized by inequivalent effective quivers. Studying the details of such subcategories is beyond the scope of the present paper. Here, we will simply comment on two interesting components. Consider first the case $N_1 = N_2$ (which implies $N_5=N_6$). In this case, by a standard argument  (see e.g. Appendix D of \cite{Cecotti:2012va}), we expect that either $X^1_{12}$ or $X^2_{12}$ is an isomorphism. In this case we can exploit the isomorphism to identify the nodes $1$ and $2$. We therefore obtain an effective quiver description for the electric subcategory of the $E_3$ theory. For the patch such that $X^1_{12}$ is an isomorphism, we have an effective quiver description:
\medskip
\be\label{eq dP3 elcat}
\begin{gathered}
\xymatrix{*++[o][F]{3} \ar@/_0.9pc/[ddrrr]\ar@/^1.2pc/[rr]&&*++[o][F]{1} \ar@{=>}[ddl] \ar@(ur,ul)[] \ar@/^1.2pc/[ll] \ar@/_1.2pc/[rr] &&*++[o][F]{4} \ar@/_1.2pc/[ll]\ar[ddl]\\
\\
&*++[o][F]{5}\ar[uul]\ar@/^0.9pc/[uurrr] && *++[o][F]{6} \ar@{=>}[uul] & }\\
\begin{aligned}
&W_{eff}= X_{14} X_{41} -X_{11} X_{13} X_{31}  +X^1_{15} X_{53} X_{31} -X^1_{15} X_{54} X_{41}\\
&\qquad\; +X_{61}^2 X_{13} X_{36} - X_{61}^2 X_{14} X_{46} + X^2_{15} X_{54} X_{46}  X^1_{61} X_{11}
  -X_{15}^2 X_{53} X_{36}  X^1_{61} ~.
  \end{aligned}
\end{gathered}
\ee
Clearly, we can use the F-term relations $X_{14} = X^1_{15} X_{54}$ and $X_{41} = X_{46} X^2_{61}$ to integrate out these two arrows, resulting in the quiver with superpotential:
\be
\begin{gathered}
\phantom{hua!}\\
\xymatrix{*++[o][F]{3} \ar@/_0.9pc/[ddrrr]\ar@/^1.2pc/[rr]&&*++[o][F]{1} \ar@{=>}[ddl] \ar@(ur,ul)[] \ar@/^1.2pc/[ll] &&*++[o][F]{4} \ar[ddl]\\
\\
&*++[o][F]{5}\ar[uul]\ar@/^0.9pc/[uurrr] && *++[o][F]{6} \ar@{=>}[uul] & }\\
\begin{aligned}
&W^{(1)}_{eff}=  -X_{11} X_{13} X_{31}  +X^1_{15} X_{53} X_{31}+X_{61}^2 X_{13} X_{36} \\
&\qquad\; - X_{61}^2 X^1_{15} X_{54} X_{46} + X^2_{15} X_{54} X_{46}  X^1_{61} X_{11}
  -X_{15}^2 X_{53} X_{36}  X^1_{61} ~.
  \end{aligned}
\end{gathered}
\ee

\noindent Similarly, for the patch such that $X^2_{12}$ is an isomorphism, we have an effective quiver description:
\medskip

\be
\begin{gathered}
\xymatrix{*++[o][F]{3} \ar@/_0.9pc/[ddrrr]\ar@/^1.2pc/[rr]&&*++[o][F]{1} \ar@{=>}[ddl] \ar@(ur,ul)[] \ar@/^1.2pc/[ll] \ar@/_1.2pc/[rr] &&*++[o][F]{4} \ar@/_1.2pc/[ll]\ar[ddl]\\
\\
&*++[o][F]{5}\ar[uul]\ar@/^0.9pc/[uurrr] && *++[o][F]{6} \ar@{=>}[uul] & }\\
\begin{aligned}
&W_{eff}= X_{11} X_{14} X_{41}  - X_{13} X_{31}  +X^1_{15} X_{53} X_{31} -X^1_{15} X_{54} X_{41}\cr
&\qquad\; +X_{61}^2 X_{13} X_{36} - X_{61}^2 X_{14} X_{46} + X_{15}^2 X_{54} X_{46}  X_{61} 
  -X_{15}^2 X_{53} X_{36}  X_{61} X_{11}~.
  \end{aligned}
\end{gathered}
\ee
We see that we can integrate out $X_{13} = X_{15}^1 X_{53}$ and $X_{31} =  X_{36} X_{61}^2$, to obtain:
\be
\begin{gathered}
\phantom{yo!}\\
\xymatrix{*++[o][F]{3} \ar@/_0.9pc/[ddrrr]&&*++[o][F]{1} \ar@{=>}[ddl] \ar@(ur,ul)[]\ar@/_1.2pc/[rr] &&*++[o][F]{4} \ar@/_1.2pc/[ll]\ar[ddl]\\
\\
&*++[o][F]{5}\ar[uul]\ar@/^0.9pc/[uurrr] && *++[o][F]{6} \ar@{=>}[uul] & }\\
\begin{aligned}
&W^{(2)}_{eff}= X_{11} X_{14} X_{41}  -X^1_{15} X_{54} X_{41}+X_{61}^2 X_{15}^1 X_{53} X_{36} \\
&\qquad\; - X_{61}^2 X_{14} X_{46} + X_{15}^2 X_{54} X_{46}  X_{61} 
  -X_{15}^2 X_{53} X_{36}  X_{61} X_{11}~.
  \end{aligned}
\end{gathered}
\ee
It is interesting to note that, on the nodes $1,3,4$, we have the full quiver for the electric category of the 4d $\CN=2$ $SU(2)$ $N_f = 2$ theory \protect\cite{Cecotti:2012va}. Indeed, the restricted induced superpotential is such that for representations supported on $1,3,4$ only we obtain two copies of the following electric category, distinct only by the superpotentials:
\be
\begin{gathered}
\phantom{hej}\\
\xymatrix{*++[o][F]{3} \ar@/^1.2pc/[rr]&&*++[o][F]{1} \ar@(ur,ul)[] \ar@/^1.2pc/[ll] \ar@/_1.2pc/[rr] &&*++[o][F]{4} \ar@/_1.2pc/[ll]}\\
\\
W^{(1)}_{eff} =  X_{11}X_{13} X_{31} - X_{14} X_{41} \qquad W^{(2)}_{eff} = X_{13} X_{31} - X_{11} X_{14} X_{41}
\end{gathered}
\ee
In the first case we integrate $X_{14}$ and $X_{41}$ out, and in the second $X_{13}$ and $X_{31}$. As a result, the only stable objects with $N_5 = N_6 = 0$ in the electric category are indeed given by the representations with dimension vectors
\be
(N_1=N_2, N_3,N_4) \in \{(1,0,0),(0,1,0),(0,0,1),(1,1,0),(1,0,1)\}
\ee
which, with the exception of $(1,0,0)$ that gives rise to the W-boson, all correspond to hypermultiplets. Combining this with \eqref{eigenvectors M dP3}, we have that:
\be\begin{aligned}
&(0,1,0)=[\CE_3] = {1 \over 2} v_{F_2} - {1\over 2} \bd \\
&(0,0,1)=[\CE_4] = {1 \over 2} v_{F_1} - {1\over 2} \bd \\
&(1,1,0)=[\CE_1]+[\CE_2]+[\CE_3] = {1 \over 2} v_{F_2} + {1\over 2} \bd\\
&(1,0,1)=[\CE_1]+[\CE_2]+[\CE_4] = {1 \over 2} v_{F_1} + {1\over 2} \bd\\
\end{aligned}
\ee
These four objects, together with the corresponding CPT conjugates are precisely the components of two full hypermultiplets transforming in the fundamental representation of $SU(2)$. These are the states we have labeled $\CH_{\bullet,\bullet}$ in equation \eqref{eq:statesdp3}. Exploiting these relations, we have the further constraint:
\be
[\CE_1] + [\CE_2] + [\CE_3] + [\CE_4] + [\CE_5] + [\CE_6] = v_{KK} \quad\Rightarrow\quad  [\CE_5] + [\CE_6] = v_{KK} - {1\over 2} v_{F_1} - {1\over 2} v_{F_2}~,
\ee
from which it follows that $[\CE_5]$ and $[\CE_6]$ must have equal and opposite $U(1)_\CI$ charges.

\medskip
\noindent Another interesting component of the electric subcategory consists of the objects such that $N_1 = N_5$ and $N_2 = N_6$. In these cases, for those objects such that the arrows $X_{15}$ and $X_{61}$ are isomorphisms we obtain an effective quiver description consisting of a preprojective version of the affine $\widehat{A}_3$ quiver.

\subsection{The 5d $SU(2)$, $N_f=1$ gauge theory: local $dP_2$}
All the other ``rank-one'' toric singularities can be obtained by partial resolution of the $dP_3$ singularity discussed above. A partial resolution corresponds to removing points in a toric diagram, and its effect on the brane tiling is well-understood \cite{Franco:2005rj, Franco:2017jeo}.  

\begin{figure}[t]
\begin{center}
\subfigure[\small $dP_2$ toric diagram.]{
\includegraphics[height=4cm]{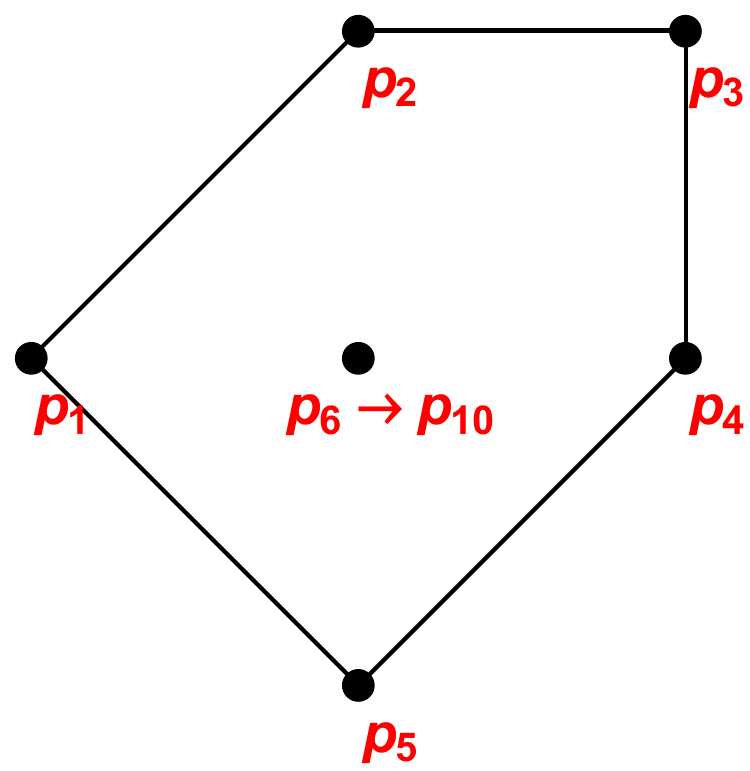}\label{fig:dP2 toric diag}}\;
\subfigure[\small  higgsed $dP_3$ brane tiling.]{
\includegraphics[height=4cm]{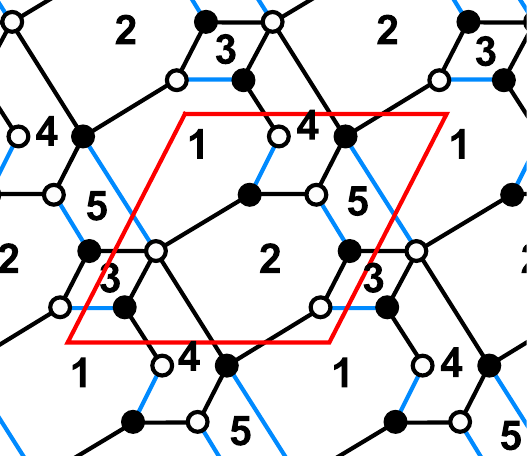}}\label{fig:dP2 bt from Higgs}\;
\subfigure[\small  $dP_2$ brane tiling.]{
\includegraphics[height=4cm]{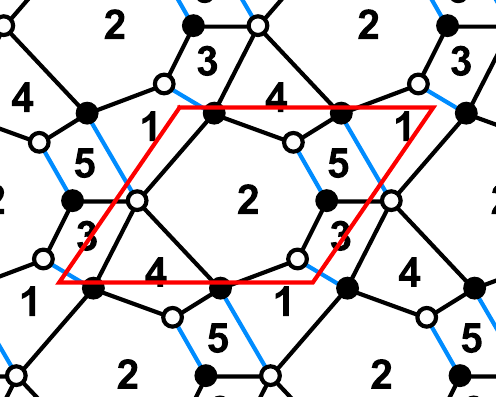}}\label{fig:dP2 bt}
\caption{\textsc{Left}: Toric diagram of the $dP_2$ singularity.  \textsc{Middle and right}:  $dP_2$ brane tiling obtained by higgsing the $dP_3$ brane tiling (with the faces relabelled). The perfect matching $p_{10}$ is shown in light blue. The edge $X_{41}$, shown in the middle figure, can be integrated out together with $X_{14}$, giving us the brane tiling on the right.  \label{fig:dP2 TD and BT}}
 \end{center}
 \end{figure}

\begin{figure}[t]
\begin{center}
\subfigure[\small $dP_2$, triangulated.]{
\raisebox{0.5cm}{\includegraphics[width=5.2cm]{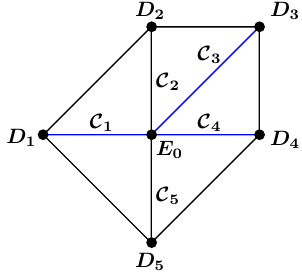}}\label{fig:dP2 TD with curves}}
\subfigure[\small  $\CQ_{\MG(dP_3)}$ higgsed.]{
\includegraphics[width=5cm]{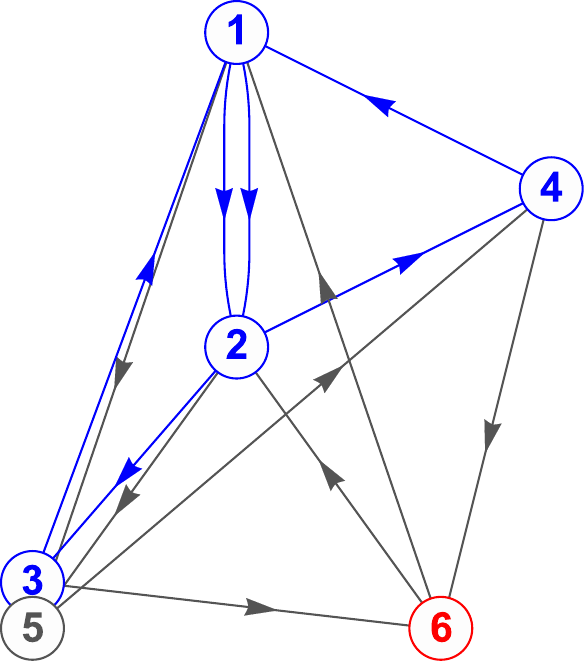}\label{fig:dP3 Q fused}}\hspace{-0.2cm}
\subfigure[\small  $\CQ_{\MG(dP_2)}$.]{
\includegraphics[width=5cm]{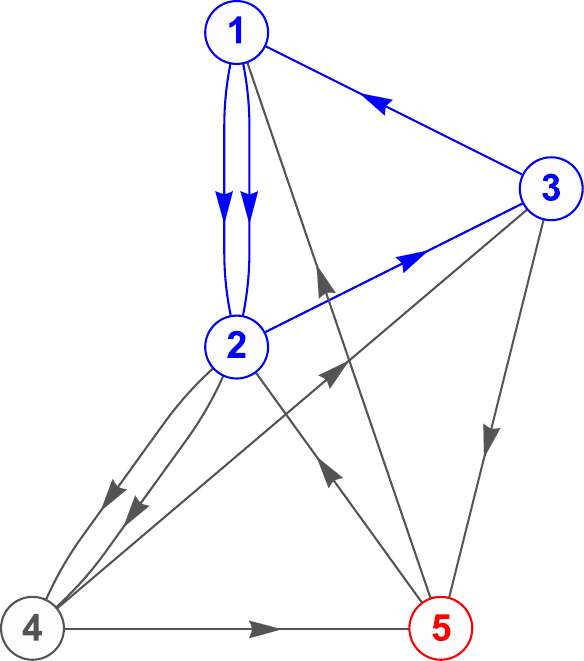}\label{fig:dP2 Q}}
\caption{\textsc{Left}: Triangulated toric diagram.  
\textsc{Middle}:  higgsing of the $dP_3$ quiver with $\langle X_{53}\rangle\neq 0$ corresponds to ``fusing'' two nodes. The pair of fields between $e_1$ and the new node $e_3=e_5$ are now massive and can be integrated out. 
\textsc{Right}: The resulting $dP_2$ quiver (with the nodes relabelled). The subquiver corresponding to the 4d $\CN=2$ $SU(2)$ $N_f=1$ gauge theory is indicated in blue. 
\label{fig:dP3 quiver higgsed to dP2}}
 \end{center}
 \end{figure}

Consider the partial resolution of the $dP_3$ singularity corresponding to removing the point $p_6$ in Fig.~\ref{fig:dP3 TD pms}. This gives us the $dP_2$ toric diagram shown in Fig.~\ref{fig:dP2 toric diag}. It corresponds to giving a VEV to the field $X_{53}$ in the $dP_3$ quiver. The effect is to ``fuse'' the nodes $e_3$ and $e_5$ together, as indicated in Figure~\ref{fig:dP3 Q fused}. The resulting quiver (upon integrating out massive arrows, and with the nodes relabelled) is shown in Fig.~\ref{fig:dP2 Q}.
Its superpotential is given by:
 \bea\label{W dP2}
&W_{dP_2}={\color{blue} X_{12}^1 X_{23} X_{31} }  -X_{23} X_{35} X_{52} +X_{24}^1 X_{45} X_{52}-X_{12}^2 X_{24}^1 X_{43}X_{31} \cr
&\qquad\quad +X_{12}^2 X_{24}^2 X_{43} X_{35}X_{51}-X_{12}^1 X_{24}^2 X_{45} X_{51}~.
 \eea
 This corresponds to the brane tiling of Figure~\ref{fig:dP2 bt}.
 
The smaller $dP_2$ singularity is obtained by first flopping the curve $\CC_6$ in Fig.~\ref{fig:dP3 curves}, and then sending its volume to infinity. In the 5d $SU(2)$ $N_f=2$ gauge theory, this correspond to turning on a large negative mass for the second hypermultiplet, $m_2 \rightarrow -\infty$, thus flowing to the $SU(2)$ $N_f=1$ gauge-theory description.  The higgsing of the BPS quiver implies the existence of a new simple object at the partially resolved singularity, which is the bound state $\CE_3\oplus \CE_5$ in the $dP_3$ theory.

We now have a Beilinson quiver associated to the perfect matching:
\be\label{p10 dP2}
p_{10}= \{X_{31}, X_{51}, X_{52} \}
\ee
of the $dP_2$ quiver. Note that \eqref{p10 dP2} descends from \eqref{p13 dP3} on the $dP_3$ brane tiling. The corresponding tilting line bundles are:
\bea
&L_1= \CO~, \qquad \qquad&& L_4= \CO(D_3+D_4+D_5)~,\cr
&L_2=\CO(D_3+D_4)~,\qquad\qquad\qquad && L_5=\CO(2D_3+2D_4+D_5)~.\cr
& L_3=\CO(D_3+2D_4+D_5)~,&&
\eea
Here, we pick the basis $\{ D_3, D_4, D_5\}$ for the non-compact toric divisors, with the exceptional curves $\CC_i \cong D_i \cdot \bE_0$ in the resolution of Fig.~\ref{fig:dP2 TD with curves}. We then have:
\be
S^{-1}= \smat{ 1 & 2 & 5 & 4 & 6 \\
 0 & 1 & 3 & 2 & 4 \\
 0 & 0 & 1 & 0 & 1 \\
 0 & 0 & 1 & 1 & 2 \\
 0 & 0 & 0 & 0 & 1}~,\quad
 M=\smat{1 & -2 & 1 & 0 & 1 \\
 2 & -3 & 1 & -2 & 3 \\
 5 & -7 & 2 & -5 & 6 \\
 4 & -6 & 1 & -3 & 5 \\
 6 & -8 & 1 & -6 & 8}~,\quad
 Q^\vee= \smat{1 & 0 & 0 & 0 & 0 \\
 -1 & 1 & 1 & 0 & 0 \\
 0 & -1 & 0 & 0 & 0 \\
 -1 & -1 & -1 & 1 & 0 \\
 1 & 1 & 0 & -1 & 1}~,
\ee
in the K-theory basis:
\be
\big([\bE_0]~, \; [\CC_3]~, \; [\CC_4]~, \; [\CC_5]~, \; [{\rm pt}] \big)~.
\ee
In particular, the simple objects are now assigned the brane charges:
\bea\label{K of B dP2}
&K[\CE_1] \cong [\bE_0]~, \qquad && K[\CE_4] \cong  -[\bE_0]- [\CC_3]- [\CC_4]+ [\CC_5]~, \cr
&K[\CE_2] \cong -[\bE_0]+ [\CC_3]+ [\CC_4]~, \qquad && K[\CE_5] \cong [\bE_0]+ [\CC_3]- [\CC_5]+[{\rm pt}]~. \cr
&K[\CE_3] \cong -[\CC_3]~,  \qquad\quad &&
\eea
This perfectly matches the 5d gauge-theory description \cite{Closset:2018bjz}, similarly to the $dP_3$ example. The fractional branes $\CE_1, \CE_2, \CE_3$ give rise to the 4d $\CN=2$ BPS quiver for the $SU(2)$ $N_f=2$ gauge theory, while the fractional branes $\CE_4$ and $\CE_5$ carry the instanton and KK charge. 
We can also derive \eqref{K of B dP2} directly from \eqref{K of B dP3}. Consider flopping the curve $\CC_6$ in the resolved $dP_3$ singularity of Fig.~\ref{fig:dP3 curves}. The new curves, denoted by $\CC'$, satisfy:
\be
[\CC_6']\cong -[\CC_6]~, \qquad 
[\CC_1']\cong [\CC_1]+[\CC_6]~, \qquad 
[\CC_5']\cong [\CC_1]+[\CC_6]~, \quad 
[\CC_{2,3,4}']\cong [\CC_{2,3,4}]~.
\ee
Sending the size of $\CC_6'$ to infinity, we obtain the geometry of Fig.~\ref{fig:dP2 TD with curves}, with $\CC'$ renamed to  $\CC$. From \eqref{K of B dP3}, we then see that:
\bea
&K[\CE_{1,2}]\big|_{dP_3}= K[\CE_{1,2}]\big|_{dP_2}~,  \qquad
&&K[\CE_{4}]\big|_{dP_3}= K[\CE_{3}]\big|_{dP_2}~,  \\
&K[\CE_3\oplus \CE_5]\big|_{dP_3}= K[\CE_4]\big|_{dP_2}~,  \qquad
&&K[\CE_{6}]\big|_{dP_3}= K[\CE_{5}]\big|_{dP_2}~,
\eea
thus reproducing \eqref{K of B dP2}. In particular, the dependence on $[\CC_6']$ cancels out, as expected.

\paragraph{Higgsing as a control.} At the level of the BPS categories, this Higgsing procedure is realized by a subcategory as well: the BPS category of the $E_2$ theory is obtained from that of the $E_3$ theory by imposing that the arrow $X_{53}$ is an isomorphism. At the level of the corresponding abelian heart this can be implemented by the control function:
\be
\lambda_H(\CO) \equiv N_3 - N_5~.
\ee
The RG flow from the $E_3$ theory to the $E_2$ theory is then realized by deforming the central charge for the BPS category of the $E_3$ theory by:
\be\label{eq stabat}
Z(\cdot) \mapsto \CZ(\cdot) \equiv Z(\cdot) + \alpha \lambda_H(\cdot)
\ee
and sending $\alpha \to \infty$. For this regime, whenever $X_{53}$ is not an isomorphism, this induces a destabilizing subrepresentation for the deformed stability condition in equation \eqref{eq stabat}. Indeed, consider a representation $\CO$ for the 5d BPS quiver of the $D_{S^1} E_3$ theory. Any subrepresentation with support on $\CO_5$ is destabilizing in this case: in the limit $\alpha \to \infty$, $\arg \CZ_5 \to \pi$. The subrepresentations of $\CO$ generated by $\text{ker } X_{53}$ and by the complement of $\text{coker } X_{53}$ are indeed supported on $X_5$ and therefore are destabilizing. As a result, the stable objects in the controlled subcategory are effectively described exploiting the quiver in Figure  \ref{fig:dP3 quiver higgsed to dP2}.

\paragraph{Electric subcategories and 5d duality.} It is interesting to note that, since the composition of two control functions is still a control, the Higgsing procedure is compatible with the process of taking the electric categories, and therefore our analysis of the previous section can be exploited to study the electric subcategories for the BPS category of the $E_2$ SCFT as well. In particular, we have that the representation with dimension vector:
\be
\bd = (1,1,0,0,0) 
\ee
gives the $SU(2)$ W-boson. The associated magnetic charge of an object $\CO$ with charge $\gamma_\CO = (N_1,N_2,N_3,N_4,N_5)$ is:
\be
\mathbf{m}(\CO) = \langle [\CO], \mathbf{q} \rangle_D =  N_1 - N_2 + N_5 - N_4~.
\ee
An effective quiver for the associated electric category can be obtained by identifying the nodes 1 and 2 exploiting the fact that either $X_{12}^1$ or $X_{12}^2$ is an isomorphism, we obtain:
\be\label{eq dP2 elcat}
\begin{gathered}
\xymatrix{
&*++[o][F]{1} \ar@{=>}[dd]\ar@(ul,dl)[]\ar@/^0.9pc/[rr]&&*++[o][F]{3} \ar@/^0.9pc/[ll]\ar[dd]\\
\\
&*++[o][F]{4}\ar[rr]\ar[uurr]&&*++[o][F]{5}\ar@{=>}[uull]\\
}
\end{gathered}
\ee
which can be obtained from the electric category associated to the $E_3$ SCFT in \eqref{eq dP3 elcat} by removing node 3 and relabelling. Of course, the associated superpotentials are also compatible. We stress that the control function also requires that $N_4=N_5$. It is clear that objects satisfying the above constraint are mutually local.

\medskip

Let us remark that, for the quiver in Fig.~\ref{fig:dP2 Q}, there is an inequivalent $SU(2)$ $N_f = 1$ 4d BPS full subquiver with superpotential supported on the nodes 2,4,5. The associated W-boson has charge:
\be
\widehat{\bd} = (0,1,0,1,0)~.
\ee
Note also that:
\be
\langle\bd,\widehat{\bd}\rangle_D = -4~,
\ee
and therefore the two are not mutually local. This is an interesting example of a 5d duality in action. The $E_2$ SCFT has two inequivalent gauge theory phases which are both characterized in terms of a 5d $SU(2)$ $N_f=1$ model. The two W-bosons we have just obtained correspond to these dual phases, that are respectively associated to the vertical and to the horizontal IIA reductions associated to the toric diagram in figure  \ref{fig:dP2 TD with curves}---for a discussion, see \cite{Closset:2018bjz}. 

\medskip

There is an alternative electric category associated to the magnetic charge:
\be
\widehat{\mathbf{m}}(\CO) = \langle [\CO], \widehat{\mathbf{q}}\rangle_D = N_1 + N_2 - N_3 - N_4
\ee
whose effective quiver description can be obtained exploiting patches associated to using $X_{24}^1$ or $X_{24}^2$ as isomorphisms. The resulting effective quiver is:
\be
\begin{gathered}
\xymatrix{
&*++[o][F]{1} \ar@{=>}[dd]&&*++[o][F]{3} \ar[ll]\ar[dd]\\
\\
&*++[o][F]{4}\ar@(ul,dl)[]\ar@/^0.9pc/[rr] \ar@{=>}[uurr]&&*++[o][F]{5}\ar@/^0.9pc/[ll]\ar[uull]\\
}
\end{gathered}
\ee
Remarkably, this is the opposite quiver compared to the one in equation \eqref{eq dP2 elcat}, and the resulting categories are therefore derived equivalent, which is expected since the electric category is associated to the genuinely 5d BPS states.

\medskip
\subsection{The 5d $SU(2)_0$ or  $SU(2)_\pi$ gauge theory: local $\mathbb{F}_0$ or $\mathbb{F}_1\cong dP_1$}
  \begin{figure}[t]
\begin{center}
\subfigure[\small $dP_2$ quiver.]{
\includegraphics[height=4.5cm]{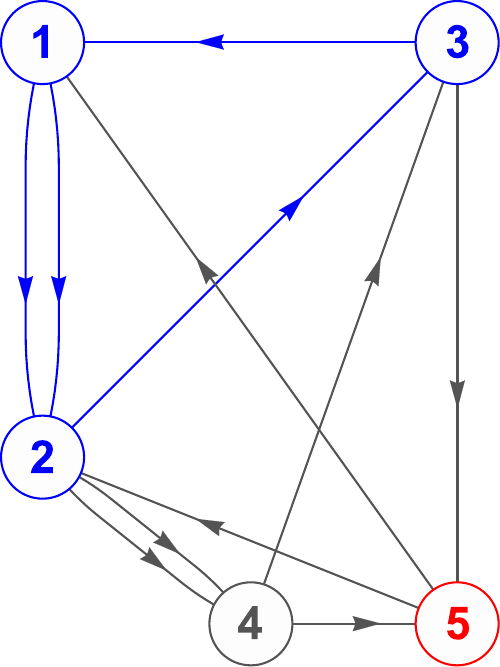}\label{fig:dP2 Q bis}} \;\;
\subfigure[\small $dP_2 \rightarrow \mathbb{F}_0$.]{
\includegraphics[height=4.5cm]{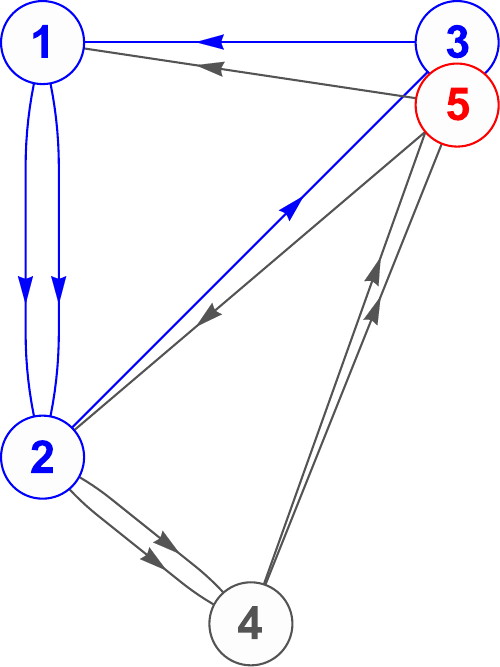}\label{fig:dP2 fused to F0}}\;\;
\subfigure[\small $dP_2 \rightarrow \mathbb{F}_1$.]{
\includegraphics[height=4.5cm]{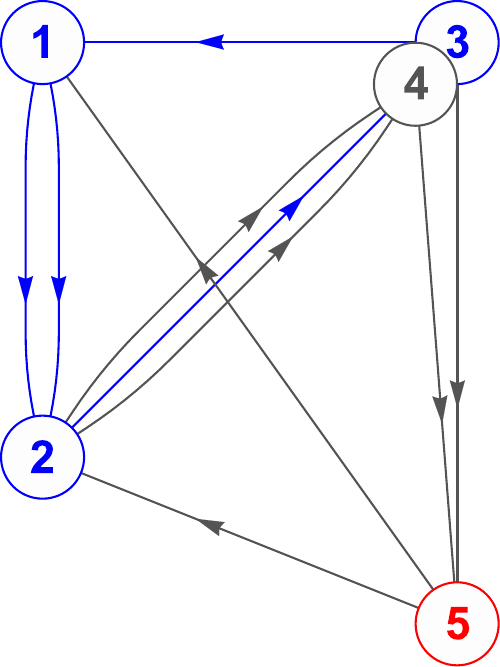}\label{fig:dP2 fused to F1}}\;\;
\subfigure[\small $dP_2 \rightarrow \mathbb{F}_1$, bis.]{
\includegraphics[height=4.5cm]{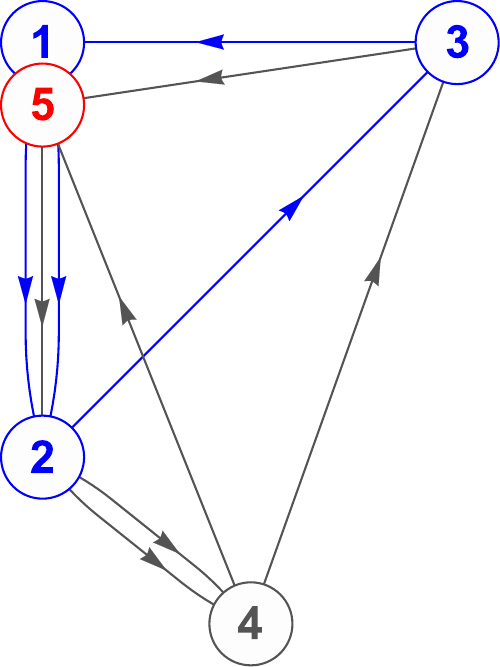}\label{fig:dP2 fused to F1 dual}}
\caption{higgsing patterns from the $dP_2$ quiver to the  $\mathbb{F}_0$ and $ \mathbb{F}_1\cong dP_1$ quivers, respectively. 
\label{fig:dP2 quiver higgsed to F0 F1}}
 \end{center}
 \end{figure} 
We can perform a further partial resolution of the $dP_2$ singularity to either the $\mathbb{F}_0$ or the $dP_1\cong \mathbb{F}_1$ singularity. This corresponds to removing  the point $p_3$ or $p_4$, respectively, in the toric diagram of Figure~\ref{fig:dP2 toric diag}. By studying the perfect matchings of the $dP_2$ tiling,  one finds that the two partial resolutions corresponds to the following simple Higgssing pattern on the $dP_2$ quiver:
\bea\label{partial res dP2}
& \text{remove} \; p_3: \quad && dP_2 \longrightarrow \mathbb{F}_0 \; \qquad &\leftrightarrow&\qquad \langle X_{43} \rangle \neq 0~,\cr
& \text{remove} \; p_4: \quad && dP_2 \longrightarrow \mathbb{F}_1 \; \qquad &\leftrightarrow&\qquad \langle X_{35} \rangle \neq 0~.
\eea
The corresponding fusions of nodes on the quiver are shown in Figure~\ref{fig:dP2 quiver higgsed to F0 F1}.
This reproduces the well-known toric quivers for these singularities.
From the point of view of the 5d $SU(2)$ $N_f=1$ gauge theory, the partial resolutions \eqref{partial res dP2} correspond to integrating out the single flavor hypermultiplet with a large positive or negative real mass, respectively. The $m_1{\rightarrow}\infty$ limit gives rise to the parity-preserving $SU(2)_0$ theory, while the  $m_1{\rightarrow}-\infty$ limit gives rise to the parity-violating $SU(2)_\pi$.

Note that another way to obtain the $dP_1$ singularity from $dP_2$ is by removing the point $p_2$ in Fig.~\ref{fig:dP2 toric diag}. This correspond to $\langle X_{51} \rangle \neq 0$ in the quiver, as shown in Fig.~\ref{fig:dP2 fused to F1 dual}, giving rise to an equivalent $dP_1$ quiver. From the point of view of the gauge-theory description (in our ``vertical'' $S$-duality frame), this corresponds to integrating out an instanton particle, and therefore looks non-perturbative from the 5d low-energy point of view.~\footnote{Of course, this can also be seen as an ordinary ``perturbative flow'' in some S-dual description (the 5d $SU(2)$ $N_f=1$ theory  is ``self-dual'', with the S-duality corresponding to rotating the $dP_2$ toric diagram by 90 degrees).}
Incidentally, the two ``perturbative'' partial resolutions~\eqref{partial res dP2} are the only ones that preserve the perfect matching $p_{10}$ in \eqref{p10 dP2}, in the sense that $p_{10}$ is mapped to a perfect matching of the brane tiling of the higgsed theory; instead, we have $X_{51}=p_2 p_{10}$ in terms of perfect matching variables, and therefore the higgsing of Fig.~\ref{fig:dP2 fused to F1 dual} removes that perfect matching from the description.

\begin{figure}[t]
\begin{center}
\subfigure[\small $\mathbb{F}_0$, triangulated.]{
\includegraphics[height=5cm]{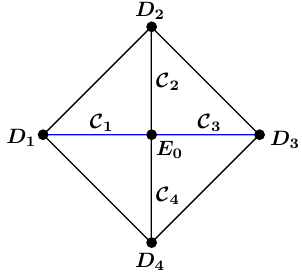}\label{fig:F0 with curv}} \;\qquad\qquad
\subfigure[\small $\CQ_{\MG(\mathbb{F}_0)}$.]{
\includegraphics[height=5cm]{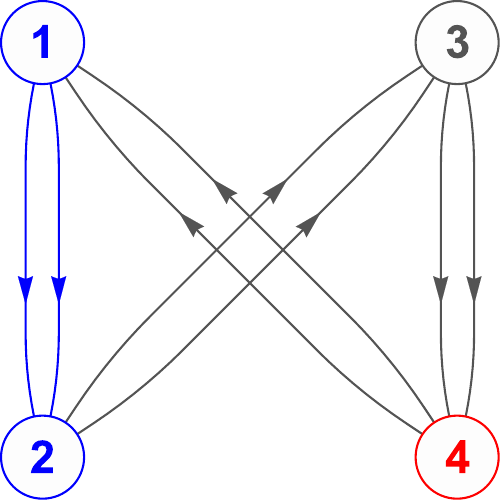}\label{fig:F0 quiver}}
\caption{Triangulated toric diagram and BPS quiver for the $\mathbb{F}_0$ singularity.
\label{fig:F0 quiver and toric diag}}
 \end{center}
 \end{figure}

\paragraph{The $\mathbb{F}_0$ BPS quiver.}  Let us first consider the resolved $\mathbb{F}_0$ singularity, with the toric diagram shown in Fig.~\ref{fig:F0 with curv}. 
The quiver is shown in Figure~\ref{fig:F0 quiver}.
It has superpotential:
\be
W_{\mathbb{F}_0}= X_{12}^1 X_{23}^1 X_{34}^2 X_{41}^2 - X_{12}^1 X_{23}^2 X_{34}^2 X_{41}^1+X_{12}^2 X_{23}^2 X_{34}^1 X_{41}^1-X_{12}^2 X_{23}^1 X_{34}^1 X_{41}^2~.
\ee
 
 The toric geometry of Fig.~\ref{fig:F0 with curv} has the relations $D_1 \cong D_3$ and $D_2 \cong D_4$ amongst the toric divisors, and we pick a basis $\{ D_1, D_2\}$. The $\mathbb{F}_0$ brane tiling has 4 external and 4 internal perfect matchings.
From the $dP_2$ brane tiling, we inherit the internal perfect matching:
\be
p_8= \{ X_{41}^1, X_{41}^2\}~,
\ee 
  of the $\mathbb{F}_0$ tiling. This gives us to the tilting line bundles:
\be
  L_1 = \CO~, \qquad L_2 = \CO(D_1)~, \qquad L_3 = \CO(D_1+ D_2)~, \qquad L_4=\CO(2D_1 +D_2)~.
\ee
 One can then compute:
 \be
 S^{-1}= \smat{ 1 & 2 & 4 & 6 \\
 0 & 1 & 2 & 4 \\
 0 & 0 & 1 & 2 \\
 0 & 0 & 0 & 1}~, \qquad
 M= \smat{ 1 & -2 & 0 & 2 \\
 2 & -3 & -2 & 4 \\
 4 & -6 & -3 & 6 \\
 6 & -8 & -6 & 9}~, \qquad
 Q^\vee= \smat{1 & 0 & 0 & 0 \\
 -1 & 1 & 0 & 0 \\
 -1 & -1 & 1 & 0 \\
 1 & 0 & -1 & 1}~.
 \ee
 We use the K-theory basis $([\bE_0], [\CC_1], [\CC_2], [{\rm pt}])$ for the brane charge, so that:
  \bea\label{K of E F0}
& K(\CE_1)= [\bE_0]~, \qquad 
  &&K(\CE_3)=  -[\bE_0]-[\CC_1]+[\CC_2]~,  \cr
  &  K(\CE_2)= -[\bE_0]+[\CC_1]~, \qquad 
   &&K(\CE_4)= [\bE_0]-[\CC_2]+ [{\rm pt}]~. 
 \eea
Indeed, the M2-brane wrapped over the curve $\CC_1$ gives rise to the W-boson in this S-duality frame, and the curve $\CC_2$ gives the instanton particle. The subquiver formed by $\CE_1$ and $\CE_2$ is the 4d $\CN=2$ BPS quiver for the pure $SU(2)$ gauge theory.

The naive analysis of electric subcategory for this model is very similar to the one done above for the $E_2$ and the $E_3$ theories. It is discussed in more detail in Appendix~\ref{app:F0ele}.

\begin{figure}[t]
\begin{center}
\subfigure[\small $dP_1$, triangulated.]{
\includegraphics[height=5cm]{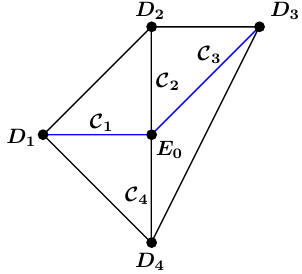}\label{fig:F1 with curv}} \;\qquad\qquad
\subfigure[\small $\CQ_{\MG(dP_1)}$.]{
\includegraphics[height=5cm]{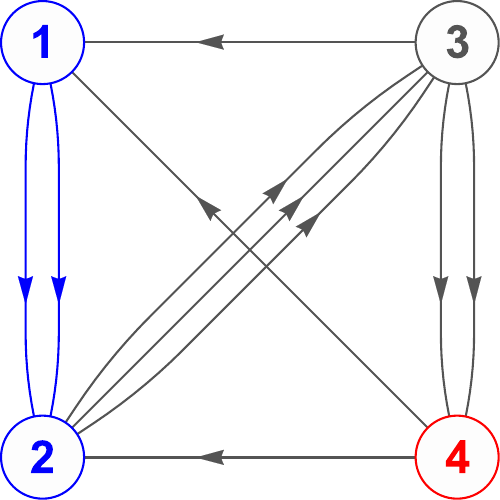}\label{fig:F1 quiver}}
\caption{Triangulated toric diagram and BPS quiver for the $\mathbb{F}_0$ singularity.
\label{fig:dP1 quiver and toric diag}}
 \end{center}
 \end{figure}

\paragraph{The $\mathbb{F}_1$ BPS quiver.}  Let us next consider the resolved $\mathbb{F}_1 \cong dP_1$ singularity, with the toric diagram shown in Fig.~\ref{fig:F1 with curv}. 
The quiver is shown in Figure~\ref{fig:F1 quiver}.
It has superpotential:
\bea
&W_{dP_1}&=&\; X_{12}^2 X_{23}^2 X_{31} - X_{12}^2 X_{23}^1 X_{31} + X_{23}^1 X_{34}^2 X_{42} - X_{23}^2 X_{34}^1 X_{42} \cr
&&&+ X_{12}^2 X_{23}^3 X_{34}^1 X_{41}- X_{12}^1 X_{23}^3 X_{34}^2 X_{41}~.
\eea
For this singularity, we have the relations $D_1 \cong D_3$ and $D_4 \cong D_1 + D_2$ amongst toric divisors. We also have the linear equivalences:
\be
\CC_3 \cong \CC_1~, \qquad \CC_4 \cong \CC_1+ \CC_2~,
\ee
amongst the curves in the resolution of Fig.~\ref{fig:F1 with curv}. Now, the perfect matching inherited from $dP_2$ is:
\be\label{p8 dP1}
p_8= \{ X_{42}, X_{41}, X_{31} \}~, 
\ee
which gives us the line bundles:
\be
  L_1 = \CO~, \qquad L_2 = \CO(D_1)~, \qquad L_3 = \CO(2D_1+ D_2)~, \qquad L_4=\CO(3D_1 +D_2)~.
\ee
 One can then compute:
 \be
 S^{-1}= \smat{ 1 & 2 & 5 & 7 \\
 0 & 1 & 3 & 5 \\
 0 & 0 & 1 & 2 \\
 0 & 0 & 0 & 1}~, \qquad
 M= \smat{  1 & -2 & 1 & 1 \\
 2 & -3 & -1 & 3 \\
 5 & -7 & -3 & 6 \\
 7 & -9 & -6 & 9}~, \qquad
 Q^\vee= \smat{1 & 0 & 0 & 0 \\
 -1 & 1 & 0 & 0 \\
 -1 & -1 & 1 & 0 \\
 1 & 0 & -1 & 1}~.
 \ee
Here, we picked the K-theory basis $([\bE_0], [\CC_1], [\CC_2], [{\rm pt}])$, in which case the  brane charges of the fractional branes take exactly the same form as in \eqref{K of E F0}. Note that the local geometry $\mathbb{F}_1$ is the simplest non-trivial example of a ruled surface:
\be
\mathbb{P}^1 \cong \CC_1 \longrightarrow  \mathbb{F}_1  \longrightarrow  \CC_2~.
\ee
It is well known that the degeneration of the fiber curve, $\CC_1$, gives rise to an $SU(2)$ gauge group in 5d, while the M2-brane on the base curve gives us the instanton particle.

\subsection{The $E_0$ SCFT and local $\mathbb{P}^2$}\label{subsec: E0 rk1}
\begin{figure}[t]
\begin{center}
\subfigure[\small $dP_0$, triangulated.]{
\includegraphics[height=4.5cm]{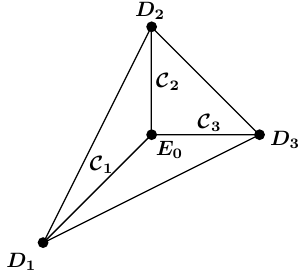}\label{fig:dP0 with curv}} \;
\subfigure[\small $dP_1 \rightarrow dP_0$.]{
\includegraphics[height=4.5cm]{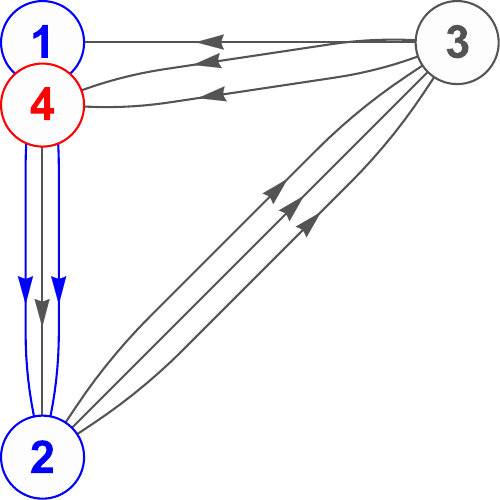}\label{fig:dP1 higgsed}}\;
\subfigure[\small $\CQ_{\MG(dP_0)}$.]{
\includegraphics[height=4.5cm]{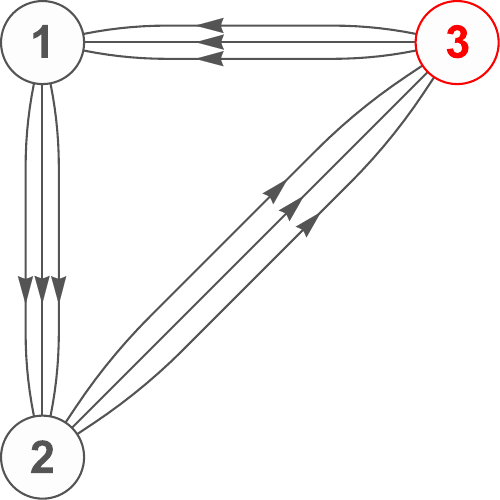}\label{fig:dP0 quiver}}
\caption{Triangulated toric diagram and BPS quiver for the $dP_0$ singularity. 
\label{fig:dP0 quiver and toric diag}}
 \end{center}
 \end{figure}

To conclude this discussion of rank-one theories with 5d BPS quiver of toric type, we should discuss the $E_0$ singularity, corresponding to the collapse of a $\mathbb{P}_2$, whose corresponding 5d SCFT has no gauge-theory phase \cite{Morrison:1996xf}. The $\mathbb{P}_2\cong dP_0$ geometry can be obtained by partial resolution of the $dP_1$ singularity, removing the point $p_2$ in Figure~\ref{fig:F1 with curv}. This correspond to higgsing the $dP_1$ quiver of Fig.~\ref{fig:F1 quiver} with  $\langle X_{41} \rangle \neq 0$, as shown in Fig.~\ref{fig:dP0 with curv}. This gives us the well-studied $dP_0$ quiver of Fig.~\ref{fig:dP0 quiver}, with the superpotential:
\be
W_{dP_0} = \epsilon_{abc}  X_{12}^a X_{23}^b X_{31}^c~,
\ee
with $a,b,c\in \{1,2,3\}$. 

Note that we have $X_{41}=p_2 p_8$ in the $dP_1$ theory, so the present higgsing does not preserve the prefect matching \eqref{p8 dP1}. Instead, to study the brane charge, we can pick, for instance, the perfect matching:
\be
p_6= \{X_{31}^1, X_{31}^2, X_{31}^3 \}~.
\ee
The corresponding Beilinson quiver reassigns the KK charge to the third node, $e_3$, by convention. In this geometry, we have $D_1 \cong D_2\cong D_3$. The tilting line bundles are:
\be
L_1= \CO~, \qquad L_2= \CO(D_1)~,\qquad L_2 = \CO(D_2)~,
\ee
corresponding to the exceptional collection $\{\CO, \CO(1), \CO(2)\}$ on $\mathbb{P}^2$. We then find:
\be
S^{-1} = \smat{1 & 3 & 6 \\
 0 & 1 & 3 \\
 0 & 0 & 1 }~, \qquad
 M= \smat{ 1 & -3 & 3 \\
 3 & -8 & 6 \\
 6 & -15 & 10}~, \qquad Q^\vee=  \smat{1 & 0 & 0 \\
 -2 & 1 & 0 \\
 1 & -1 & 1}~,
\ee
 in the K-theory basis $([\bE_0], [\CC_1], [{\rm pt}])$. The K-theory charges are:
 \be\label{dP0 brane charges}
 K(\CE_1) = [\bE_0]~, \qquad  K(\CE_2) =-2 [\bE_0]+[\CC_1]~, \qquad 
  K(\CE_3) = [\bE_0]-[\CC_1]+[{\rm pt}]~.
 \ee
In this case, it is well known that  the fractional branes at the (resolved) singularity can be taken to be
$\CE_1 \cong i_\ast \CO_{\mathbb{P}^2}$, $\CE_2 \cong i_\ast \Omega_{\mathbb{P}^2}(1)[1]$ and $\CE_3 \cong i_\ast  \Omega^2_{\mathbb{P}^2}(2)[2]$ \cite{Douglas:2000qw, Cachazo:2001sg, Aspinwall:2004jr}.~\footnote{See {\it e.g.} Appendix B of \protect\cite{Closset:2017yte} for a recent review.}

\paragraph{Electric category for $D_{S^1}E_0$.} Naively the $E_0$ theory does not have an electric category as defined above, since it does not admit gauge theory phases. Indeed the $E_0$ theory is a close 5d analogue of the simplest 4d $\CN=2$ AD theories. However, there is still a notion of magnetic charge one the Coulomb branch, anda corresponding controlled subcategory. We have seen above that the BPS category of the $E_0$ theory is obtained by higgsing from that of the $SU(2)_\pi$ theory. The latter has a well-defined magnetic charge given by:
\be
\mathbf{m}(\CO) = {1\over 2} \langle [\CO], \bd\rangle_D  = N_1 - N_2 - N_3 + N_4~.
\ee
Exploiting it as a control, we can easily describe the corresponding effective quivers for the electric subcategory of $D_{S_1}\widehat{E}_1$. Now, to obtain the electric category associated to the $E_0$ theory, we simply have to exploit the control: 
\be\lambda_{\widehat{E}_1 \to E_0} = N_1 - N_4~,\ee 
which implements the higgsing RG flow $\widehat{E}_1 \to E_0$, by imposing that the arrow $X_{41}$ is an isomorphism. The magnetic charge for the $E_0$ theory is therefore: 
\be
\mathbf{m}_{E_0} \equiv \mathbf{m}_{\widehat{E}_1} \Big|_{\lambda_{\widehat{E}_1 \to E_0} = 0} = 2N_1 - N_2 - N_3
\ee
and the controlled subcategory with $\mathbf{m}_{E_0}$ as a control function gives the genuine 5d BPS particles. Taking into account the relabelling of the nodes from Fig.~\ref{fig:dP1 higgsed} to Fig.~\ref{fig:dP0 quiver}, this agrees with the charge assignment \eqref{dP0 brane charges}.


\section{Rank-two BPS quivers and UV duality}\label{sec:rank2}
 \begin{figure}[t]
\begin{center}
\subfigure[\small $SU(3)_0^{N_f{=}2}$]{
\includegraphics[width=2.6cm]{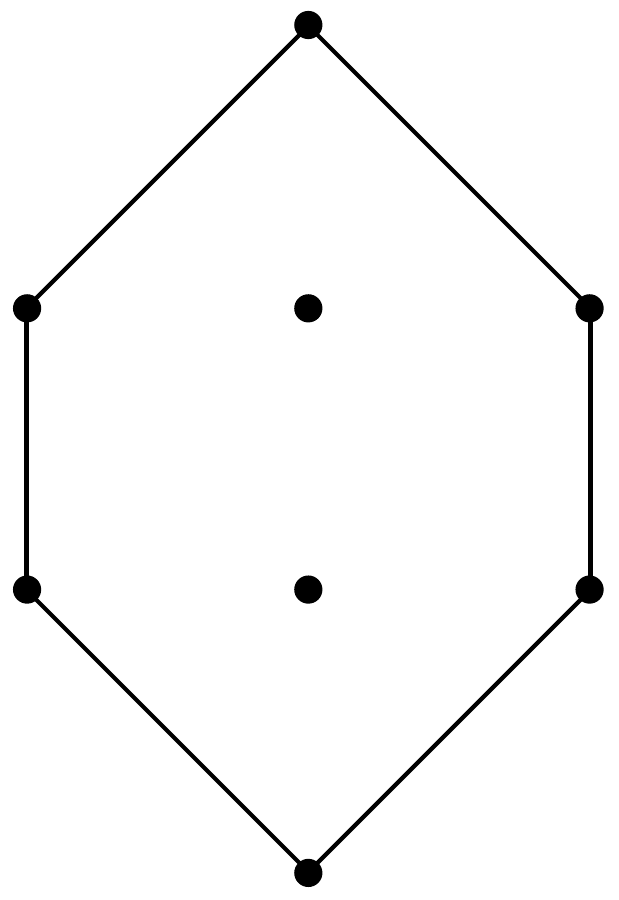}\label{fig: SU3 0 Nf2 sing}}\quad
\subfigure[\small $SU(3)_1^{N_f{=}2}$]{
\includegraphics[width=2.6cm]{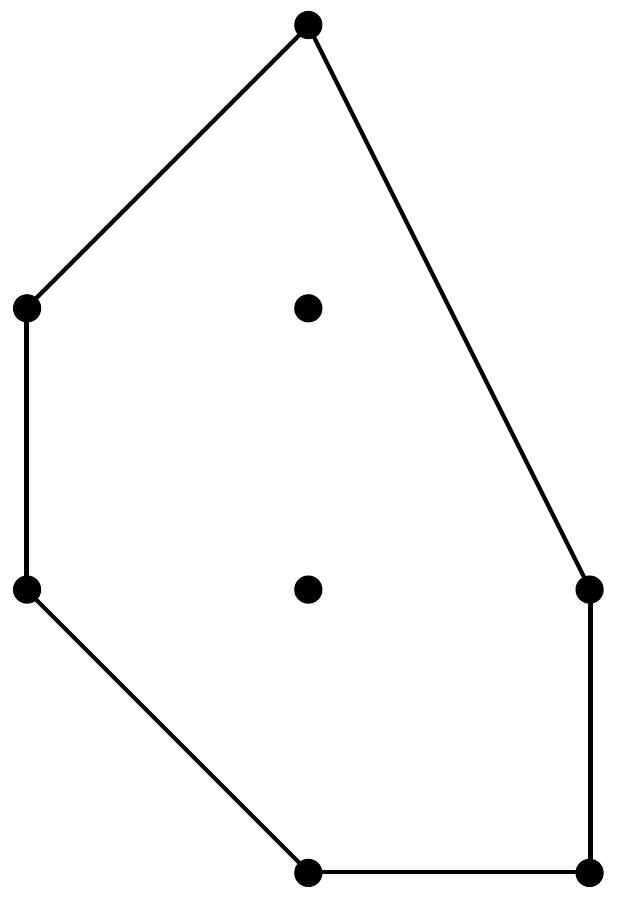}\label{fig: SU3 1 Nf2 sing}}\quad
\subfigure[\small $SU(3)_{1\ov 2}^{N_f{=}1}$]{
\includegraphics[width=2.6cm]{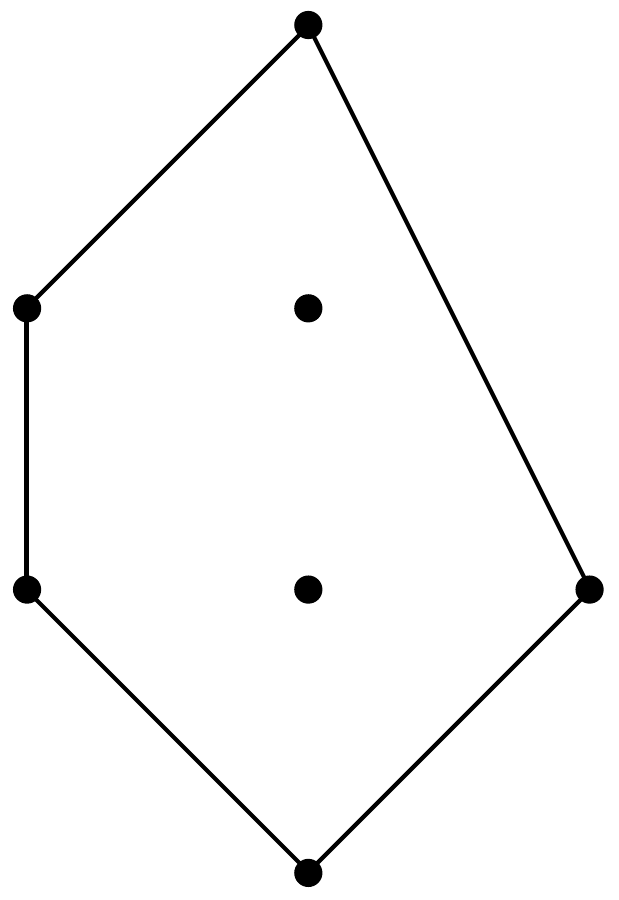}\label{fig: SU3 12 Nf1 sing}}\quad
\subfigure[\small $SU(3)_{3\ov 2}^{N_f{=}1}$]{
\includegraphics[width=2.6cm]{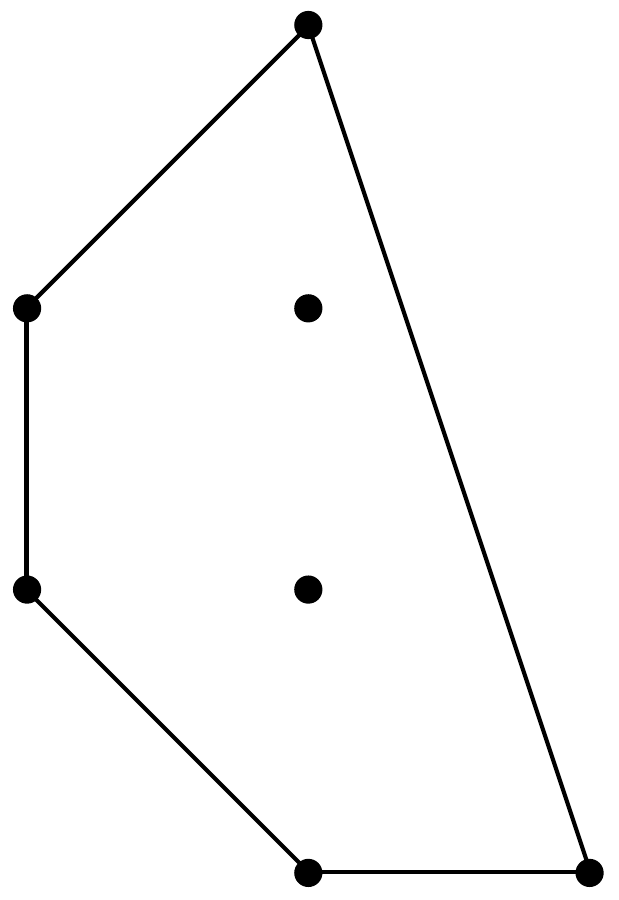}\label{fig: SU3 32 Nf1 sing}}\quad
\subfigure[\small $SU(3)_0$]{
\includegraphics[width=2.6cm]{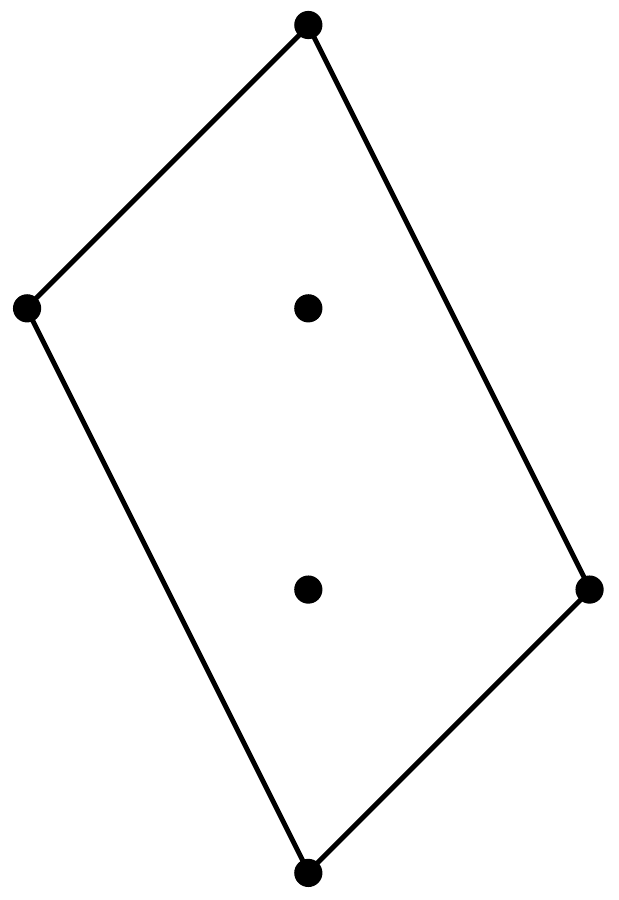}\label{fig: SU3 0 Nf0 sing}}
\subfigure[\small $SU(3)_1$]{
\includegraphics[width=2.6cm]{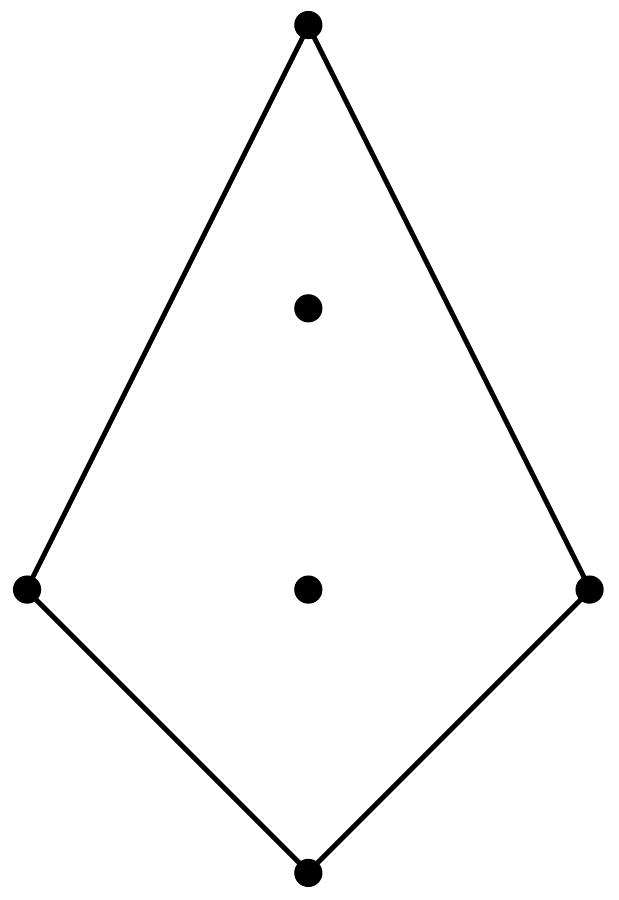}\label{fig: SU3 1 Nf0 sing}}\quad
\subfigure[\small $SU(3)_2$]{
\includegraphics[width=2.6cm]{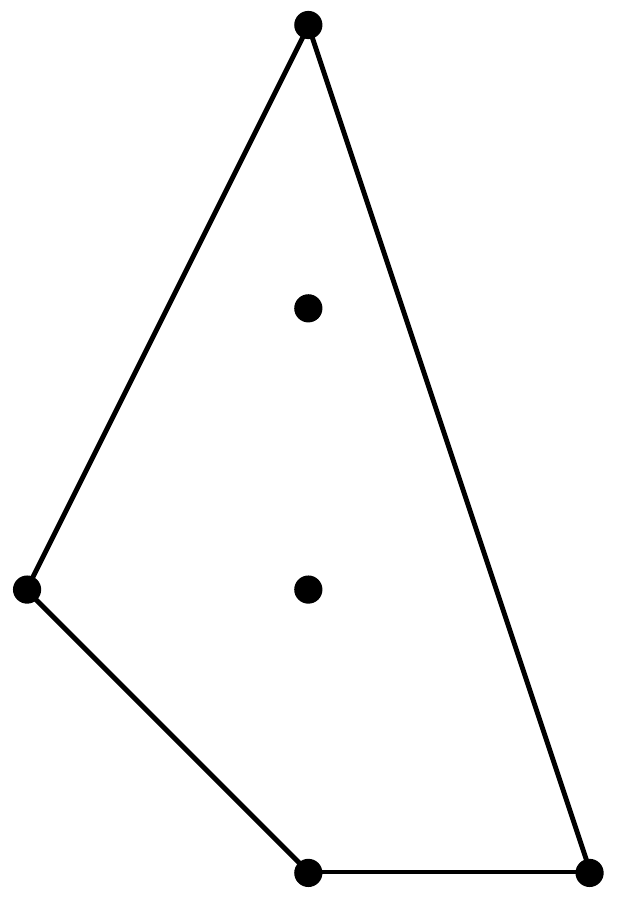}\label{fig: SU3 2 Nf0 sing}}\quad
\subfigure[\small ${\rm NL}^{N_f{=}2}$]{
\includegraphics[width=2.6cm]{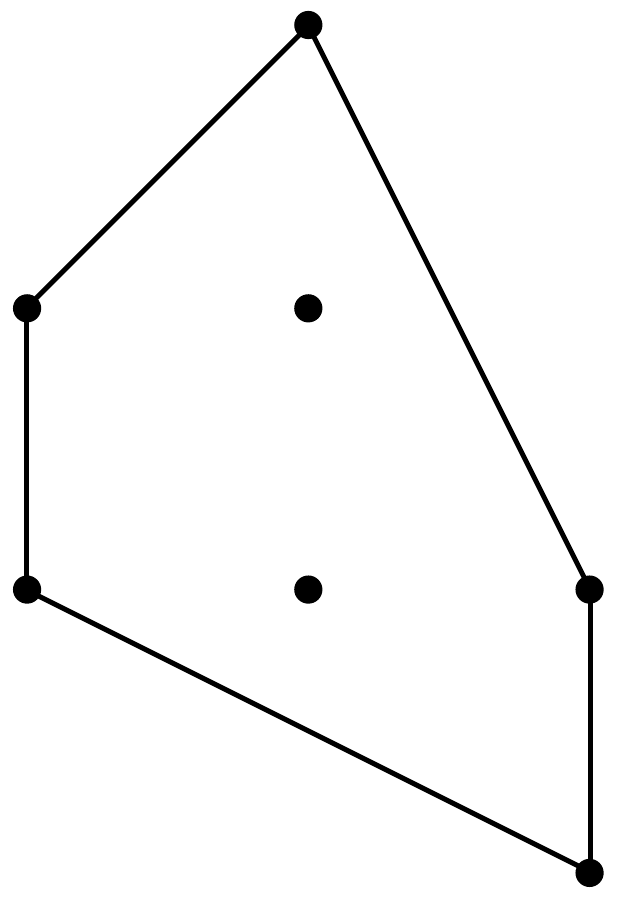}\label{fig: NL Nf2 sing}}\quad
\subfigure[\small ${\rm NL}^{N_f{=}1}$]{
\includegraphics[width=2.6cm]{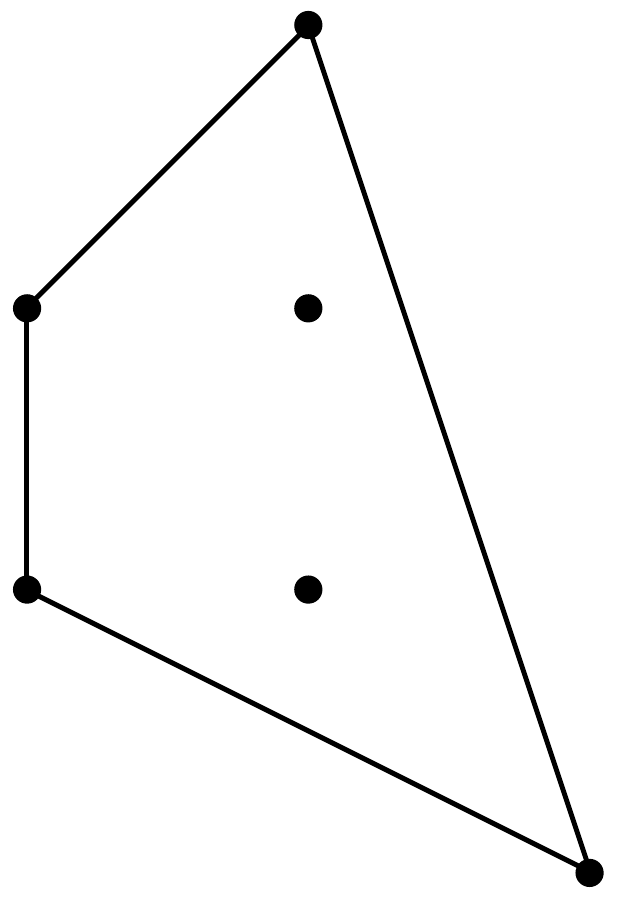}\label{fig: NL Nf1 sing}}\quad
\subfigure[\small ${\rm NL}^{N_f{=}0}$]{
\includegraphics[width=2.6cm]{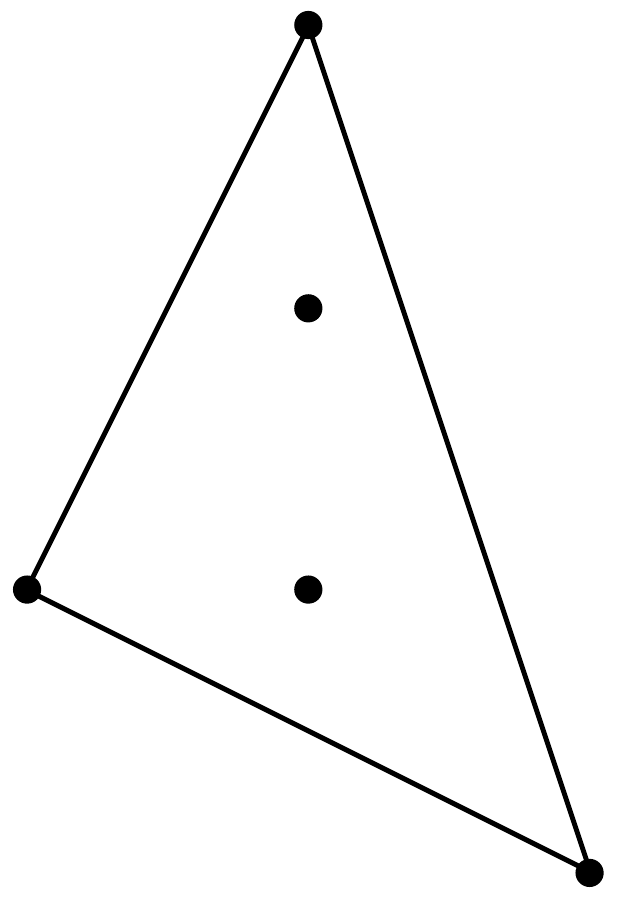}\label{fig: NL Nf0 sing}}
\caption{The 10 rank-two isolated toric singularities, labelled by  their $SU(3)$ gauge-theory phases. The last three singularties are ``non-Lagrangian''---they do not have any weakly-coupled gauge-theory description.  \label{fig:rk 2 geoms}}
 \end{center}
 \end{figure}

In this section, we consider some rank-two 5d SCFTs and their associated 5d BPS quivers. 
There are only ten isolated rank-two toric singularities, out of which seven have a 5d gauge-theory phase with an $SU(3)$ gauge group, as summarized on Figure~\ref{fig:rk 2 geoms}. See  \cite{Saxena:2019wuy} for a detailed discussion. Note that all these singularities can be obtained by partial resolution from the first two, in Fig.~\ref{fig: SU3 0 Nf2 sing} and Fig.~\ref{fig: SU3 1 Nf2 sing}, which correspond to $SU(3)_k$ with two flavors and CS level $k=0$ or $k=1$, respectively. The 5d RG flows correspond to integrating out flavors and/or instanton particles.

  \begin{figure}[t]
\begin{center}
\subfigure[\small $SU(3)$ $N_f=2$ ]{
\includegraphics[height=5cm]{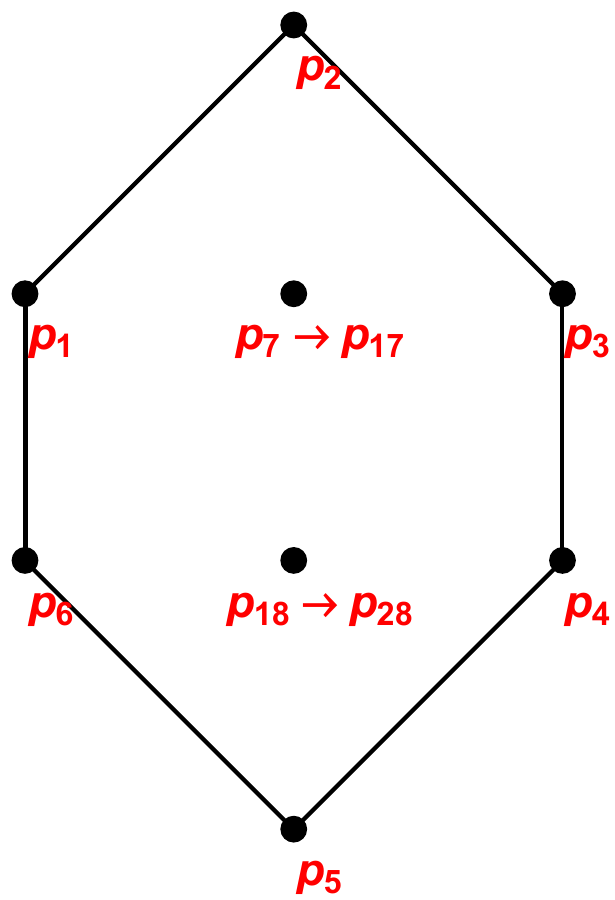}\label{fig:toric diag SU3Nf2}}\;\;\;\qquad\qquad
\subfigure[\small $SU(2)\times SU(2)$]{
\raisebox{0.5cm}{\includegraphics[width=5cm]{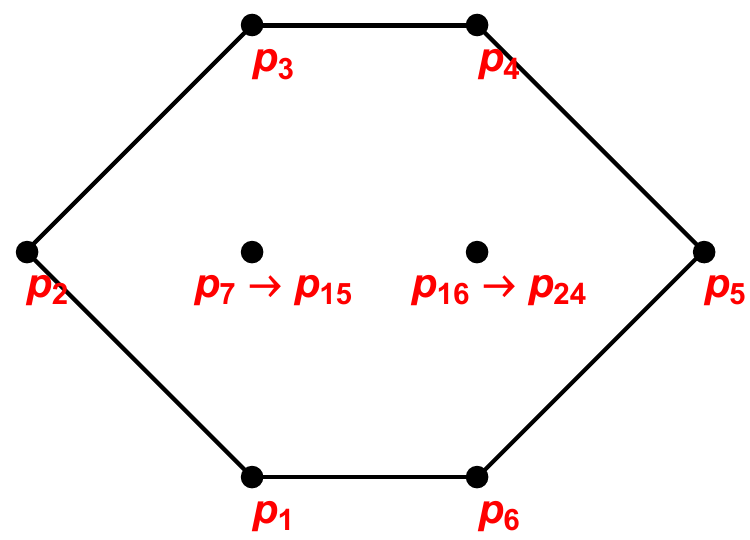}\label{fig:toric diag SU2SU2}}}
\caption{Toric diagram for the beetle geometry, with the external and internal perfect matchings as indicated for either gauge theory phase, corresponding to ``vertical reductions''~\protect\cite{Closset:2018bjz}. \label{fig:beetle toric diag}}
 \end{center}
 \end{figure} 
 
 \subsection{The 5d $SU(3)_0$, $N_f=2$ gauge theory and its BPS quiver}\label{section:beetle}
A particularly interesting rank-two 5d SCFT is obtained from the toric singularity of Fig.~\ref{fig: SU3 0 Nf2 sing}. Its gauge-theory phases were recently discussed in detail in \cite{Closset:2018bjz}.  It has two distinct 5d gauge-theory descriptions, either as an $SU(3)$ gauge theory with $N_f=2$ hypermultiplets in the fundamental representation, or as an $SU(2) \times SU(2)$ quiver gauge theory \cite{Bergman:2013aca}; this follows from the ``vertical reduction'' of the toric diagrams as shown in Figure~\ref{fig:beetle toric diag}. 

\begin{figure}[t]
\begin{center}
\subfigure[\small $\CQ_\MG\left(SU(3)_0, N_f=2\right)$.]{
\includegraphics[height=7cm]{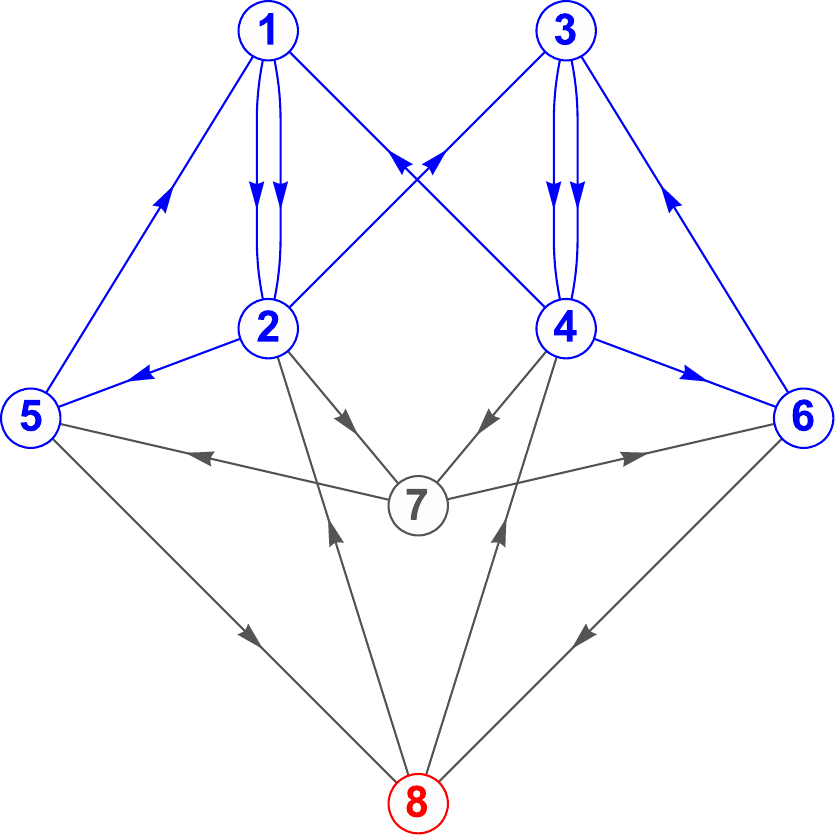}\label{fig:QSU3Nf2}}\;\;\;
\subfigure[\small $\CQ_\MG\left(SU(2)\times SU(2)\right)$.]{
\includegraphics[height=7cm]{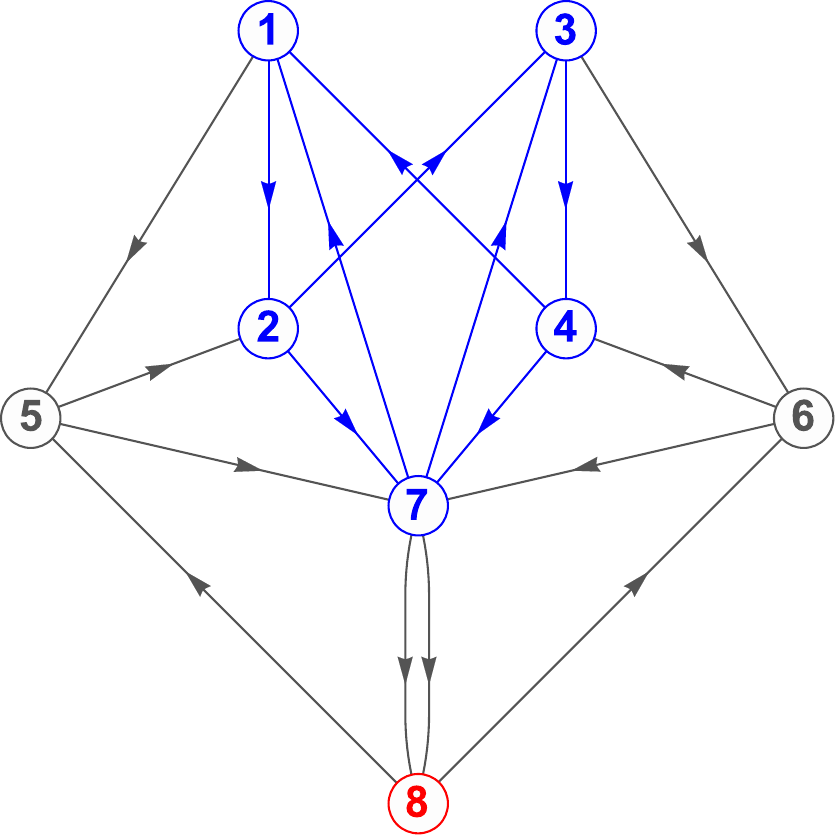}\label{fig:QSU2SU2}}
\caption{BPS quivers for the 5d   $SU(3)$,  $N_f=2$ and $SU(2){\times}SU(2)$ gauge theories.\label{fig:beetle quivers}}
 \end{center}
 \end{figure}

We can easily derive toric quivers which make these gauge-theory descriptions manifests. The quivers are shown in Figure~\ref{fig:beetle quivers}. They are simply related by a pair of mutations, on node 5 and on node 6 (in either order, the mutations commute in this case):
\be
\CQ_\MG\left(SU(3)_0, N_f=2\right) \qquad \overset{5, 6}{ \longleftrightarrow} \qquad \CQ_\MG\left(SU(2)\times SU(2)\right)
\ee
Therefore, at least in this example, the 5d ``UV duality'' (that is, S-duality) is realised at the level of the 5d BPS quiver by quiver mutations, which induces a tilting autoequivalence for the BPS category $\CAT$.

\medskip

\begin{figure}[t]
\begin{center}
\includegraphics[height=3.8cm]{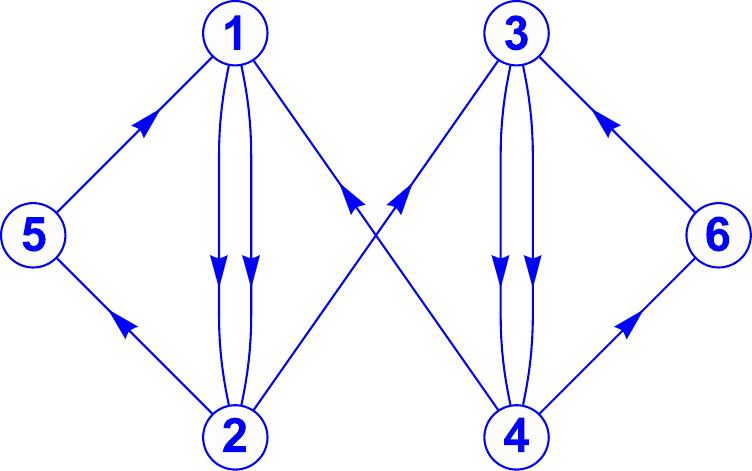}
\caption{BPS quiver for the 4d $\CN=2$  $SU(3)$, $N_f=2$ gauge theory. \label{fig:su3Nf2 quivers 4d}}
 \end{center}
 \end{figure}

Let us first look at the 5d $SU(3)$ gauge theory coupled to $N_f=2$ fundamental hypermultiplets. 
The 5d BPS quiver is shown in Fig.~\ref{fig:QSU3Nf2}. The superpotential reads:
\bea\label{W SU3 Nf2}
&W_{[SU(3)_0, N_f=2]}^{{\rm 5d} \;  \CN=1}= \; {\color{blue} X_{12}^1X_{25}X_{51}+ X_{34}^2X_{46}X_{63}+ X_{23}X_{34}^1X_{41}X_{12}^2-X_{23}X_{34}^2X_{41}X_{12}^1 }  \\
&\qquad\qquad\qquad\quad - X_{25}X_{58}X_{82}
+ X_{27}X_{76}X_{68}X_{82}- X_{27}X_{75}X_{51}X_{12}^2\\
&\qquad\qquad \qquad\quad  -X_{46}X_{68}X_{84}+ X_{47}X_{75}X_{58}X_{84}
 - X_{47}X_{76}X_{63}X_{34}^1~.
\eea
It contains the 4d $\CN=2$ $SU(3)$, $N_f=2$ quiver as a subquiver, shown in Fig.~\ref{fig:su3Nf2 quivers 4d}, which is obtained by deleting the nodes $7$ and $8$ in Fig.~\ref{fig:QSU3Nf2}; in particular, the brane tiling reproduces the correct quiver for the 4d BPS quiver \cite{Alim:2011kw}. Its superpotential  is given by the first line in \eqref{W SU3 Nf2}, namely:
\be
{\color{blue} W_{[SU(3), N_f=2]}^{{\rm 4d} \;  \CN=2}= \; X_{12}^1X_{25}X_{51}+ X_{34}^2X_{46}X_{63}+ X_{23}X_{34}^1X_{41}X_{12}^2-X_{23}X_{34}^2X_{41}X_{12}^1~.}
\ee

\begin{figure}[t]
\begin{center}
\subfigure[\small $B\CQ_{\MG}(p_{17})$.]{
\includegraphics[height=8cm]{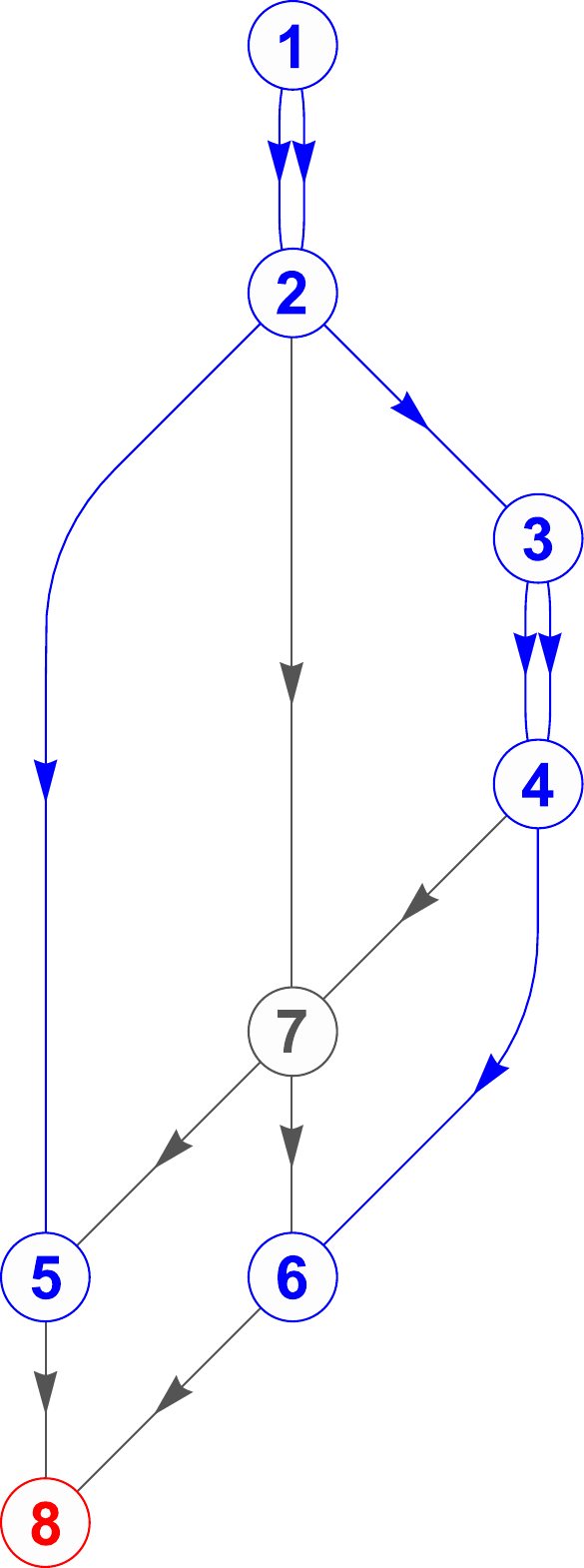}\label{fig:BQ beetle p17}} \;\;\;
\subfigure[\small $B\CQ_{\MG}(p_{28})$.]{
\includegraphics[height=8cm]{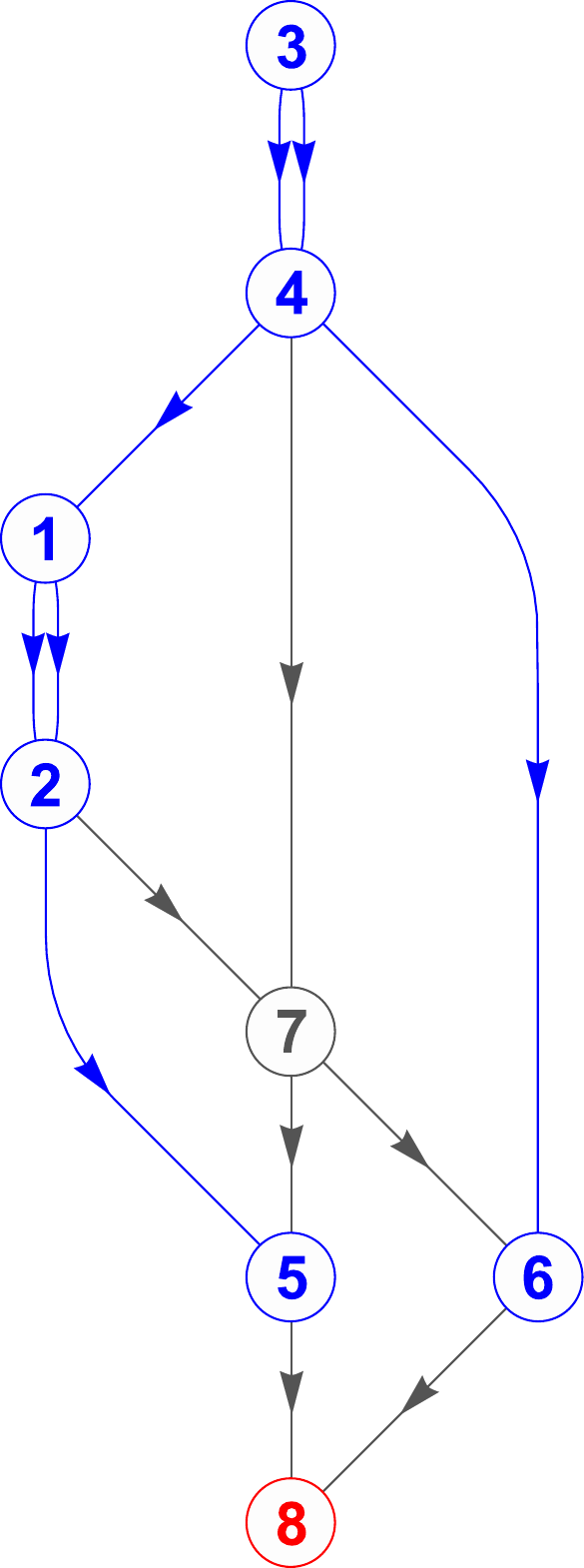}\label{fig:BQ beetle p28}}\qquad
\subfigure[\small Triangulated toric diagram.]{
\raisebox{0.5cm}{\includegraphics[height=6.5cm]{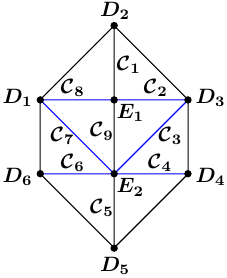}\label{fig:SU2Nf2 curves}}}
\caption{\textsc{Left and middle:} Two Beilinson quivers obtained from the brane tiling, corresponding to the 5d $SU(3)$ $N_f=2$ brane-charge assignment.   \textsc{Right:} Triangulated toric diagrams with our conventions for the toric divisors and curves. 
\label{fig:BQ and curves SU2Nf2}}
 \end{center}
 \end{figure}

\medskip
\noindent
The toric quiver of Fig.~\ref{fig:QSU3Nf2} has two perfect matchings:
\be
p_{17}= \{ X_{41}, X_{51}, X_{63}, X_{62}, X_{84}\}~, \qquad \qquad 
p_{28}= \{ X_{23}, X_{51}, X_{63}, X_{62}, X_{84}\}~.
\ee
with Beilinson quivers shown in Fig.~\ref{fig:BQ beetle p17}-\ref{fig:BQ beetle p28}, which make the gauge-theory phase manifest. There is an obvious symmetry between the two choice, which corresponds to taking either of the two compact divisors as the K-theory charge associated to the first node of the partially-ordered Beilinson quiver.

We choose to work with $p_{17}$.  The corresponding matrices $S^{-1}$ and $M$ are:
\be
S^{-1}= \smat{1 & 2 & 2 & 3 & 6 & 6 & 5 & 7 \\
 0 & 1 & 1 & 2 & 4 & 4 & 3 & 5 \\
 0 & 0 & 1 & 2 & 2 & 3 & 2 & 3 \\
 0 & 0 & 0 & 1 & 1 & 2 & 1 & 2 \\
 0 & 0 & 0 & 0 & 1 & 0 & 0 & 1 \\
 0 & 0 & 0 & 0 & 0 & 1 & 0 & 1 \\
 0 & 0 & 0 & 0 & 1 & 1 & 1 & 2 \\
 0 & 0 & 0 & 0 & 0 & 0 & 0 & 1}~, \qquad
 M= \smat{1 & -2 & 0 & 1 & 1 & 0 & 0 & 0 \\
 2 & -3 & -1 & 2 & 1 & 0 & -1 & 1 \\
 2 & -3 & 0 & 0 & 1 & 1 & -1 & 1 \\
 3 & -4 & 0 & 0 & 1 & 1 & -3 & 3 \\
 6 & -8 & -2 & 3 & 2 & 0 & -4 & 4 \\
 6 & -8 & -1 & 2 & 1 & 1 & -5 & 5 \\
 5 & -7 & -1 & 2 & 1 & 0 & -3 & 4 \\
 7 & -9 & -2 & 3 & 1 & 0 & -5 & 6}~.
\ee
The monodromy matrix has four eigenvectors with eigenvalue $1$, which read:
\bea\label{eigenvectors M SU3Nf2}
&v_{\rm KK} = (1,1,1,1,1,1,1,1)~, \qquad\qquad && v_{F_1}= (2,2,1,1,3,0,0,0)~, \cr
&v_{\CI} = (4,1,5,2,0,3,3,0)~, \quad &&v_{F_2}= (1,1,2,2,0,3,0,0)~,
\eea
and one eigenvector with eigenvalue $-1$, given by:
\be\label{vW vec su3Nf2}
v_W= (1,1,-1,-1,1,1,1,1)~.
\ee
Let us consider $\h \MG$ the partial resolution of the singularity of Fig~\ref{fig:SU2Nf2 curves}, with the divisors and curves as indicated. We have the linear relations:
\be
D_1 +D_6\sim D_3+D_4~, \quad \bE_1\sim -D_1-2D_2 -D_3+D_5~, \quad 
\bE_2\sim D_2-D_4-2D_5-D_6~,
\ee
\be
\CC_1\cong\CC_9~, \quad \CC_2 \cong \CC_8~, \quad \CC_2+ \CC_4\cong \CC_6 + \CC_7~, \quad \CC_5\cong \CC_3+ \CC_7 + \CC_9~,
\ee
amongst divisors and curves, respectively.
The tilting line bundles on $\h \MG$ are given by:
\bea
&L_1= \CO~, \qquad \qquad&& L_5= \CO(D_2+2D_3+2D_4)~,\cr
&L_2=\CO(D_3+D_4)~,\qquad\qquad\qquad && L_6=\CO(D_1+D_2+2D_3+D_4)~.\cr
& L_3=\CO(D_1+D_2-D_4-D_5)~,&&  L_7=\CO(D_2+2D_3+D_4)~,\cr
& L_4=\CO(D_1+D_2+D_3-D_5)~,&& L_8=\CO(D_2+3D_3+2D_4)~.
\eea
In the K-theory basis:
\be
\left([\bE_1]~, \,[\bE_1]~,\, [\CC_2]~, \,  [\CC_3]~, \,  [\CC_6]~, \,  [\CC_7]~,  [\CC_9]~, \, [{\rm pt}]\, \right)~,
\ee
we claim that the brane charges of the simple objects are:
\bea\label{eq KTHCHG}
&K(\CE_1)=  [\bE_1]~,  &&K(\CE_3)=  [\bE_2]~, &K(\CE_5)= - [\CC_2]- [\CC_3]~,\cr
&K(\CE_2)=  -[\bE_1]+ [\CC_2]~,  &&K(\CE_4)=  -[\bE_2]+ [\CC_6]+ [\CC_7]~, &K(\CE_6)= - [\CC_6]~,\cr
\eea
and:
\bea\label{eq KTHCHG2}
&K(\CE_7)=  -[\bE_1]-[\bE_2]-[\CC_2]+[\CC_3]-[\CC_6]-[\CC_7]+[\CC_9]~,\cr
&K(\CE_8)=  [\bE_1]+[\bE_2]+ [\CC_2]+[\CC_6]-[\CC_9]+ [{\rm pt}]~.
\eea
This brane-charge assignment is fully consistent with the 5d analysis of \cite{Closset:2018bjz}.~\footnote{Here, we follow the notation of section 6.2 of \protect\cite{Closset:2018bjz}  for the gauge theory and for the geometry, with our curves $\CC_k$ corresponding to $\CC_k^a$ there.}
The dimension vector $v_{\rm KK}$ corresponds to the D0-brane charge, $[\rm pt]$, and the other 3 dimensions vectors in \eqref{eigenvectors M SU3Nf2} correspond to the three curves in $\h \MG$ which have vanishing inersection with $\bE_1$ and $\bE_2$---and thus correspond to electrically neutral objects:
\bea
& [v_{\CI}] \cong -2 [\CC_2]+3 [\CC_3]-4 [\CC_6]-[\CC_7]+3 [\CC_9]~, \qquad &&(M= 3 h_0)~,\cr
& [v_{F_1}] \cong -[\CC_2]-3 [\CC_3]+[\CC_6]+[\CC_7]~, \qquad &&(M=-3 \t m_1)~,\cr
& [v_{F_2}] \cong [\CC_2]- [\CC_6]+2[\CC_7]~, \qquad &&(M=3 \t m_2)~.
\eea
Starting from this quiver, one can easily study its various decoupling limits to other 5d $SU(3)$  gauge theories, corresponding to integrating out hypermultiplets. 
\medskip

\paragraph{Consistency check.} Consider the quiver in figure \ref{fig:QSU3Nf2}. We immediately recognize the $SU(3)$ W-bosons corresponding to the simple roots of $SU(3)$. They correspond to the representations with dimension vectors:
\be
\bd_1 = [\CE_1] + [\CE_2] \qquad \bd_2 = [\CE_3] + [\CE_4]~.
\ee
The corresponding magnetic charges are:
\bea
\mathbf{m}_1(\CO) &= {2\over 3} \langle [\CO], \bd_1\rangle_D + {1\over 3} \langle [\CO], \bd_2\rangle_D\\
&=N_1 - N_2 - N_7 + N_8~,\\
\mathbf{m}_2(\CO) &= {1\over 3} \langle [\CO], \bd_1\rangle_D + {2\over 3} \langle [\CO], \bd_2\rangle_D\\
&=N_3 - N_4 - N_7 + N_8~.\\
\eea
Note that this agrees with the assignments of charges in \eqref{eq KTHCHG} and \eqref{eq KTHCHG2}. The electric category is the category controlled by $\mathbf{m}_1$ and $\mathbf{m}_2$.  In this case, its structure is much more complicated than in the rank-one examples considered in the previous section. We plan to return to this topic in the future \cite{PROG}.

   \begin{figure}[t]
\begin{center}
\subfigure[\small 4d $\CN{=}2$ subquiver.]{
\includegraphics[height=3.8cm]{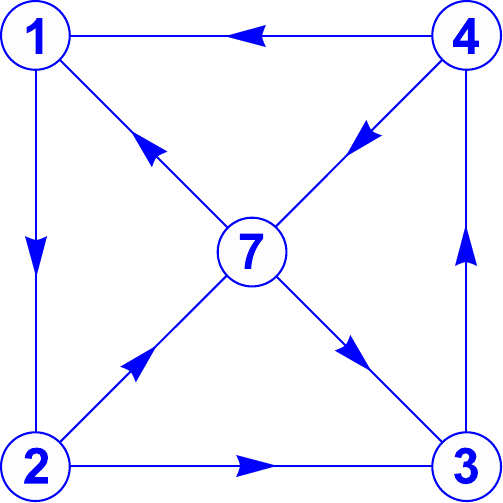}\label{fig:su2su2 A}}\qquad \qquad\qquad
\subfigure[\small Mutated quiver.]{
\includegraphics[height=3.8cm]{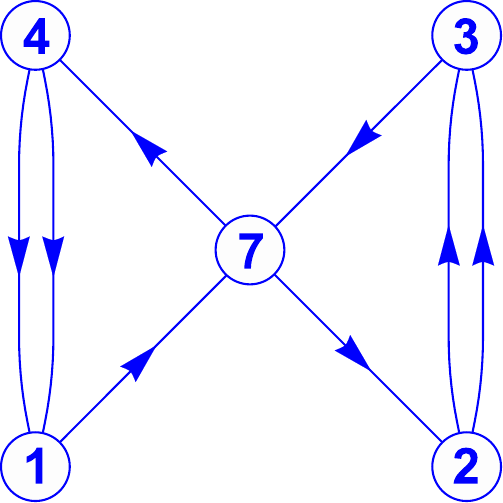}\label{fig:su2su2 B}}
\caption{BPS quivers for the 4d $\CN=2$  $SU(2)\times SU(2)$ gauge theory. \label{fig:su2su2 quivers 4d}}
 \end{center}
 \end{figure} 

\subsection{The 5d $SU(2)\times SU(2)$ gauge theory and its BPS quiver}
Let us now look at the 5d $\CN=1$ $SU(2) \times SU(2)$ gauge theory with a single bifundamental hypermultiplet. The 5d BPS quiver is shown in Fig.~\ref{fig:QSU2SU2}. The superpotential reads:
\bea\label{W SU2SU2}
&W^{{\rm 5d} \;  \CN=1}_{[SU(2){\times}SU(2)]}&= &\; {\color{blue} X_{12} X_{23}  X_{34} X_{41}- X_{12} X_{27} X_{71}- X_{34} X_{47} X_{73} }  \\
&&&+X_{15} X_{57}X_{71}+X_{36}X_{67}X_{73}-X_{57}X_{78}^1X_{85}-X_{67}X_{78}^2X_{86}\\
&&&+X_{47}X_{78}^1X_{86}X_{64}+ X_{27}X_{78}^2X_{85}X_{52} - X_{15}X_{52}X_{23}X_{36}X_{64}X_{41}~.
\eea
The 5d BPS quiver contains the 4d $\CN=2$ quiver shown in Figure~\ref{fig:su2su2 A}, with the superpotential given by the first line in \eqref{W SU2SU2}:
\be
{\color{blue}  W^{{\rm 4d} \;  \CN=2}_{[SU(2){\times}SU(2)]}= X_{12} X_{23}  X_{34} X_{41}- X_{12} X_{27} X_{71}- X_{34} X_{47} X_{73}~.}
\ee
This is mutation-equivalent (by performing a mutation on node 7) to the 4d $\CN=2$ BPS quiver  \cite{Alim:2011kw} shown in Fig.~\ref{fig:su2su2 B} with a  sextic superpotential:
\be
{\color{blue}  W^{{\rm 4d} \;  \CN=2}_{[SU(2){\times}SU(2)]'}= X_{12} X_{72}  X_{23}^1 X_{37}X_{74} X_{41}^2+ X_{17} X_{74} X_{41}^1+ X_{37} X_{72} X_{23}^2~,}
\ee
in agreement with \cite{Alim:2011ae}. We could also perform a mutation on node 7 of the full 5d BPS quiver of Fig.~\ref{fig:QSU2SU2}. The resulting 5d BPS quiver contains the 4d BPS quiver of Fig~\ref{fig:su2su2 B} as a subquiver, but it is non-toric---it does not correspond to a brane tiling.

  \begin{figure}[t]
\begin{center}
\subfigure[\small Toric diagram.  ]{
\raisebox{0.7cm}{\includegraphics[height=5cm]{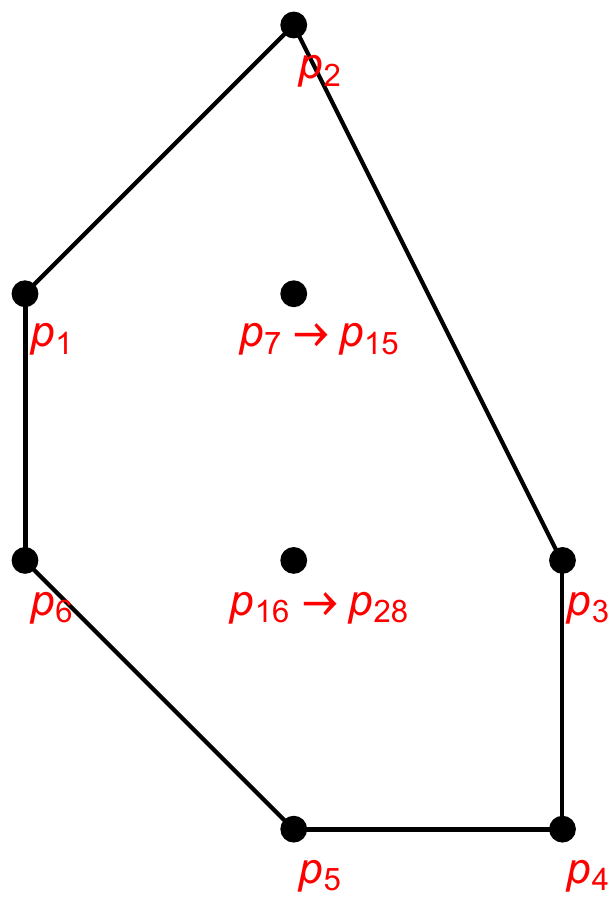}}\label{fig:toric diag SU3 k1 Nf2}}\;\;\;\qquad
\subfigure[\small $\CQ_\MG\left(SU(3)_1, N_f{=}2\right)$.]{
\includegraphics[height=7cm]{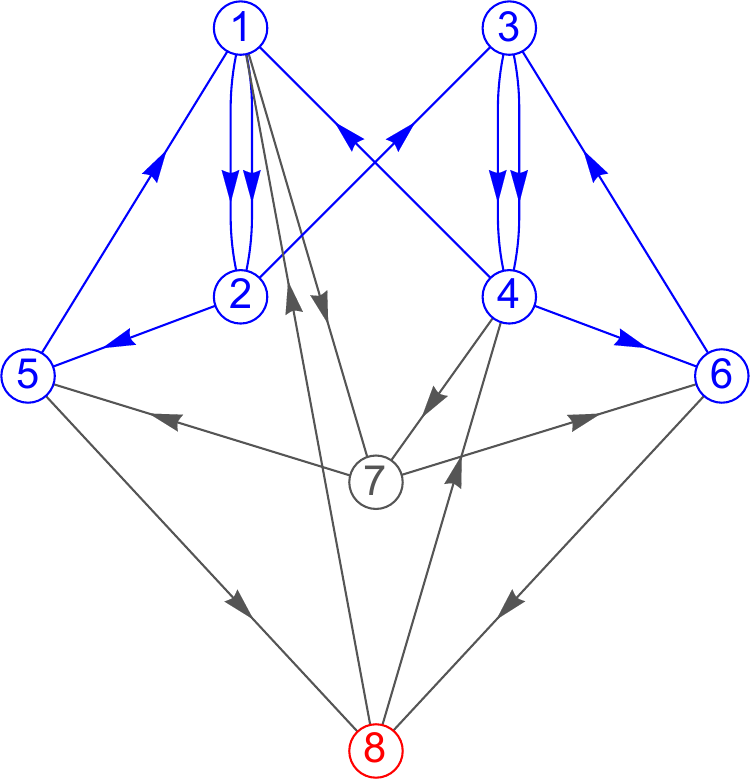}\label{fig: quiver SU3 k1 Nf2}}
\caption{Toric diagram and BPS quivers for the 5d   $SU(3)_1$  $N_f{=}2$ gauge theory. \label{fig:SU31 Nf2 quiver}}
 \end{center}
 \end{figure} 

 \subsection{The $SU(3)_1$, $N_f=2$  gauge theory}\label{sec: SU3 k1 Nf2}

Next, consider the toric geometry of Fig.~\ref{fig:toric diag SU3 k1 Nf2}. It has a single gauge-theory phase, the 5d $SU(3)_1$ gauge theory with $N_f=2$. Note that this theory, unlike the one of section~\ref{section:beetle}, breaks parity due to the non-zero CS level. The toric quiver is shown in Fig.~\ref{fig: quiver SU3 k1 Nf2}. Its superpotential reads:
\bea\label{W SU3 k1 Nf2}
&W_{[SU(3)_1, N_f=2]}^{{\rm 5d} \;  \CN=1}= \;  {\color{blue} X_{12}^1X_{25}X_{51}+ X_{34}^2X_{46}X_{63}+ X_{23}X_{34}^1X_{41}X_{12}^2-X_{23}X_{34}^2X_{41}X_{12}^1 }  \\
&\qquad\qquad\qquad\quad - X_{17}X_{75}X_{51}
+ X_{17}X_{76}X_{68}X_{81}- X_{25}X_{58}X_{81}X_{12}^2\\
&\qquad\qquad \qquad\quad  -X_{46}X_{68}X_{84}+ X_{47}X_{75}X_{58}X_{84}
 - X_{47}X_{76}X_{63}X_{34}^1~.
\eea
Note that this similar to the superpotential \eqref{W SU3 Nf2} for the $k=0$ theory; only the second line in \eqref{W SU3 k1 Nf2} is different from  \eqref{W SU3 Nf2}, corresponding to the fact that only the ``left-hand-side'' of the quiver of Fig.~\ref{fig: quiver SU3 k1 Nf2} differs from the one of Fig.~\ref{fig:QSU2SU2}.

  \subsection{The $SU(3)_k$, $N_f=1$ gauge theory ($k=\half, {3\ov 2}$)}
  Starting from the results of sections~\ref{section:beetle} and \ref{sec: SU3 k1 Nf2}, we can derive 5d BPS quivers for all the other singularities of Fig.~\ref{fig:rk 2 geoms}, by partial resolution, just as for the rank-one examples. If we remove a single point of the toric diagram, decreasing the rank of the flavor symmetry from $f=3$ to $f=2$, we have:
\bea
  &SU(3)_0^{N_f{=}2}\quad \overset{\langle X_{58}\rangle }{\longrightarrow} \quad   SU(3)_{\half}^{N_f{=}1}~, \qquad\quad    &&SU(3)_1^{N_f{=}2}\quad \overset{\langle X_{75}\rangle }{\longrightarrow}  \quad  SU(3)_{\half}^{N_f{=}1}~, \cr
  &SU(3)_1^{N_f{=}2}\quad \overset{\langle X_{23}\rangle }{\longrightarrow} \quad   {\rm NL}^{N_f{=}2}~,  \qquad \quad&&SU(3)_1^{N_f{=}2}\quad \overset{\langle X_{58}\rangle }{\longrightarrow} \quad   SU(3)_{{3\ov 2}}^{N_f{=}1}~, \cr
\eea
Here, we indicated which arrow of the right-hand-side quiver must be ``Higgsed'' to flow to the new quiver. The first line give two distinct way to obtain the same 5d BPS quiver for $SU(3)_{\half}$ with a single flavor. 
  \begin{figure}[t]
\begin{center}
\subfigure[\small $\CQ_\MG\left(SU(3)_{1\ov 2}, N_f{=}1\right)$ ]{
\includegraphics[height=7cm]{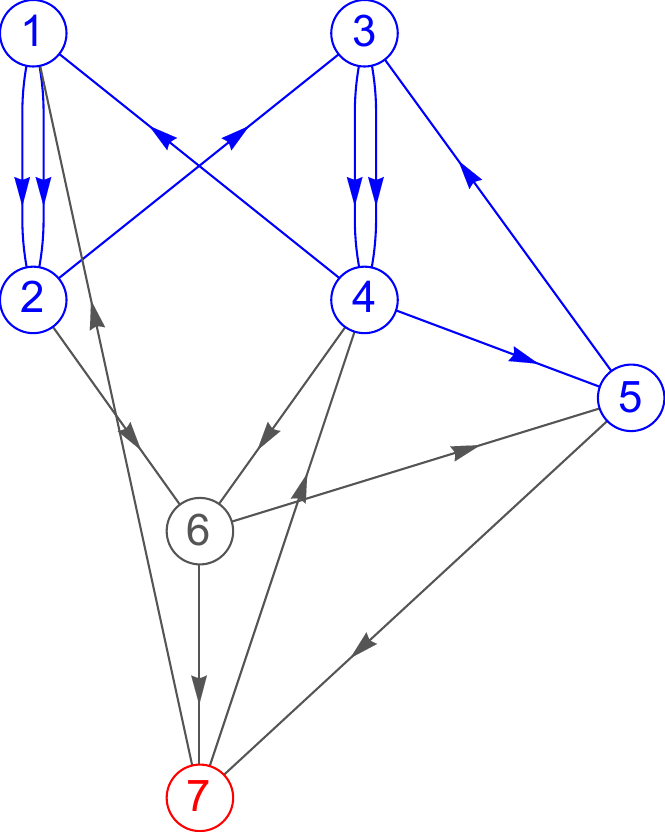}\label{fig:toric diag SU3 k12 Nf1}}\;\;\;\qquad
\subfigure[\small $\CQ_\MG\left(SU(3)_{3\ov 2}, N_f{=}1\right)$.]{
\includegraphics[height=7cm]{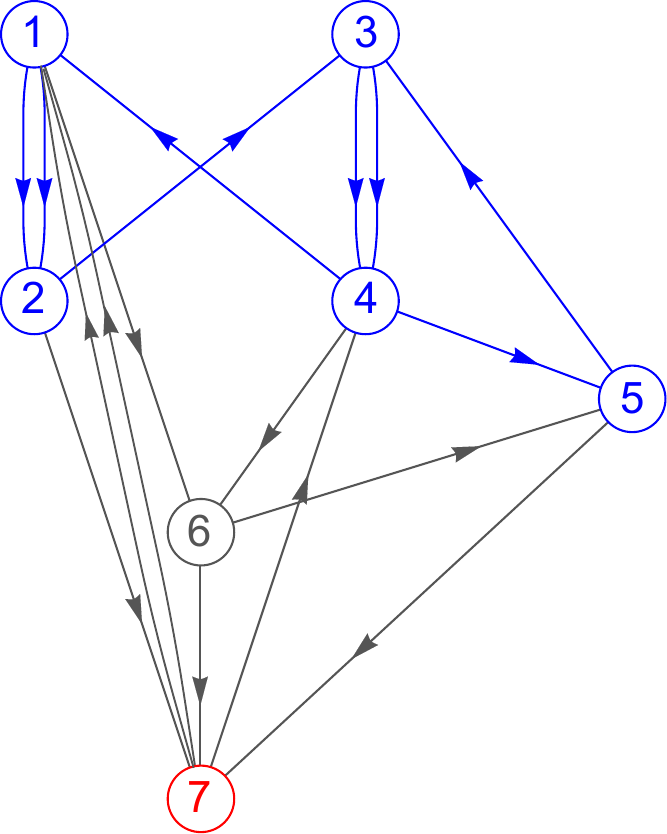}\label{fig: quiver SU3 k32 Nf1}}
\caption{BPS quivers for the 5d   $SU(3)_k$  $N_f{=}1$ gauge theories. \label{fig:SU3 k Nf1 quivers}}
 \end{center}
 \end{figure} 
  
The BPS quivers for the two $SU(3)$ $N_f=1$ gauge theories are shown in Figure~\ref{fig:SU3 k Nf1 quivers}. Their superpotentials are given by:
\bea
&W_{[SU(3)_{1\ov 2}, N_f=1]}^{{\rm 5d} \;  \CN=1}&=& \;  {\color{blue}  X_{34}^2X_{45}X_{53}+ X_{23}X_{34}^1X_{41}X_{12}^2-X_{23}X_{34}^2X_{41}X_{12}^1 }  \\
&&&\; X_{26}X_{65}X_{57}X_{71}X_{12}^1 - X_{26}X_{67}X_{71}X_{12}^2\\
&&&\;- X_{46}X_{65}X_{53}X_{34}^1+ X_{46}X_{67}X_{74}-X_{45}X_{57}X_{74}~,
\eea
and:
\bea
&W_{[SU(3)_{3\ov 2}, N_f=1]}^{{\rm 5d} \;  \CN=1}&=& \;  {\color{blue}  X_{34}^2X_{45}X_{53}+ X_{23}X_{34}^1X_{41}X_{12}^2-X_{23}X_{34}^2X_{41}X_{12}^1 }  \\
&&&\;  + X_{16} X_{65} X_{57} X_{71}^2- X_{16} X_{67} X_{71}^1+X_{12}^1 X_{27} X_{71}^1- X_{12}^2 X_{27} X_{71}^2 \\
&&&\;- X_{46}X_{65}X_{53}X_{34}^1+ X_{46}X_{67}X_{74}-X_{45}X_{57}X_{74}~,
\eea
respectively. One could similarly derive the quiver for the ``non-Lagrangian theory'' $ {\rm NL}^{N_f{=}2}$.

\subsection{The $SU(3)_k$ 5d Chern-Simons gauge theory ($k=0,1,2$)}
  \begin{figure}[t]
\begin{center}
\subfigure[\small $\CQ_\MG\left(SU(3)_0\right)$.]{
\includegraphics[height=7cm]{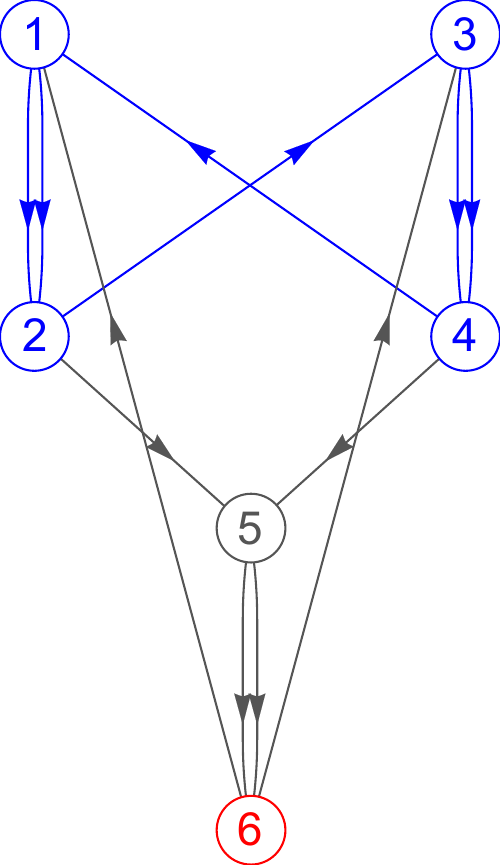}\label{fig:Q SU3k0}}\;\;\;
\subfigure[\small $\CQ_\MG\left(SU(3)_1\right)$.]{
\includegraphics[height=7cm]{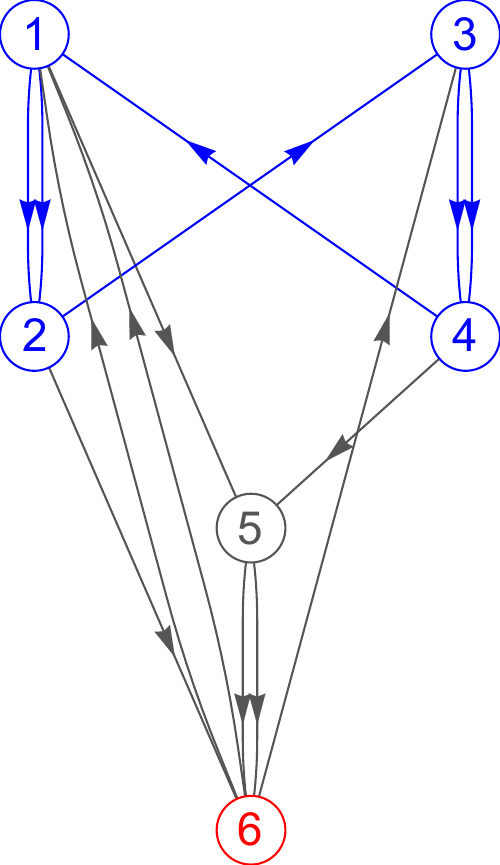}\label{fig:Q SU3k1}}\;\;\;
\subfigure[\small $\CQ_\MG\left(SU(3)_2\right)$.]{
\includegraphics[height=7cm]{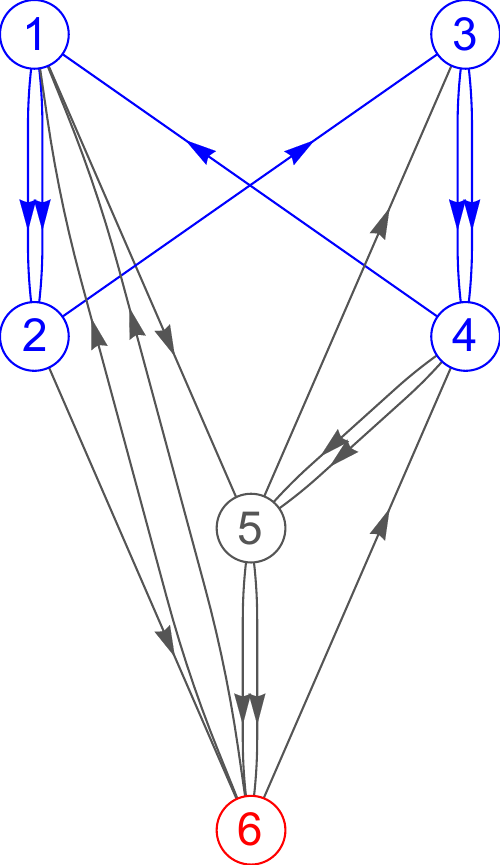}\label{fig:Q SU3k2}}
\caption{BPS quivers for the 5d   $SU(3)_k$ gauge theories. \label{fig:SU3k}}
 \end{center}
 \end{figure} 
With an additional partial resolution, we reduce the flavor group rank to $f=1$, thus obtaining the 5d BPS quivers for the $SU(3)_k$ theories, with $k=0$, $1$ or $2$, as well as the non-Lagrangian theory $ {\rm NL}^{N_f=1}$. The corresponding Higgsing patterns are:
\bea
  &SU(3)_\half^{N_f{=}1}\quad \overset{\langle X_{57}\rangle }{\longrightarrow} \quad   SU(3)_0~, \qquad\quad    &&&SU(3)_\half^{N_f{=}1}\quad \overset{\langle X_{65}\rangle }{\longrightarrow} \quad   SU(3)_0~, \cr
    &SU(3)_{3\ov2}^{N_f{=}1}\quad \overset{\langle X_{57}\rangle }{\longrightarrow} \quad   SU(3)_1~, \qquad\quad    &&&SU(3)_{3\ov2}^{N_f{=}1}\quad \overset{\langle X_{65}\rangle }{\longrightarrow} \quad   SU(3)_2~, \cr
     &SU(3)_{3\ov2}^{N_f{=}1}\quad \overset{\langle X_{23}\rangle }{\longrightarrow} \quad   {\rm NL}^{N_f=1}~.
\eea
The BPS quivers for the $SU(3)_k$ theories are shown in Figure~\ref{fig:SU3k}. The superpotentials take the form:
\be
W_{[SU(3)_0]}^{{\rm 5d} \;  \CN=1}={\color{blue}  W_{[SU(3)]}^{{\rm 4d} \;  \CN=2}}+ \Delta W_{[SU(3)_0]}^{{\rm 5d} \;  \CN=1}~,
\ee
with the 4d $\CN=2$ BPS quiver superpotential for the pure $SU(3)$ gauge theory given by:
\be
{\color{blue}  W_{[SU(3)]}^{{\rm 4d} \;  \CN=2}= \epsilon_{\alpha\beta}  X_{23}X_{34}^\alpha X_{41}X_{12}^\beta}~,
\ee
and additional terms that read:
\bea
&\Delta W_{[SU(3)_0]}^{{\rm 5d} \;  \CN=1}&=&  \epsilon_{\alpha\beta} \left(X_{34}^\alpha X_{45} X_{56}^\beta X_{63} + X_{56}^\alpha X_{61} X_{12}^\beta X_{25}\right)~,\cr
&\Delta W_{[SU(3)_1]}^{{\rm 5d} \;  \CN=1}&=& \;   \epsilon_{\alpha\beta} \left(X_{34}^\alpha X_{45} X_{56}^\beta X_{63} + X_{56}^\alpha X_{61}^\beta X_{15}+X_{61}^\alpha X_{12}^\beta X_{26}\right)~,\cr
&\Delta W_{[SU(3)_2]}^{{\rm 5d} \;  \CN=1}&=& \;   \epsilon_{\alpha\beta} \left(X_{34}^\alpha X_{45}^\beta X_{53}+X_{45}^\alpha X_{56}^\beta X_{64} + X_{56}^\alpha X_{61}^\beta X_{15 }+X_{61}^\alpha X_{12}^\beta X_{26}\right)~.
\eea

  \begin{figure}[t]
\begin{center}
\subfigure[\small Rank-two $E_0$.]{
\includegraphics[height=5.5cm]{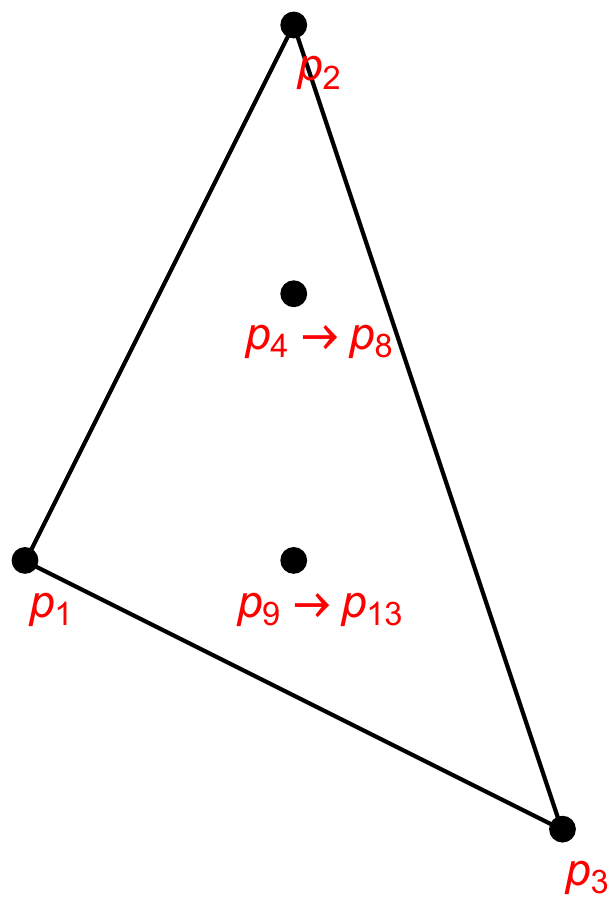}\label{fig:sing E2r0}}\;\;
\subfigure[\small $SU(3)_2 \rightarrow {\rm NL}^{N_f=0}$.]{
\includegraphics[height=5.5cm]{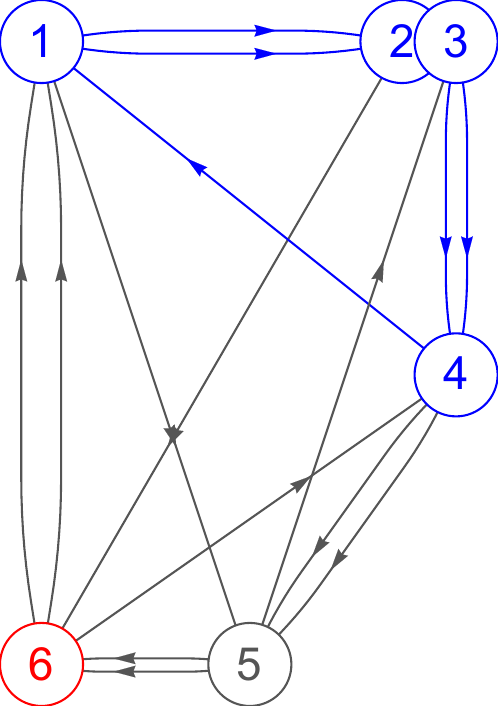}\label{fig:Q E2r0 from Higgs}}\;\;
\subfigure[\small $\CQ_\MG\left({\rm NL}^{N_f=0}\right)$.]{
\includegraphics[height=5.5cm]{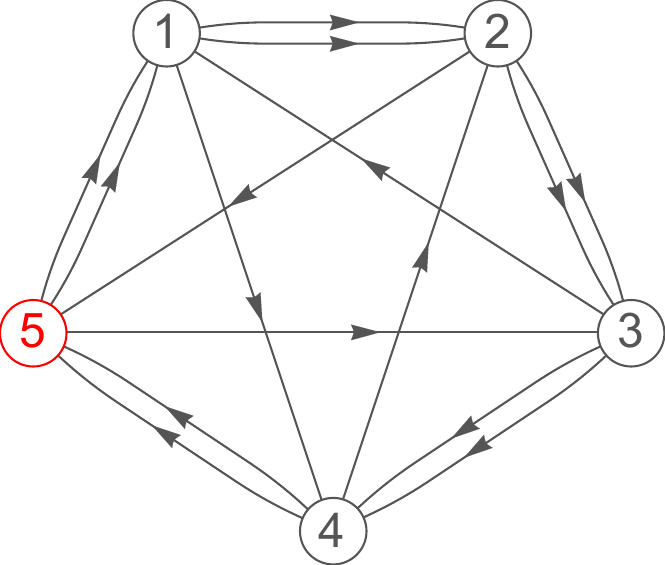}\label{fig:Q E2r0}}
\caption{The rank-two $E_0$ ($f=0$) singularity and its 5d BPS quiver. \label{fig:E0r2}}
 \end{center}
 \end{figure} 

\subsection{The rank-two $E_0$ theory}
There are three rank-two toric singularities which do not admit any gauge-theory phase, and whose quivers are easily obtained by following the relevant RG flows. As a curiosity, let us write down the BPS quiver for the simplest of such theories, ${\rm NL}^{N_f=0}$, whose toric singularity is shown in Fig.~\ref{fig:sing E2r0}. It has trivial flavor symmetry, $f=0$, and it can be obtained by a non-perturbative RG flow from the $SU(3)_2$ theory, corresponding to integrating out an instanton particle, exactly like the rank-one $E_0$ theory discussed in section~\ref{subsec: E0 rk1}. We have:
\be
SU(3)_2 \quad \overset{\langle X_{23}\rangle }{\longrightarrow} \quad  {\rm NL}^{N_f=0}~.
\ee
The Higgsing pattern is shown in Fig.~\ref{fig:Q E2r0 from Higgs}, and the resulting 5d BPS quiver is drawn in Fig.~\ref{fig:Q E2r0}. Its superpotential reads:
\be
W_{[{\rm NL}, {N_f=0}]}^{{\rm 5d} \;  \CN=1}  =\sum_{i=1}^5 \epsilon_{\alpha\beta} X_{i, i+1}^\alpha X_{i+1, i+2}^\beta X_{i+2, i}~,
\ee
with cyclic subscripts. This naturally generalizes the $dP_0$ quiver of the rank-one $E_0$ theory.

\section{Higher-rank example: $Y^{p,q}$ and the 5d $SU(p)_q$ gauge theory}\label{sec:rankn}
One of the most studied infinite families of toric CY$_3$ singularities is that of the real cones over the Sasaki-Einstein five-manifolds $Y^{p,q}(\mathbb{P}^1)$, with $p$ and $q$ integers \cite{Gauntlett:2004yd, Martelli:2005wy}---the $Y^{p,q}$ singularities, for short. Their fractional-brane quivers are well-known \cite{Benvenuti:2004dy, Franco:2005rj, Franco:2005sm}.

  \begin{figure}[t]
\begin{center}
\subfigure[\small $Y^{4,0}$ quiver.]{
\includegraphics[height=6cm]{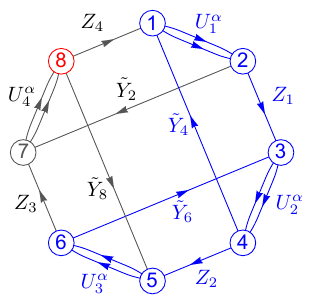}\label{fig:Q Y40}}\;\;\;
\subfigure[\small $Y^{4,1}$ quiver.]{
\includegraphics[height=6cm]{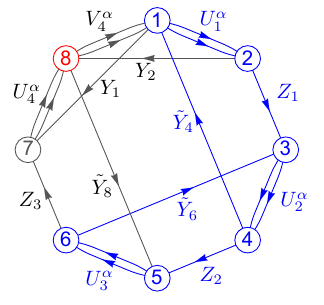}\label{fig:Q Y41}}\;\;\;
\subfigure[\small $Y^{4,2}$ quiver.]{
\includegraphics[height=6cm]{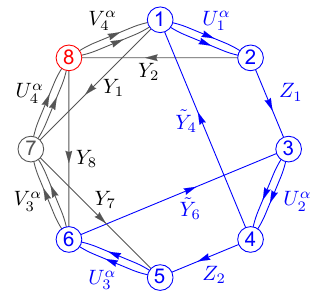}\label{fig:Q Y42}}\;\;\;
\subfigure[\small $Y^{4,3}$ quiver.]{
\includegraphics[height=6cm]{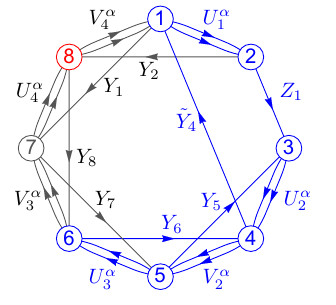}\label{fig:Q Y43}}
\caption{The $Y^{4,q}$ quivers. \label{fig:Q Y4q}}
 \end{center}
 \end{figure} 
  \begin{figure}[t]
\begin{center}
\subfigure[\small $\CQ_\MG\left(SU(4)_0\right)$.]{
\includegraphics[height=6cm]{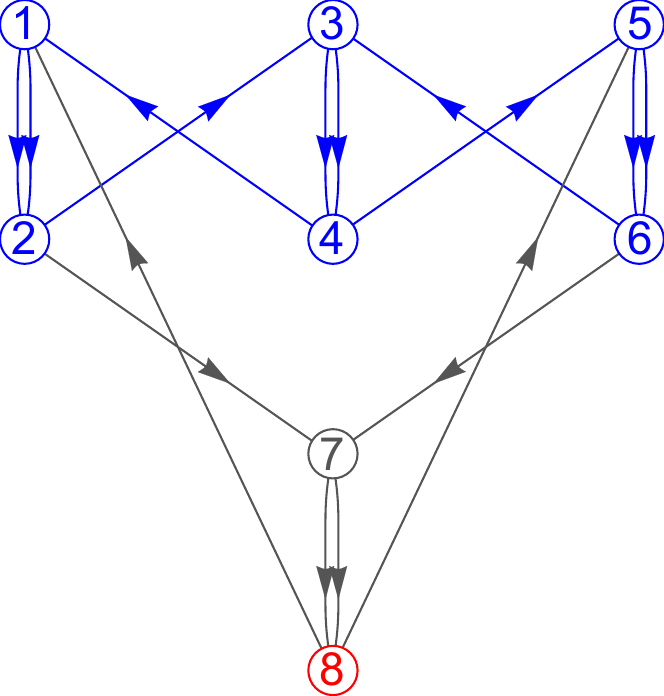}\label{fig:Q SU4k0}}\;\;\;\qquad
\subfigure[\small $\CQ_\MG\left(SU(4)_1\right)$.]{
\includegraphics[height=6cm]{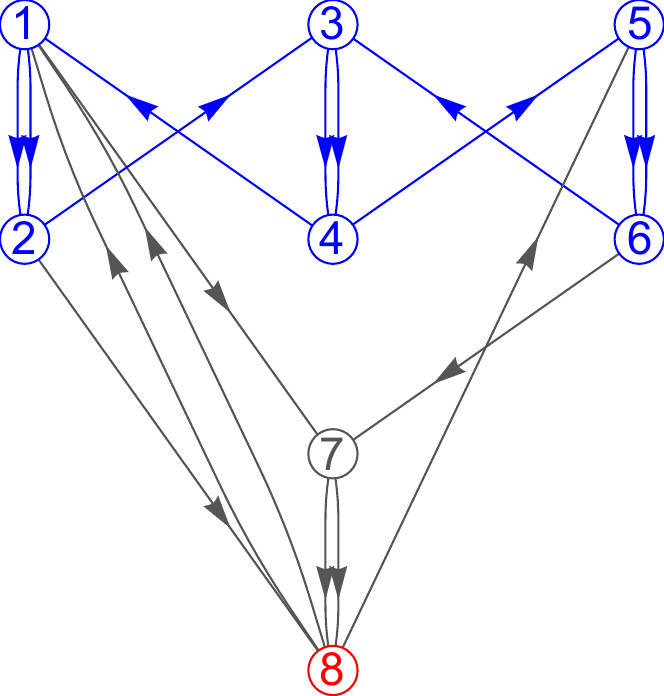}\label{fig:Q SU4k1}}\;\;\;
\subfigure[\small $\CQ_\MG\left(SU(4)_2\right)$.]{
\includegraphics[height=6cm]{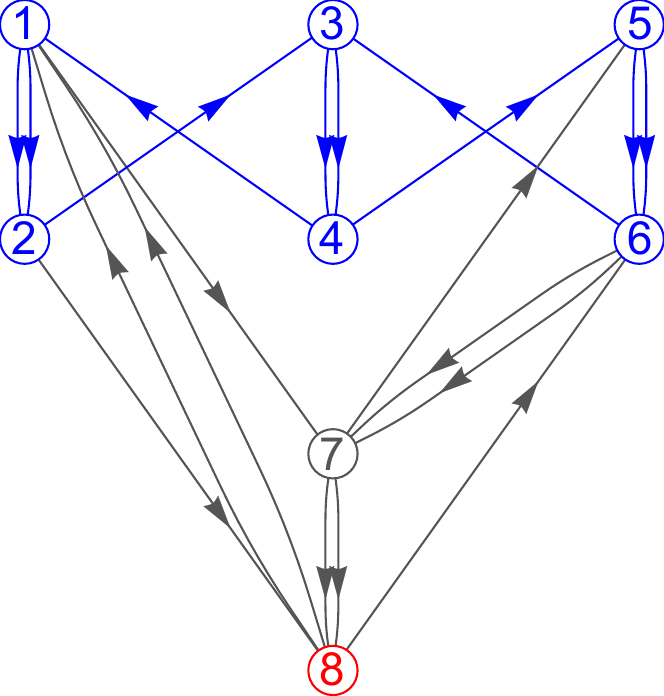}\label{fig:Q SU4k2}}\;\;\;\qquad
\subfigure[\small $\CQ_\MG\left(SU(4)_3\right)$.]{
\includegraphics[height=6cm]{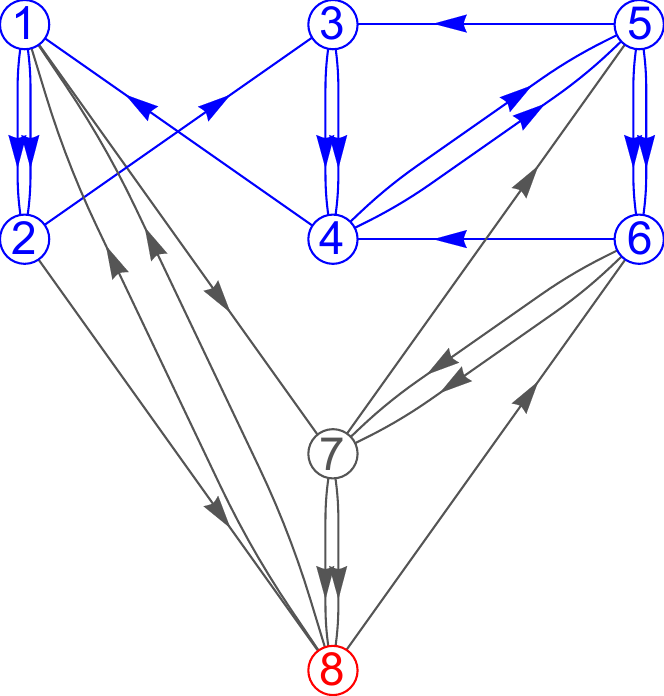}\label{fig:Q SU4k3}}
\caption{BPS quivers for the 5d  $SU(4)_k$ gauge theories. \label{fig:SU4k}}
 \end{center}
 \end{figure} 

The $Y^{p,q}$ singularity is isolated if $p>0$ and $0\leq q<p$. The toric diagram has four external points:
\be
w_1=(-1,p-q)~, \quad w_2=(0,p)~, \quad w_3=(1,0)~, \quad w_4= (0,0)~,
\ee
and $p-1$ internal points $w^I_j = (0, j)$, $j=1, \cdots, p-1$. We therefore have:
\be
f=1~, \qquad\qquad r= p-1~,\qquad \qquad n_G= 2p~.
\ee
Particular examples are the conifold ($Y^{1,0}$), the $\mathbb{F}_0$ and $dP_1$ singularities ($Y^{2,0}$ and $Y^{2,1}$, respectively), and the rank-two singularities realizing $SU(3)_k$ for $k=0,1,2$ ($Y^{3,k}$). In general, the 5d SCFT associated to the $Y^{p,q}$ singularity has a single gauge theory phase, as an $SU(p)$ gauge theory at CS level $k=q$:
\be
Y^{p,q} \quad \longleftrightarrow \quad \CT_{Y^{p,q}} \qquad \longrightarrow \qquad SU(p)_q~.
\ee
The relation of the $Y^{p,q}$ geometry to 5d physics has previously been discussed in \cite{Hanany:2005hq}. In passing, let us note that the infinite family $X^{p,q}$ of \cite{Hanany:2005hq} similarly corresponds to 5d $SU(p)$ gauge theories with a single fundamental flavor.


As expected on general grounds, the 5d BPS quiver for $SU(p)_q$ consists of $2p$ nodes: $2p-2$ nodes for electric and magnetic charges (giving us the 4d $\CN=2$ $SU(p)$ quiver), plus two nodes carrying the instanton and the KK charge.  
The $Y^{p,q}$ quiver contains $4p+2q$ arrows, which can be organised as follows \cite{Benvenuti:2004dy}:
\bea
&U_i^\alpha  &\equiv & \;  X_{2i-1,2i} \qquad && i=1, \cdots, p~, \qquad && \alpha=1,2~,\cr
&Z_k  &\equiv & \;  X_{2k,2k+1} \qquad && k=1, \cdots, p-q~,\cr
&V_{m}^\alpha  &\equiv& \;  X_{2m,2m+1} \qquad && m=p-q+1, \cdots, p~,  \quad&& \alpha=1,2~,\cr
&\t Y_{2k+2}  &\equiv& \;  X_{2k+2,2k-1} \qquad && k=1, \cdots, p-q~,\cr
&Y_{2m+1}  &\equiv& \;  X_{2m+1,2m-1} \qquad && m=p-q+1, \cdots, p~,\cr
&Y_{2m+2}  &\equiv& \;  X_{2m+2,2m} \qquad && m=p-q+1, \cdots, p~,
\eea
  \begin{figure}[t]
\begin{center}
\subfigure[\small $Y^{5,0}$.]{
\includegraphics[height=6cm]{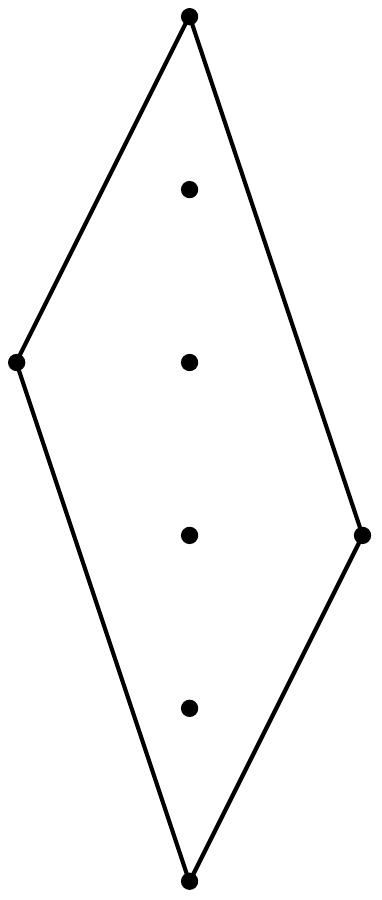}\label{fig:td SU50}}\qquad\quad
\subfigure[\small $\CQ_\MG\left(SU(5)_0\right)$.]{
\includegraphics[height=6cm]{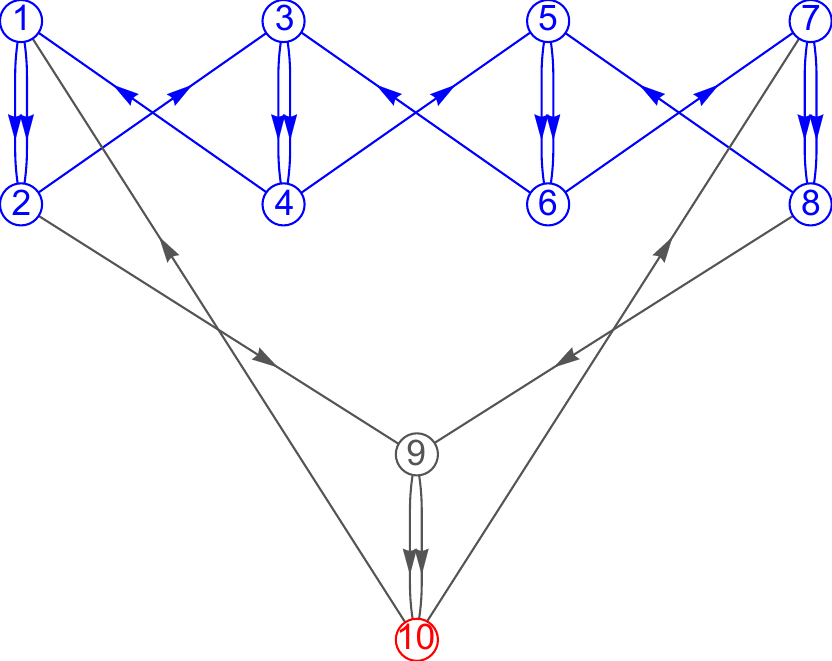}\label{fig:Q SU50}}
\caption{Toric diagram for the $Y^{5,0}$ singularity and its 5d BPS quiver. \label{fig:SU50}}
 \end{center}
 \end{figure} 
Here, the subscripts in $X_{i,j}$ are understood to be cyclic. 
It is useful to first consider the case $q=0$, since the 5d BPS quiver for the  $SU(p)_0$ theory only has arrows of type $U_i^\alpha$, $Z_k$ and $\t Y_{2k+1}$. One can then increase $q$ step by step, with each step converting one single arrow $Z_k$ to a pair of arrows $V_{m}^\alpha$; this is shown in Fig.~\ref{fig:Q Y4q} for the  $SU(4)_q$ BPS quivers.   
 The $Y^{p,q}$ quiver superpotential reads:
 \be
W = \sum_{l=1}^{p-q} \epsilon_{\alpha\beta} U_l^\alpha Z_l U_{l+1}^\beta \t Y_{2l+2} + \sum_{m=p-q+1}^p  \epsilon_{\alpha\beta}\left( U_m^\alpha V_m^\beta Y_{2m+1}+ V_m^\alpha U_{m+1}^\beta Y_{2m+2} \right)~.
 \ee
One can find a brane charge assignment compatible with the $SU(p)$ gauge-theory interpretation. As an example, in Figure~\ref{fig:SU4k}, the $SU(4)_q$ BPS quivers are displayed in a way that makes the 4d $\CN=2$ $SU(4)$ subquivers manifests.

The general pattern is obvious. For any $SU(p)_0$ theory, the 4d $\CN=2$ subquiver is the $SU(p)$ BPS quiver  studied in \cite{Fiol:2000pd,Cecotti:2010fi,Alim:2011kw}; see Fig.~\ref{fig:SU50} for another example. In general, for higher $q$, the 4d $\CN=2$ BPS subquiver will be a mutation-equivalent quiver, as for instance in the case of the quiver for $SU(4)_3$ of Fig.~\ref{fig:Q SU4k3}.
We can always mutate the full 5d BPS quiver to obtain whatever form of the 4d $\CN=2$ subquiver we want; in general,  the resulting 5d quiver may not correspond to a brane tiling, however.

\section{BPS spectroscopy of 4d KK theories}\label{sec:BPSspectrum}
In this section, we further comment on the BPS spectrum using the 5d BPS quivers. In particular, for the rank-one $E_1$ theory, we argue for the existence of a (formal) tame chamber in which the spectrum takes an especially simple form. 

\subsection{Finite, tame and wild BPS chambers} 
The spectrum of stable BPS states of any four-dimensional $\CN=2$ field theory is organized in BPS chambers, separated by walls of marginal stability where wall-crossing transitions occur. This is the case in particular for the 4d $\CN=2$ KK theories we are considering in this paper. It can happen that the charges of the BPS states in a given chamber are organized according to the roots of a triangular algebra ${\,\mathfrak T}\subset \CJ$: these chambers are called \textit{triangular BPS chambers} (see section 2.2 of \cite{Cecotti:2015qha} for a more detailed definition). An important feature of triangular BPS chambers is that there is an underlying quadratic Tits form $q_{\,\mathfrak T}(\cdot)$ which controls the properties of the stable representations: given $\CO \in \text{mod-}{\mathfrak T}$, we have that:
$$\dim_\C \mathcal M(\CO) = 1 - q_{\,\mathfrak T} (\dim \CO)\,.$$
Since the dimension of a K\"ahler manifold is also the value of its highest Lefschetz spin, we have that, for triangular chambers, the spin of the coresponding BPS particle is controlled by the Tits form. Recall that stable objects are always bricks for the corresponding module category, i.e. $\text{End}(\CO) \simeq \C$. A given triangular chamber is called: 
\begin{itemize}
\item {\bf Finite} iff $q_{\mathfrak T}(\CO) = 1$ for all bricks $\CO \in \text{mod } {\mathfrak T}$;
\item {\bf Tame} iff $0\leq q_{\mathfrak T}(\CO) \leq 1$ for all bricks $\CO \in \text{mod } {\mathfrak T}$;
\item {\bf Wild} iff $q_{\mathfrak T}(\CO) \leq 1$, but unbounded below for all bricks $\CO \in \text{mod } {\mathfrak T}$.
\end{itemize}
In a wild triangular chamber, the spectrum of BPS states is still organized by the roots of $q_{\, \mathfrak T}$, which determines which states can be stable, but the degeneracies of states are typically not under control: typically, BPS states are organized in Regge trajectories with arbitrarily high spin \cite{Galakhov:2013oja,CDZunp,Cordova:2015vma}. 
\begin{figure}
\includegraphics[scale=0.6]{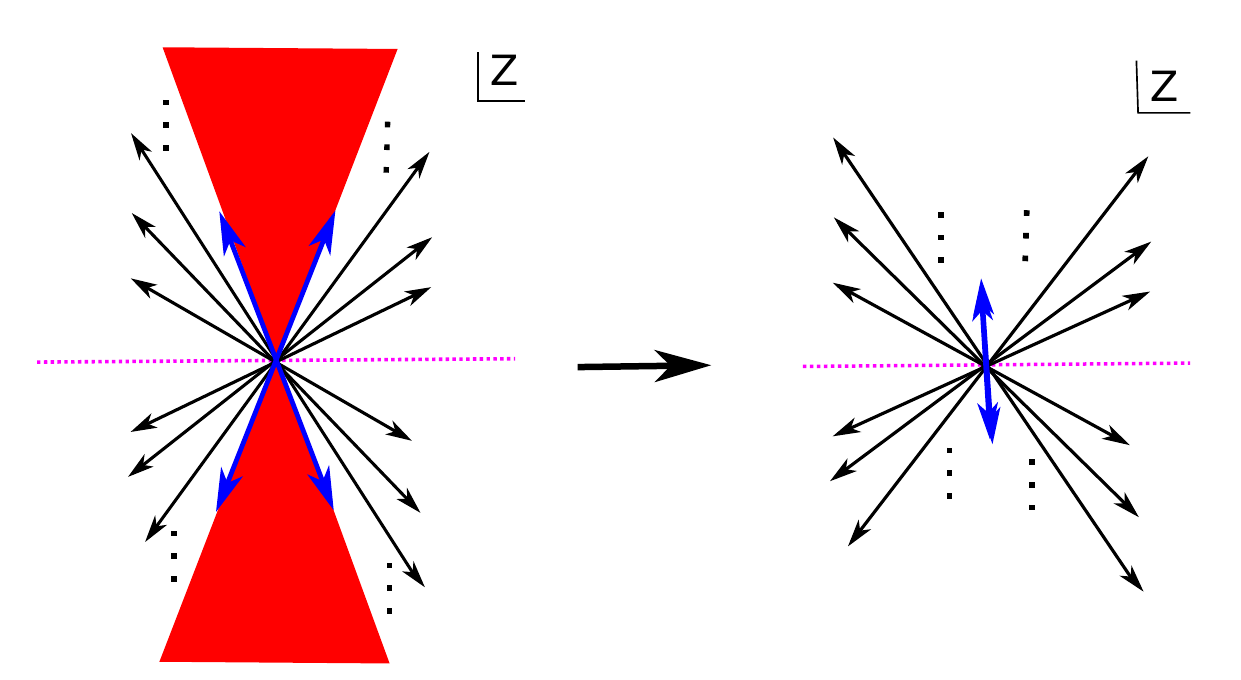}
\caption{Creation of a tame BPS chamber by tuning. \textsc{Left}: a wild BPS spectrum at a generic point of the extended Coulomb branch. We have a sequence of hypermultiplets (black) which accumulate on the edge of the red region, where a vector multiplet is located. Across that boundary the spectrum becomes dense within the red area. \textsc{Right:} By tuning the moduli, all wild BPS states become semi-stable and decay: we can obtain a (finely tuned) tame chamber. Provided the vectormultiplets corresponding to the blue arrows in the figure are mutually local, this chamber is a genuine BPS chamber.}\label{fig:tame}
\end{figure}

\medskip
\noindent
In the context of 4d BPS quivers, several examples of finite BPS chambers are known, which prove particularly useful in computing wall-crossing invariants. In particular, for genuine four-dimensional theories, a well-known conjecture relates a partition function on $S^1 \times S^3$ with the trace of the Kontsevich-Soibelman quantum monodromy operator \cite{Iqbal:2012xm,Cordova:2015nma,Cecotti:2015lab,Cordova:2016uwk,Cordova:2017ohl,Cordova:2017mhb}. Computing said trace becomes a simple exercise for models which admit a finite BPS spectrum (up to subtleties related to regularization --- see the discussion in \cite{Cordova:2016uwk}). For 5d BPS quivers, finite BPS chambers seem to be rather more complicated to obtain, which is related to the presence of the winding modes of the 5d BPS strings on $S^1$. However, the 4d KK theories arising from 5d SCFTs have \textit{tame} BPS chambers, where the spectrum is controlled by roots of (not necessarily simple) affine Lie algebras.\footnote{ In fact, we expect that extended affine Lie algebras will play a role in describing the tame BPS chambers of five-dimensional theories.} In this section, we discuss a tame BPS chamber for the $\KK E_1$ theory, and we present a conjecture about the structure of tame BPS chambers for gauge theories with gauge theory phases characterized by a simple gauge group.
\medskip

We stress that the tame BPS chambers we obtain are extremely fine tuned, and it is possible that they belong to unphysical regions of the space of stability conditions. Nevertheless, their importance stems from an application to determine wall-crossing invariants. Exploiting tame BPS chambers it is a straightforward exercise to compute the Kontsevich-Soibelman quantum monodromy operator for the 4d KK theories of interest. Along the lines of \cite{Iqbal:2012xm,Cordova:2015nma}, we conjecture that the trace of said operator correspond to a partition function for the 5d SCFTs on $T^2 \times S^3$, which we will study in future work.

\subsection{A tame BPS chamber for the $\KK E_1$ theory}\label{subsec:tameE1}

The main aim in this section is to exploit the mutation algorithm of \cite{Alim:2011kw} to compute the BPS spectrum of the $\KK E_1$ theory. For this type of chamber however the mutation algorithm is not sufficient to fully constrain the spectrum, and on top of that a bit of explicit representation theory is needed --- we devote appendix \ref{app:endomorfo} to the relevant proofs of the claim presented here. The reason for this is that the mutation algorithm is very effective for finite chambers. Since finite chambers are populated only by hypermultiplets, one can factor the CPT functor ({\it i.e.} the shift functor, $[1]$) in a sequence of elementary mutations/tilting equivalences. For chambers with higher spins, the mutation method is effective to determine the spectrum of hypers up to the first BPS particle with spin greater than 0. The generic situation (wild chamber) is depicted in Figure \ref{fig:tame} (left): the mutation algorithm cannot access the charges of the particles in the red region, because tilting a vector multiplet leads necessarily to a heart of $\CAT$ that cannot have a quiver description. However, under favorable circumstances, it is possible to fine-tune the stability conditions in such a way that the whole red region of Figure~\ref{fig:tame} (left) becomes marginally stable and decays. If such a stability condition can be found, the image of the BPS spectrum on the $Z$-plane becomes that in Figure \ref{fig:tame} (right) and the chamber is tame. Let us proceed constructing one such example for the 4d KK $E_1$ theory.
\medskip

\noindent The 5d BPS quiver of $\KK E_1$  is given by:
\be
\begin{gathered}
\xymatrix{
B_1 \ar@{=>}[rr]&& W_2\ar@{=>}[dd]\\
\\
W_1\ar@{=>}[uu] &&\ar@{=>}[ll]B_2
}
\end{gathered}
\ee
Let us argue for the existence of a tame chamber of the form:
$$\Delta A_1^{(1)} \oplus \Delta A_1^{(1)}~,$$
exploiting the mutation algorithm. The above notation refers to the fact that this chamber is also Coxeter factorized in the sense of \cite{Cecotti:2011gu}: the resulting spectrum is organized as two copies of the root lattice of the affine Lie algebra  $A_1^{(1)}$. It is easy to see that the sequence of mutations:\footnote{ In our conventions mutation arising from clockwise rotations of the upper half plane are given by
\be
\mu_i(\gamma_k) = \begin{cases}-\gamma_k &\text{if } i = k\\ \gamma_k + [B_{ik}]_+ \gamma_i &\text{otherwise} \end{cases}
\ee
Mutation arising from counterclockwise rotations of the upper half plane are given by
\be
\hat\mu_i(\gamma_k) = \begin{cases}-e_k &\text{if } i = k\\ e_\gamma + [-B_{ik}]_+ \gamma_i &\text{otherwise} \end{cases}
\ee
}
\be
\mathbf{m} = \mu_{W_1}\mu_{W_2}\mu_{B_2}\mu_{B_1}
\ee
maps this quiver back to itself and the corresponding charges to:
\be
\xymatrix{
B_1 + 2\d_u \ar@{=>}[rr]&& - (B_1 + \d_u)\ar@{=>}[dd]\\
\\
-(B_2 + \d_d)\ar@{=>}[uu] &&\ar@{=>}[ll] B_2 + 2 \d_d
}
\ee
where we denote by:
$$\d_u = B_1 + W_2 \qquad\qquad \d_d = B_2 + W_1$$
the minimal imaginary roots of the two Kronecker quivers on the corresponding nodes. These give the two $A_1^{(1)}$ factors of $\mathfrak T$. This corresponds to rotating out of the upper half plane clockwise the charges:
$$B_1, \quad B_2, \quad B_1 + \d_u, \quad B_2 + \d_d$$
The $n$-fold iteration of this mutation sequence corresponds to mapping the quiver to itself with charges
\be
\xymatrix{
B_1 + 2n\d_u \ar@{=>}[rr]&& - (B_1 + (2n-1)\d_u)\ar@{=>}[dd]\\
\\
-(B_2 + (2n-1)\d_d)\ar@{=>}[uu] &&\ar@{=>}[ll] B_2 + 2n \d_d
}
\ee
in correspondence with rotating out a tower of hypermultiplets:
\be
\cdots ,B_1 + (2n-2)\d_u , \quad B_2 + (2n-2)\d_d, \quad B_1+ (2n-1)\d_u, \quad B_2 + (2n-1)\d_d, \cdots
\ee
Note that all these hypers are pairwise mutually local because of linearity of the Dirac pairing and the fact that $\langle\d_u,\d_d \rangle_D = 0$.
Similarly, iterating the sequence of left mutations:
\be
\hat{\mathbf{m}} = \hat\mu_{B_1}\hat\mu_{B_2}\hat\mu_{W_2}\hat\mu_{W_1}
\ee
gives rise to the tower of hypers:
\be
\cdots, W_1 + (2m - 2) \d_d, \quad W_2 + (2m-2)\d_u, \quad W_1 + (2m - 1) \d_d, \quad W_2 + (2m-1)\d_u, \cdots
\ee
This suggests this theory has a tame chamber organized as two copies of the maximal chamber of the Kronecker, i.e. two copies of the weak coupling chamber for pure $SU(2)$ SYM. Clearly the triangular algebra $\mathfrak T$ is encoded in the following biquiver (the numbering on the nodes refers to the ordering for the Tits $S$-matrix):\footnote{ For the relevant definitions, see \cite{Cecotti:2015qha}.}
\be
\begin{gathered}
\xymatrix{
1 \ar@{=>}[rr]&& 4 \ar@{..>}[dd]\ar@<-0.2pc>@{..}[dd]\\
\\
3\ar@{..>}[uu]\ar@<-0.2pc>@{..}[uu] &&\ar@{=>}[ll] 2 
}
\end{gathered}
\ee
has a quadratic form:
\be q_{\mathfrak T} = (b_1-w_2)^2 + (b_2-w_1)^2 + 2 (b_1 w_1+  b_2 w_2)~.
\ee
The roots satisfy $q_{\mathfrak T} \leq 1$. It is clear that:
\bea
&q_{\mathfrak T}(\d_u) = q_{\mathfrak T}(\d_d) = 0\\
&q_{\mathfrak T}(B_1 + k \d_u)= q_{\mathfrak T}(W_2 + k \d_u)= q_{\mathfrak T}(B_2 + k \d_d)= q_{\mathfrak T}(W_1 + k \d_d) = 1\\
\eea
are the only roots. This indeed coincides with two copies of the $A^{(1)}_1$ lattice, as claimed. This chamber has two vector multiplets of charges $\d_u$ and $\d_v$, on top of two towers of hypers with charges as above (and their CPT conjugates). This gives precisely two copies of the 4d $\CN=2$ pure $SU(2)$ SYM weakly coupled spectrum. This is also in agreement with the known properties of the 5d SW curve \cite{Nekrasov:1996cz}. In order to verify that this is the case, we need to exhibit a (formal) set of central charges which is such that the dashed arrows in the biquiver vanish. This is readily done: we can choose:
\be
Z_1 = Z_3 \qquad Z_2 = Z_4 \qquad \arg Z_1 = \arg Z_3 > \arg Z_2 = \arg Z_4~.
\ee
Note that these are severely fine-tuned, but that no BPS particle is on a wall of marginal stability: all the particles with aligned charges are mutually local. In Appendix \ref{app:endomorfo}, we complete the proof that the arrows $A_2,A_4$ as well as $B_2,B_4$ always vanish for a representation to be stable in this chamber.

\subsection{Tame chambers for the 4d KK $G$ SYM theories}
Exploiting mirror symmetry, one can show that the 5d BPS quivers for the pure $SU(N)_k$ gauge theories have the form  \cite{PROG}:\footnote{We refer to \protect\cite{Cecotti:2012jx} for the construction of these quivers from a IIB geometric engineering perspective. For the definition of the $\boxtimes$ operation our readers can consult \cite{Cecotti:2010fi}.}
\be
A(1,1)\boxtimes A(p,q)~,
\ee
when $N = p+q$ and $k = q-p$. Moreover, for the pure 5d $G$ gauge theory with simple simply laced $G$ the 5d BPS quiver takes the form:
\be
A(1,1) \boxtimes \widehat G\,.
\ee
Along the lines of \cite{Cecotti:2010fi} we are lead to conjecture that these gauge theories all have at least two (Coxeter factorized) tame chambers with spectrum encoded in the roots of the affine Lie algebras:
\be
\Delta G^{(1)} \oplus \Delta G^{(1)}~, \qquad \qquad \underbrace{\Delta A^{(1)} \oplus\Delta A^{(1)} \oplus \cdots \Delta A^{(1)}}_{r_G + 1 \text{ times}}~,
\ee
respectively. More in general, for all the 5d BPS quivers of type $\widehat H \boxtimes \widehat G$, we expect to find Coxeter factorized tame chambers corresponding to $r_H+1$ copies of $\Delta G^{(1)}$, in analogy with the spectrum found for the $(G,G')$ 4d SCFTs in \cite{Cecotti:2010fi}.

\subsection*{Acknowledgements}  We would like to thank Fabio Apruzzi, Antoine Bourget, Stefano Cremonesi, Richard Eager, Marco Fazzi, Sebastian Franco, I\~{n}aki Garc\'ia Etxebarria, Chris Herzog, Tyler Kelly, Heeyeon Kim, Pietro Longhi, Jan Manschot, Sakura Sch\"afer-Nameki, and Mauricio Romo for interesting discussions and correspondence. MDZ especially thanks Sergio Cecotti and Cumrun Vafa for sharing their deep insights about the categorial approach to $\CN=2$ SQFTs and its relation to geometric engineering. CC is a Royal Society University Research Fellow and a Research Fellow at St John's College, Oxford.

\appendix


\section{A brief review of controlled subcategories}\label{app:control}
Here we review the concept of controlled subcategory, as presented in \cite{LenzingICTP}. Consider an abelian category $\mathcal A$. Take the Gr\"othendieck group $K_0\mathcal A$. A {\it control} is a map:
\be
\lambda : K_0 \mathcal A \longrightarrow \mathbb Z~,
\ee
 which is linear and additive with respect to exact sequences, meaning that for $A,B,C \in \mathcal A$ and $A \to B \to C$ exact, $\lambda(A) - \lambda(B) + \lambda(C) = 0$. Here and in what follows, we abuse notation and write $\lambda(A)$ for $\lambda([A])$, where $[A]$ is the class of $A$ in $K_0 \mathcal A$. Let $\CA_\lambda$ denote the full subcategory of $\CA$ consisting of all objects that satisfy $\lambda(A) =0$ and $\lambda(A') \leq \lambda(A)$ for all subobjects $A'\subseteq A$.

 Thus, $\CA_\lambda$ is the subcategory of $\CA$ controlled by $\lambda$: it is an exact subcategory, it is closed under extensions, and it is abelian. To derive these properties, one can proceed as follows. We need to show that the controlled subcategory is closed under kernels, cokernels and extensions. Consider the exact sequence:
\be
0 \to \text{Ker } u \to A \xrightarrow{u} B~,
\ee
where $A,B \in \CA_\lambda$. By definition we have that $\lambda(\text{Ker u})\leq 0$. Moreover $\lambda(\text{im } u)\leq 0$ because $\text{im } u \subseteq B$. However we have that, by exactness:
\be\label{eq:exacter}
0 \to \text{Ker } u \to A \xrightarrow{u} \text{im } u \to 0 \quad\Rightarrow\quad \lambda(\text{Ker } u) - \underbrace{\lambda(A)}_{=0} + \lambda(\text{im } u) = 0~,
\ee
which implies that $\lambda(\text{im } u) \geq 0$. Therefore $\lambda(\text{im } u) = 0$. By \eqref{eq:exacter} then, $\lambda(\text{Ker } u) = 0$, which shows that $\CA_\lambda$ is closed under kernel. At this point, we can consider the exact sequence:
\be\label{eq:exacter2}
0 \to \text{Ker } u \to A \xrightarrow{u} B \to \text{coker } u \to 0 \quad\Rightarrow\quad \lambda(\text{coker } u) = 0~,
\ee
because all other objects are in $\CA_\lambda$, therefore $\CA_\lambda$ is closed also under cokernel. Finally, by definition, $E$ is an extension of $A$ and $B$ if the sequence $0 \to A \to E \to B \to 0$ is exact, and therefore if $A,B \in \CA_\lambda$, then $E\in \CA_\lambda$ also.


\section{BPS states for the conifold}\label{app:conifBPS}
The conifold quiver is:
\be
\begin{gathered}
\xymatrix{\bullet_1 \ar@/^1.5pc/@{=>}[rr]^{A_1}_{B_1}&&\bullet_2\ar@/^1.5pc/@{=>}[ll]^{A_2}_{B_2}}\\
W=B_2 B_1 A_2 A_1 - A_2 B_1 B_2 A_1~.
\end{gathered}\ee
Here and in the following Appendices, as opposed to what we do in the main text, we are using the mathematicians' conventions, where arrows are ordered according to the composition of the corresponding maps. By symmetry of the problem the two BPS chambers for this theory have identical structure. We therefore assume that
\be\label{eq:conifstab}
\arg Z_1 > \arg Z_2~,
\ee
from now on.

\medskip
\noindent The relations for this algebra are:
\be
\begin{aligned}
&B_2 B_1 A_2 = A_2 B_1 B_2~, \qquad A_1 B_2 B_1 = B_1 B_2 A_1~,\\
&A_2 A_1 B_2 = B_2 A_1 A_2~, \qquad B_1 A_2 A_1 = A_1 A_2 B_1~.
\end{aligned}
\ee
It follows that any representation has the following endomorphisms:
\be\label{eq:endomoconifo}
\begin{aligned}
&\Xi_1 \equiv (A_2 A_1,A_1 A_2)~, \qquad \Xi_2 = (B_2 B_1,B_1 B_2)~, \\
&\Xi_3 \equiv (B_2 A_1,A_1 B_2)~, \qquad \Xi_4 = (A_2 B_1,B_1 A_2)~.
\end{aligned}
\ee
By the Schur lemma, $\Xi_i = \lambda_i \text{id}_X$ for every representation $\CO$. We have the following cases:
\begin{itemize}
\item {\bf Case 1:} $\lambda_i = 0$ for all $i=1,...,4$. We claim that stability for this class of representations implies that $A_2 = B_2 = 0$. Indeed, assume $A_2 \neq 0$. Consider a vector $0\neq v_1 = A_2 w_1$ for some $w_1 \in X_2$. Since $\lambda_1=0$, we have that $A_1 v_1 = 0$. Since $\lambda_4 = 0$, we have that $B_1 v_1 =0 $ and therefore $\C v_1 \subset X_1$ is a destabilizing subrepresentation. Similarly, assuming $B_2 \neq 0$, leads to a contradiction with stability using $\lambda_2 = \lambda_3 =0$.

Therefore this class of representations contributes to the BPS spectrum as the representations of the Kronecker quiver with arrows $A_1$ and $B_1$. We obtain hypermultiplets with charges:
\be\label{eq:hyperszcn}
\gamma_1 + n \delta~, \qquad \gamma_2 + (n-1) \delta~,  \qquad n \in\mathbb Z_{\geq 0}~,
\ee
and a vector multiplet with charge $\delta = \gamma_1 + \gamma_2$. The states obtained in \eqref{eq:hyperszcn} are the KK towers corresponding to the KK reduction of one 5d hypermultiplet on a circle. The vector multiplet with charge $\delta$ also comes with a KK tower; the corresponding states are obtained in {\bf Case 3} below.

\item {\bf Case 2:} At least one $\lambda_i \neq 0$ but not all. In this case we want to argue that there is always a destabilizing subrepresentation. 

We consider here the case $\lambda_1 =0$ therefore $A_1 A_2 = A_2 A_1 = 0$. By symmetry the other cases follows from a similar reasoning. Let $0\neq v_1 = A_2 w_1$. In this case $A_1 v_1 = 0$ but $B_1 v_1 = w_2 \neq 0$ (if zero we would have a destabilizing representation as above). However notice that $w_2 = B_1 v_1 = B_1 A_2 w_1 = \lambda_4 w_1$. Consider now $B_2 w_2 = B_2 B_1 v_1 = \lambda_2 v_1$. On one hand now, $A_1 B_2 w_2 = \lambda_3 w_2$, on the other $A_1 B_2 w_2 = \lambda_2 A_1 v_1 = 0$. This implies that $\lambda_3 = 0$ and therefore $A_1 B_2 = B_2 A_1 = 0$. 

If $\lambda_2$ and/or $\lambda_4$ are non-zero, this implies that $\dim X_1 = \dim X_2$ for every representation. Consider now $\hat w \in X_2$ such that $A_2 \hat w = \hat v$ and $B_2 \hat w = \hat v'$. $A_1 \hat v = A_1 \hat v' = 0$, while $B_1 \hat v = \lambda_4 \hat w$ and $B_1 \hat v' = \lambda_2 \hat w$ and therefore $(\C \hat v + \C \hat v',\C \hat w) \subset (X_1,X_2)$ is always a subrepresentation of X which is always destabilizing.

\item  {\bf Case 3:} All $\lambda_i \neq 0$. In this case we have that all arrows are isomorphisms and there are no destabilizing subrepresentations. We have BPS states with charges
\be
k \delta~, \qquad k\in\mathbb Z_{\geq 1}~.
\ee
Let us argue that these states are all associated to vector multiplets. It is well known that the Higgs branch moduli space for one D0-brane that probes the conifold is the conifold itself. This is because in this case the F-term relations are not constraining, and one can simply proceed by forming the gauge invariants
\be\label{eq:gaugeinvsconi}
X = A_1 A_2~, \qquad Y = B_1 B_2~, \qquad U = A_1 B_2~, \qquad V = A_2 B_1~,
\ee
and it is clear that these satisfy
\be
X Y = U V~,
\ee
which is the equation for the conifold. Turning on FI terms (and hence considering a non-zero central charge) corresponds geometrically to resolving the conifold, physically to consider a smooth moduli space, which is always easier to quantize. But how do we quantize a non-compact moduli space? A very natural proposal is to consider compactly supported cohomology (as suggested by \cite{MeinhardtReineke}). In the case of the resolved conifold, we only have one exceptional $\mathbb P^1$ in the geometry, and therefore the compactly supported cohomology in this case is just consisting of two states, corresponding to the cohomology of $\mathbb P^1$. This is a doublet for $SU(2)_\text{spin}$ and a singlet for $SU(2)_R$: tensoring this representation with the half-hypermultiplet we indeed do obtain a vector multiplet. Consider now the case of $k$ D0-branes. Naively, from string theory, one would expect to be able to choose generic values for the eigenvalues of the gauge invariants in equation \eqref{eq:gaugeinvsconi} so that to obtain a moduli space $\text{Sym}^k \h\CC_0$. However, most of the configurations occurring there would be only marginally-stable. This can be argued as follows. The gauge invariants in \eqref{eq:gaugeinvsconi} coincide with the endomorphisms in \eqref{eq:endomoconifo}. By the Schur lemma, representations such that these matrices have distinct eigenvalues are necessarily reducible, and hence cannot be stable. The requirement of irreducibility, then, reduces the moduli space to just one copy of $\h\CC_0$ and, by the same token, the compactly-supported cohomology contributes the same states as a $\mathbb P^1$.

\end{itemize}

\section{Comments on the electric subcategory of the $E_1$ SCFT }\label{app:F0ele}

Consider the $E_1$ 5d SCFT, which has a well known $SU(2)_0$ gauge theory phase. The 5d BPS quiver for this theory is the so-called $\mathbb{F}_0$ quiver 
\be\begin{gathered}
\xymatrix{\bullet_1 \ar@{=>}[rr]^{A_1}_{B_1}&& \circ_2\ar@{=>}[dd]^{A_2}_{B_2}\\\\\circ_1\ar@{=>}[uu]^{A_4}_{B_4}&&\bullet_2\ar@{=>}[ll]^{A_3}_{B_3}} 
\\
W= A_4 A_3A_2A_1 + B_4 B_3 B_2 B_1 -  A_4 B_3 A_2 B_1 - B_4 A_3 B_2A_1~.
\end{gathered}
\ee
For any quiver SQM, the dimension of the moduli space associated to a given stable representation can be estimated as:
\be
\dim \CM = \#\, \text{fields} - (\# \, \text{independent F-terms} + \text{gauge group dimension}) + 1~.
\ee
The $\mathbb{F}_0$ quiver has four obvious electric subcategories, given by a choice of Kronecker subquiver, that each corresponds to a 4d $\CN=2$ $SU(2)$ subsector.
\medskip

Let us proceed by analyzing one of such electric subcategories in more detail. Choose the subquiver:\be\label{eq:F0kruno}\xymatrix{\circ_1 \ar@{=>}[rr] &&\bullet_1}\ee
The corresponding W-boson is given by the representation with dimension vector:
\be
\boldsymbol{\d} = (\dim X_{\circ_1}, \dim X_{\bullet_1}, \dim X_{\circ_2}, \dim X_{\bullet_2})=(1,1,0,0)~,
\ee
which comes in a $\mathbb P^1$ moduli space and therefore corresponds to a vector multiplet. The elementary electric charge is given by the vector:
 $$\mathbf{q} = \tfrac{1}{2} (1,1,0,0) \in \Gamma \otimes \mathbb Q~.$$
 The associated magnetic charge is:
\be\label{eq:magnF0}
\mathbf{m}(X) = \dim X_{\bullet_1} - \dim X_{\circ_1}  + \dim X_{\circ_2} -  \dim X_{\bullet_2} = N_2-N_1 +N_3 - N_4~,
\ee
where we have set
$$(N_1,N_2,N_3,N_4) = (\dim X_{\circ_1}, \dim X_{\bullet_1}, \dim X_{\circ_2}, \dim X_{\bullet_2})~,$$
for ease of notation.

Consider the electric subcategory associated to the magnetic charge in \eqref{eq:magnF0}: we have to impose $\mathbf{m}(X) = 0$. This can be achieved in several (equivalent) ways, forcing the restriction of our representation to each pair of non-overlapping Kronecker subquivers to have the same dimensions. Consider for instance the pair given by \eqref{eq:F0kruno} and  
\be\label{eq:F0krdue}\xymatrix{\circ_2 \ar@{=>}[rr] &&\bullet_2}\ee
At this point, by a standard argument (see {\it e.g.} Appendix D of \cite{Cecotti:2012va}), we expect that the corresponding Kronecker representations are forced to be direct sums of regulars. A crucial property of regular modules is that at least one of the two arrows is always an isomorphism, which we can use to identify the two nodes. As a result, this component of the electric subcategory can be written as\footnote{$\,$ Here we are using the Ringel notation $\CA \bigvee \CB$ to denote the category generated by all the direct sums  and extensions of the objects of $\CA$ and $\CB$.}
\be
\mathscr E \simeq \bigvee_{\lambda \, \in \, \mathbb F_0} \mathscr E_\lambda
\ee
where we are using that $\mathbb F_0 \simeq \mathbb P^1 \times \mathbb P^1$ and the first $\mathbb P^1$ is parametrized by $[A_2:B_2]$, while the second is given by $[A_4:B_4]$. Some aspects of the detailed structure of the factors are discussed below.

Moreover, by symmetry of the $\mathbb F_0$ quiver, all electric categories obtained by choosing $SU(2)$ W-bosons in terms of a Kronecker subquiver are equivalent. This is a manifestation of the fact that the 5d $SU(2)_0$ theory is self-dual at the level of the spectrum of BPS states of the corresponding KK theory.

Consider first the case when $B_4$ and $B_2$ are isomorphisms. Then, we obtain the quiver for the controlled electric subcategory:
\be\begin{gathered}
\xymatrix{\ast_1 \ar@(dl,ul)[]^{A_4} \ar@/^0.9pc/@{=>}[rr]^{A_1}_{B_1}&& \ast_2 \ar@/^0.9pc/@{=>}[ll]^{A_3}_{B_3} \ar@(dr,ur)[]_{A_2}} 
\\
W= A _4 A_3 A_2 A_1 + B_3 B_1 -  A_4 B_3 A_2 B_1 - A_3 A_1~.
\end{gathered}
\ee
Note that the arrows $A_1$ and $B_1$ cannot be integrated away. The relations from this effective potential are:
$$\begin{gathered}
A_3 A_2 A_1 = B_2 A_2 B_1~, \qquad A_1 = A_2 A_1 A_4~, \qquad A_1 A_3 A_4 = B_1 A_4 B_3~, \\
A_3 = A _4 A_3 A_2~, \qquad B_1 = A_4 B_1 A_2~, \qquad B_3 = A _2 B_3 A_4~.
\end{gathered}
$$
Consider a stability condition for which $\arg Z_{\ast_1} >\arg Z_{\ast_2}$. Let us then remark that the $W$-boson belongs to this category: it is given by the state with dimension vector $(1,0)$.

Another patch for the electric category is obtained by considering the case $B_4$ and $A_2$ are isomorphism, in which case we have:
\be\begin{gathered} 
\xymatrix{\ast_1 \ar@(dl,ul)[]^{A_4} \ar@/^0.9pc/@{=>}[rr]^{A_1}_{B_1}&& \ast_2 \ar@/^0.9pc/@{=>}[ll]^{A_3}_{B_3} \ar@(dr,ur)[]_{B_2}} 
\\
W= A_4 A_3 A_1 + B_3 B_2 B_1 -  A_4 B_3 B_1 -  A_3 B_2A_1
\end{gathered}
\ee
as the effective quiver description. All other patches are equivalent.

\medskip

\noindent The objects of these electric categories have are all mutually local by construction: all stability conditions compatible with the stability of the gauge subsector identified by the conditions G1, G2 and G3  of section~\ref{subsec: electric subcat} lead to the same electric subcategory.

\section{Further comments on the tame chamber of the $E_1$ theory}\label{app:endomorfo}
In this Appendix, we collect some further details on the computation of the tame chamber of the rank-one $E_1$ theory, as discussed in section~\ref{subsec:tameE1}. Consider the 5d BPS quiver:
\be
\begin{gathered}
\xymatrix{
1 \ar@{=>}[rr]^{A_1}_{B_1}&& 2 \ar@{=>}[dd]^{A_2}_{B_2}\\
\\
4\ar@{=>}[uu]^{A_4}_{B_4} && 3\ar@{=>}[ll]^{A_3}_{B_3}}
\end{gathered}
\ee
Relabeling the arrows for notational convenience we have the superpotential:
$$ W = A_4 A_3 A_2 A_1 + B_4 B_3 B_2 B_1 - B_4 A_3 B_2 A_1 - A_4 B_3 A_2 B_1~.$$
The structure of endomorphisms of quiver representations is very similar to the ones of section~\ref{app:conifBPS}, which is related to the fact that the $\mathbb{F}_0$ geometry is a $\Z_2$ orbifold of the conifold.

\medskip
\noindent
The corresponding relations are (here and in the rest of this section $j \sim j+4$):
\be\label{eq:F1relations}
\begin{aligned}
\partial_{A_j} W &= A_{j+3} A_{j+2} A_{j+1} - B_{j+3} A_{j+2} B_{j+1} = 0~,\\
\partial_{B_j} W &= B_{j+3} B_{j+2} B_{j+1} - A_{j+3} B_{j+2} A_{j+1} = 0~.\\
\end{aligned}
\ee
\noindent {\bf Lemma.} \textit{Any representation $\CO$ of this quiver with superpotential has the following 9 endomorphisms:}
\bea
&\Psi^{(0)}_j = A_{j+3} A_{j+2} A_{j+1} A_j\\
&\Psi^{(k)}_j = \begin{cases}  A_{j+3} B_{j+2} B_{j+1} A_j & j = k \\ A_{j+3} A_{j+2} B_{j+1} B_j & j = k+1\\
 B_{j+3} A_{j+2} A_{j+1} B_j & j = k+2\\
B_{j+3} B_{j+2} A_{j+1} A_j & j = k +3 \\  \end{cases}\qquad\qquad k = 1,2,3,4\\
&\Phi^{(k)}_j = \begin{cases} A_{j+3} A_{j+2} A_{j+1} B_j & j = k \\
B_{j+3} A_{j+2} A_{j+1} A_{j} & j = k+1\\
A_{j+3} B_{j+2} A_{j+1} A_{j} & j = k+2\\
A_{j+3} A_{j+2} B_{j+1} A_{j}& j = k+3
 \end{cases}\qquad\qquad k = 1,2,3,4
\eea
\noindent{\bf Proof.} Let's prove this for $\Psi^{(0)}$ first. We have:
\be
A_j \Psi^{(0)}_j = A_j  A_{j+3} A_{j+2} A_{j+1} A_{j} = \Psi^{(0)}_{j+1} A_j \ee
and, moreover, the relations from \eqref{eq:F1relations} give:
\be
\Psi^{(0)}_j = B_{j+3} A_{j+2} B_{j+1} A_j =  A_{j+3} B_{j+2} A_{j+1} B_j  = B_{j+3} B_{j+2} B_{j+1} B_j~,
\ee
which implies:
\be
B_j \Psi^{(0)}_j =  B_j B_{j+3} B_{j+2} B_{j+1} B_j  = \Psi^{(0)}_{j+1} B_j~,
\ee
and so $\Psi^{(0)} \in \text{End } \CO$. Now, consider $\Psi^{(k)}$. The relevant commutation relations are proved as follows, exploiting that $k+4 \sim k$:
\bea
A_{k} \Psi^{(k)}_{k}  &= A_{k} A_{k+3} B_{k+2} B_{k+1} A_k = \Psi^{(k)}_{k+1} A_k~,\\
B_k \Psi^{(k)}_{k}  &= B_{k} A_{k+3} B_{k+2} B_{k+1} A_k = A_{k} A_{k+3} A_{k+2} B_{k+1} A_k \\&=A_{k} A_{k+3} B_{k+2} B_{k+1} B_k = \Psi^{(k)}_{k+1} B_k~,
\eea
\bea
A_{k+1} \Psi^{(k)}_{k+1}  &= A_{k+1} A_{k+4} A_{k+3} B_{k+2} B_{k+1} = B_{k+1} A_{k+4} B_{k+3} B_{k+2} B_{k+1}\\ &= B_{k+1} A_{k+4} A_{k+3} B_{k+2} A_{k+1} = \Psi^{(k)}_{k+2} A_{k+1}~,\\
B_{k+1} \Psi^{(k)}_{k+1}  &= B_{k+1} A_{k+4} A_{k+3} B_{k+2} B_{k+1} =\Psi^{(k)}_{k+2} B_{k+1}~,\\
\eea
\bea
A_{k+2} \Psi^{(k)}_{k+2}  &= A_{k+2} B_{k+1} A_{k+4} A_{k+3} B_{k+2} = B_{k+2} B_{k+1} B_{k+4} A_{k+3} B_{k+2}~,\\
&=B_{k+2} B_{k+1} A_{k+4} A_{k+3} A_{k+2} = \Psi^{(k)}_{k+3} A_{k+2}~,\\
B_{k+2} \Psi^{(k)}_{k+2}  &= B_{k+2} B_{k+1} A_{k+4} A_{k+3} B_{k+2} = \Psi^{(k)}_{k+3} B_{k+2}~,
\eea
\bea
A_{k+3} \Psi^{(k)}_{k+3} &=  A_{k+3} B_{k+2} B_{k+1} A_{k+4} A_{k+3} =  \Psi^{(k)}_{k} A_{k+3}~,\\
B_{k+3} \Psi^{(k)}_{k+3} &=  B_{k+3} B_{k+2} B_{k+1} A_{k+4} A_{k+3} = A_{k+3} B_{k+2} A_{k+1} A_{k+4} A_{k+3}~,\\
&=A_{k+3} B_{k+2} B_{k+1} A_{k+4} B_{k+3} = \Psi^{(k)}_{k} B_{k+3}~.
\eea
Now, consider $\Phi^{(k)}$. The relevant commutation relations are proved as follows, exploiting that $k+4 \sim k$:
\bea
A_{k} \Phi^{(k)}_{k}  &= A_{k} A_{k+3} A_{k+2} A_{k+1} B_k = B_{k} A_{k+3} B_{k+2} A_{k+1} B_k~,\\
&= B_{k} A_{k+3} A_{k+2} A_{k+1} A_k = \Phi^{(k)}_{k+1} A_k~, \\
B_{k} \Phi^{(k)}_{k}  &= B_{k} A_{k+3} A_{k+2} A_{k+1} B_k = \Phi^{(k)}_{k+1} B_k~,\\
\eea
\bea
A_{k+1} \Phi^{(k)}_{k+1}  &= A_{k+1} B_{k} A_{k+3} A_{k+2} A_{k+1} = \Phi^{(k)}_{k+2} A_{k+1}~,\\
B_{k+1} \Phi^{(k)}_{k+1}  &= B_{k+1} B_{k} A_{k+3} A_{k+2} A_{k+1} = B_{k+1} B_{k} B_{k+3} A_{k+2} B_{k+1}~,\\
&=A_{k+1} B_{k} A_{k+3} A_{k+2} B_{k+1} = \Phi^{(k)}_{k+2} B_{k+1}~,
\eea
\bea
A_{k+2} \Phi^{(k)}_{k+2}  &= A_{k+2} A_{k+1} B_{k} A_{k+3} A_{k+2} = \Phi^{(k)}_{k+3} A_{k+2}~,\\
B_{k+2} \Phi^{(k)}_{k+2} &= B_{k+2} A_{k+1} B_{k} A_{k+3} A_{k+2} = A_{k+2} A_{k+1} A_{k} A_{k+3} A_{k+2}~, \\
&=A_{k+2} A_{k+1} B_{k} A_{k+3} B_{k+2}  =  \Phi^{(k)}_{k+3} B_{k+2}~,
\eea
\bea
A_{k+3} \Phi^{(k)}_{k+3}  &= A_{k+3} A_{k+2} A_{k+1} B_{k} A_{k+3}  =  \Phi^{(k)}_{k} A_{k+3}~,\\
B_{k+3} \Phi^{(k)}_{k+3}  &= B_{k+3} A_{k+2} A_{k+1} B_{k} A_{k+3}  =   B_{k+3} A_{k+2} B_{k+1} B_{k} B_{k+3}~,\\&= A_{k+3} A_{k+2} A_{k+1} B_{k} B_{k+3}  =\Phi^{(k)}_{k} B_{k+3}~.
\eea
This concludes the proof of our lemma. $\square$
\bigskip

\noindent In order to be stable, a representation must be a brick, therefore $\text{End}(\CO) \simeq \C$, which implies that:
 $$\Psi^{(k)} = \lambda_k \, \text{id}_\CO~, \qquad \Phi^{(\ell)} = \mu_\ell \, \text{id}_\CO~,$$ for fixed $\lambda_k,\mu_\ell \in \C$ (Schur lemma). 
\medskip

\noindent A consequence of our lemma is that, for a representation to be a brick (which is a necessary condition for stability), the composition of every four consecutive arrows in this quiver is either proportional to the identity or zero. At this point, a crucial remark is in order: since the Jacobian algebra is not finite dimensional, we need to focus on the nilpotent representations of $\CJ$ only, as pointed out in \cite{Caorsi:2016ebt}. At the level of bricks, this implies that each endomorphism must equal zero.

\medskip

\begin{figure}
\begin{center}
\includegraphics[scale=0.4]{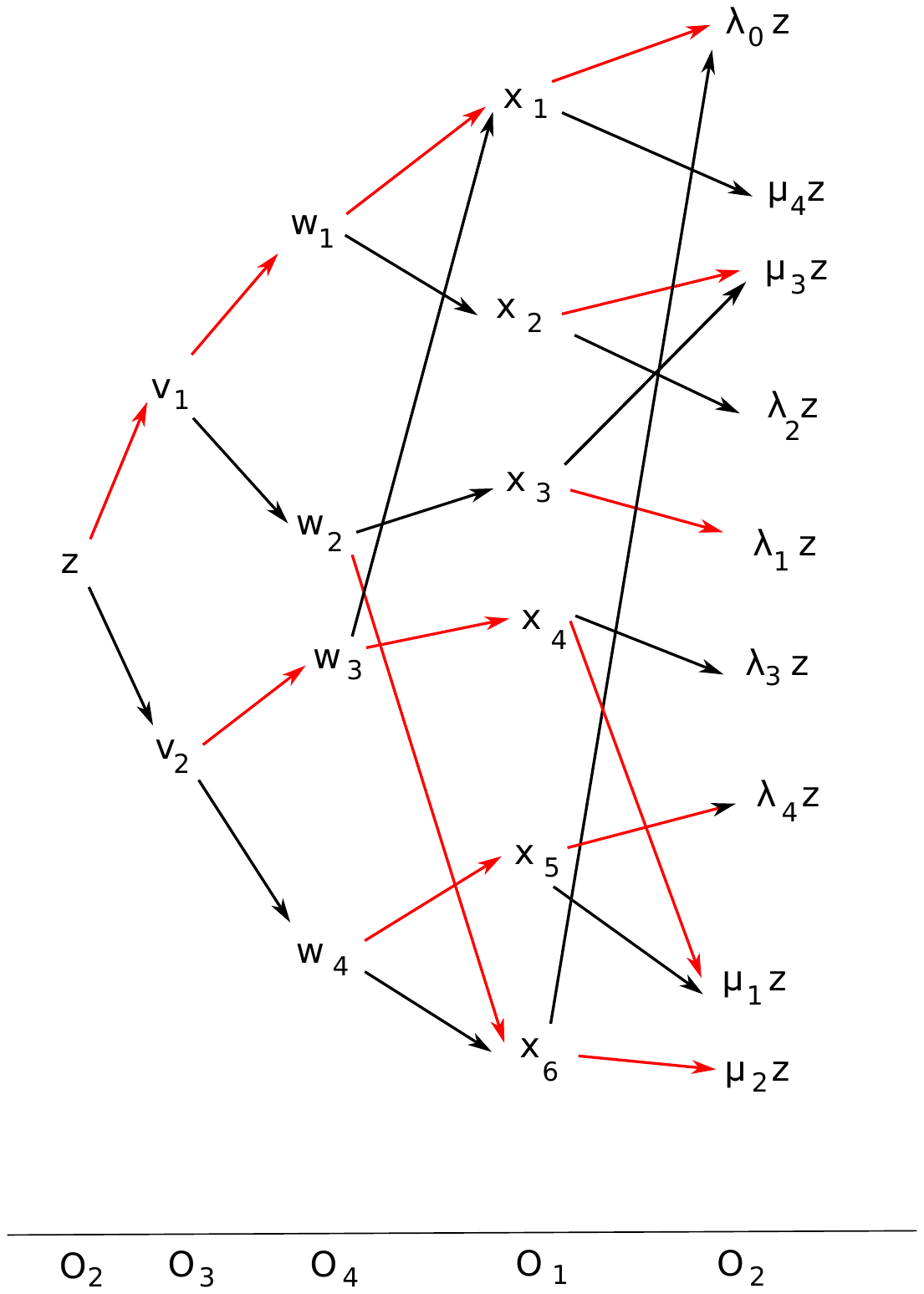}
\end{center}
\caption{The generic subrepresentation of the form $\CJ z$ for $z \in \CO_2$, imposing the Schur lemma. We draw the  arrows $A_i$ in red and the arrows $B_i$ in black. Below, we indicate the vector space $\CO_i$. We see that, for generic values of $\lambda_k$ and $\mu_\ell$, this representation has dimension vector $(6,1,2,4)$. For nilpotent modules, $\lambda_k= \mu_\ell = 0$ and we see that this induces a destabilizing subrepresentation, $\CS_1$.}\label{fig:A2destaba}
\end{figure}

\noindent The mutation sequence we constructed in the main body of the text suggests to consider the stability conditions:
\be\label{eq:stabilitah1}
\arg Z_1 = \arg Z_3 > \arg Z_2 = \arg Z_4~,
\ee
\be\label{eq:stabilitah2}
\arg{Z_1 + Z_2} = \arg{Z_3 + Z_4}~.
\ee
Our aim here is to show that with this choice of stability condition, $\CO$ is stable only if:
\be A_2 = B_2 = A_4 = B_4 = 0\,.\ee
Consider first the arrows $A_2$ and $B_2$. Suppose that at least $A_2$ is non-zero. Then there is at least one element $z \in \CO_2$ such that $0 \neq v_1 = A_2 z \in \CO_3$. Because of the Schur lemma combined with the endomorphism of each representation one can construct the submodule generated by $\C z \in \CO_2$, let us denote it by $\CS$. The structure of $\CS$ is in shown in Figure~\ref{fig:A2destaba}. It is clear that:
\be
\dim \CS \leq (6,1,2,4)~,
\ee
where the lower values can be obtained for less generic representations. Since we restrict to nilpotent modules, we impose:
\be
\lambda_k = \mu_\ell = 0~.
\ee
If that is the case, then $\CS_1$, the projection of $\CS$ on node 1, is always a subrepresentation by itself and it is destabilzing. 

By symmetry, we can similarly argue that also $B_4$ and $A_4$ have to vanish. In this case, we obtain a destabilizing subrepresentation supported on node 3. This completes the proof alluded to in the main text.

\bibliographystyle{utphys}
\bibliography{bib5d}{}

\end{document}